\documentclass[authoryear,preprint,12pt]{elsarticle}

\usepackage{amssymb}
\usepackage{amscd}
\usepackage{amsmath}
\usepackage{amsfonts}
\usepackage[all,knot,poly]{xy}
\xyoption{arc}
\usepackage{latexsym}
\usepackage[dvipsnames,usenames]{color}
\usepackage[colorlinks=true,
pdfstartview=FitV, linkcolor=cyan,
citecolor=cyan, urlcolor=cyan]{hyperref}
\journal{}
\catcode`\=\active \def {\`e}          % option `, then  e
\catcode`\Ž=\active \def Ž{\'e}          % option «, then  e
\catcode`\ˆ=\active \def ˆ{\`a}          % option `, then  a
\catcode`\"=\active \def "{\`{\i}}          % option `, then  i
\catcode`\˜=\active \def ˜{\`o}          % option `, then  o
\catcode`\=\active \def {\`u}          % option `, then  u
\catcode`\"=\active \def "{\^{\i}}         % option i,  then  i
\def\EVE{\textsc{E}}

\def\SPA{\textsc{sp}}
\def\metric{\Bg}
\def\emf{\emph{emf}}
\def\moto{{\Bvarphi}}
\def\ether{\emph{aether}}
\def\meas{\textsc{meas}}
\def\perm{\textsl{p}}
\def\Perm{\cP}
\def\ele{\textsc{ele}}
\def\mag{\textsc{mag}}
\def\inner{{\textsc{in}}}
\def\outer{{\textsc{out}}}
\def\simp{s}
\def\Simp{\textsc{Simp}}

\def\signum{\textsc{sign}}

\def\orient{\textsc{Or}}

\def\orientpairx{\set{\orient^+_\Bx,\orient^-_\Bx}}
\def\volform{\Bmu}
\def\volformn{\Bmu^n}

\def\Faraduno{\Boo^1_{\MMM,\BF}}

\def\projE{\BP_{\EU}}
\def\projI{\BP_{\II}}

\def\surf{\BSigma}
\def\surfout{\BSigma}
\def\CID{\persone{CID}}
\def\VPB{\persone{VPB}}
\def\EXForms{\BLambda}
\def\Exter{\Boo^k}

\def\Current{\Boo^3_{\MMM,\BA}}
\def\Magnet{\Boo^3_{\MMM,\BF}}
\def\Faraduno{\Boo^1_{\MMM,\BF}}
\def\Faraddue{\Boo^2_{\MMM,\BF}}
\def\Amper{\Boo^2_{\MMM,\BA}}
\def\lightvel{c}
\def\polar#1{{#1}^\circ}
\def\SIMUL{\BN}
\def\obs{\metricamma}
\def\metricM{\metric_\MMM}

\def\relat{\Bzeta}

\def\vecf{\Bh}
\def\identvec{\Br}
\def\homotopy{\Bh}
\def\nocharge{\BZ}
\def\potmag{\BF}
\def\potspost{\BA}
\def\elecurr{{\BJ_\BE}}
\def\elecurrform{\Boo^2_{\elecurr}}
\def\elecurrformttt{\Boo^2_{\elecurr,\ttt}}
\def\magind{{\BB}}
\def\magindttt{{\BB_\ttt}}
\def\eledisp{\BD}
\def\eledispttt{{\BD_\ttt}}
\def\magvector{{\BH}}
\def\elevector{\BE}
\def\elefield{\Boo^1_{\elevector}}
\def\elefieldttt{\Boo^1_{\elevector,\ttt}}

\def\eleflux{\Boo^2_{\eledisp}}
\def\elefluxttt{\Boo^2_{\eledisp,\ttt}}

\def\elefluxtau{\Boo^2_{\eledisp,\tau}}
\def\magwind{\Boo^1_\BH}
\def\magwindttt{\Boo^1_{\BH,\ttt}}
\def\magcurl{\Boo^2_{\BB}}
\def\magcurltau{\Boo^2_{\BB,\tau}}
\def\magcurlttt{\Boo^2_{\BB,\ttt}}
\def\faradpot{\Boo^1_{\potmag}}
\def\faradpottau{\Boo^1_{\potmag,\tau}}
\def\faradpotttt{\Boo^1_{\potmag,\ttt}}

\def\amperpotttt{\Boo^1_{\potspost,\ttt}}

\def\chargescalarttt{{\rho_{\BE,\ttt}}}
\def\chargescalartau{{\rho_{\BE,\tau}}}

\def\elechargeform{{\Brho^3_\BE}}
\def\elechargeformttt{\Brho^3_{\BE,\ttt}}
\def\elechargeformtau{{\Brho^3_{\BE,\tau}}}
\def\elechargedue{{\Brho^2_{\BE,\ttt}}}
\def\elecurrttt{\BJ_{\BE,\ttt}}

\def\TAN{\textsc{Tan}}
\def\CTAN{\textsc{Cotan}}
\def\projb{\Bpi}

\def\partic{\texttt{p}}
\def\trajectory{{\cT_{\EVE}}} %
\def\track{{\cT_{\EU}}}
\def\CORPO{\textsc{B}}
\def\conf{\BOmega}
\def\conft{\conf_\ttt}
\def\control{\BC}
\def\controlt{\control_\ttt}
\def\controltout{\control_\ttt^\outer}

\def\Bvt{\Bv_{\moto,\ttt}}

\def\surft{{\surf_\ttt}}
\def\surfin{{\surf^\inner_\ttt}}
\def\surfout{{\surf^\outer_\ttt}}
\def\Path{\BGamma}

\def\Bus{{\Bu^*}}

\def\TENS{{\textsc{Tens}}}
\def\FUN{{\textsc{Fun}}}

\def\MIX{{\textsc{Mix}}}
\def\VOL{{\textsc{Vol}}}
\def\COV{{\textsc{Cov}}}
\def\CON{{\textsc{Con}}}
\def\ALT{{\textsc{Alt}}}

\def\dual#1{#1^*}
\def\coppia#1#2{(#1\,,#2)}
\def\terna#1#2#3{(#1,#2,#3)} 
\def\brack#1#2{[#1\,,#2]}

\def\BS o{\BS_o}

\def\endprova{\phantom{A}\hfill{$\blacksquare$}\goodbreak\medskip}

\def\CT{\TT^*}

\def\srif{{\Bxi}}
\def\ID#1{\mathbf{id}_{\,#1}}

\def\Bus{{\Bu^*}}

\def\EU{{\cal{S}}}
\def\TEU{\TANG\EU}
\def\TI{\TANG\II}
\def\CTEU{\CTANG\EU}
\def\TORS{\textsc{Tors}}
\def\Lieder{\cL}
\def\forw{{\Uparrow\,}}
\def\back{{\Downarrow\,}}
\def\push{{\uparrow}}
\def\pull{{\downarrow}}

\def\Bvt{\Bv_\ttt}

\def\proof{\goodbreak\par\noindent\textbf{Proof.} }

\def\ID#1{\textbf{id}_{#1}}

\def\unmezzo{{\scriptstyle\onehalf}}

\def\parder#1#2{\partial_{#1=#2}\,}

\def\MMM{\mathbb{M}}
\def\NNN{\mathbb{N}}

\def\TM{\TANG\MMM}

\def\rotor{\mathrm{rot}\,}
\def\diverg{\mathrm{div}\,}
\def\grad{\mathrm{grad}\,}
\def\inv#1{#1^{-1}}
\let\aa=\alpha

\def\uu{u}
\def\xx{x}
\def\ii{i}

\def\mathscr{\mathcal}
\def\Re{\mathscr{R}}

\def\CORPO{\mathscr{B}}

\def\TANG{\mathbb{T}}
\def\CTANG{\mathbb{T}^*}

\def\cont{\mathrm{C}}

\def\di#1{(#1)}
\def\pair#1#2{\{#1\,,#2\}}
\def\onehalf{\frac{1}{2}}

\def\Baa{\boldsymbol{\alpha}}
\def\Bomega{\boldsymbol{\omega}}
\let\Boo=\Bomega
\def\BOmega{\boldsymbol{\Omega}}
\def\BLambda{\boldsymbol{\Lambda}}
\def\BGamma{\boldsymbol{\Gamma}}
\def\BSigma{\boldsymbol{\Sigma}}

\def\Bzeta{\boldsymbol{\chi}}
\def\BPi{\boldsymbol{\Pi}}
\def\Bpi{\boldsymbol{\pi}}

\def\Bss{\boldsymbol{\sigma}}
\def\Bmu{\boldsymbol{\mu}}
\def\metricamma{\boldsymbol{\gamma}}

\def\Bvarphi{\boldsymbol{\varphi}}

\def\moto{{\boldsymbol{\moto}}}
\def\Bzeta{{\boldsymbol{\zeta}}}

\def\Brho{\boldsymbol{\rho}}

\def\Bxi{\boldsymbol{\xi}}
\def\Bbeta{\boldsymbol{\beta}}
\def\torstensor{\textsc{tors}}

\def\Ba{\mathbf{a}}
\def\Bb{\mathbf{b}}
\def\Bc{\mathbf{c}}

\def\Be{\mathbf{e}}

\def\metric{\mathbf{g}}
\def\Bh{\mathbf{h}}
\def\Bi{\mathbf{i}}

\def\BI{\mathbf{I}}

\def\Bn{\mathbf{n}}
\def\Bo{\mathbf{o}}

\def\Br{\mathbf{r}}
\def\Bs{\mathbf{s}}
\def\Bt{\mathbf{t}}
\def\Bu{\mathbf{u}}
\def\Bv{\mathbf{v}}
\def\Bw{\mathbf{w}}
\def\Bx{\mathbf{x}}

\def\Bl{\mathbf{l}}
\def\BB{\mathbf{B}}
\def\BC{\mathbf{C}}
\def\BP{\mathbf{P}}
\def\BR{\mathbf{R}}
\def\BS{\mathbf{S}}

\def\BA{\mathbf{A}}
\def\BE{\mathbf{E}}
\def\BF{\mathbf{F}}

\def\BN{\mathbf{N}}
\def\BD{\mathbf{D}}

\def\BZ{\mathbf{Z}}
\def\BW{\mathbf{W}}

\def\BJ{\mathbf{J}}
\def\BH{\mathbf{H}}

\def\sub#1{{}_{\lower2pt\hbox{$\scriptstyle#1$}}}
\def\cP{\mathcal{P}}

\def\cL{\mathcal{L}}

\def\cT{\mathcal{T}}

\def\II{I}
\def\VV{V}

\def\UU{U}

\def\JJ{J}

\def\TT{T}

\def\HH{H}

\def\ff{f}

\def\ttt{t}

\def\equaldef{:=}
\def\perogni{\quad\forall\,}
\def\Linmap#1{\textrm{\textit{BL}}\,\di{#1}}
\def\sp{\,;}

\def\equi{\quad\Longleftrightarrow\quad}
\def\implies{\quad\Longrightarrow\quad}
\def\scalar#1#2{{\langle}\kern.1em#1,#2\kern.1em{\rangle}}
\def\inde{\,\mathit{d}}
\def\integrale#1#2{\int_{#1}^{#2}}
\def\ointegrale#1#2{\oint_{#1}^{#2}}
\let\punto=\cdot

\def\set#1{\{\,#1\,\}}
\def\Ker{\mathrm{Ker}\,}
\def\Im{\mathrm{Im}\,}
\def\der{d}
\def\barra#1#2{ \bigg\vert _{\hbox{$\scriptstyle#1$}}^{\raise8.2pt\hbox{$\scriptstyle#2$}}  \, }
\def\lineare#1{[#1]}
\def\talechesia{\, : \,}

\def\BBa{\mathbb{A}}
\def\BBb{\mathbb{B}}
\def\inv#1{#1^{-1}}
\def\Re{\mathscr{R}}

\def\CORPO{\mathscr{B}}

\def\TANG{\mathbb{T}}

\def\cont{\mathrm{C}}

\def\di#1{(#1)}
\def\pair#1#2{\{#1\,,#2\}}
\def\onehalf{\frac{1}{2}}
\def\sub#1{{}_{\lower2pt\hbox{$\scriptstyle#1$}}}

\def\II{I}
\def\VV{V}

\def\UU{U}

\def\HH{H}

\def\ff{f}
\def\ttt{t}

\def\equaldef{:=}
\def\perogni{\quad\forall\,}
\def\Linmap#1{\textrm{\textit{BL}}\,\di{#1}}
\def\sp{\,;}

\def\equi{\quad\Longleftrightarrow\quad}
\def\implies{\quad\Longrightarrow\quad}
\def\integrale#1#2{\int_{#1}^{#2}}
\let\punto=\cdot

\def\set#1{\{\,#1\,\}}
\def\der{d}
\def\lineare#1{[#1]}
\def\talechesia{\, : \,}

\def\persone#1{\textcolor{BrickRed}{\textsc{#1}}}
\def\Oersted{\persone{{\O}rsted}}

\def\Green{\persone{Green}}
\def\Riemann{\persone{Riemann}}
\def\Leibniz{\persone{Leibniz}}
\def\Lie{\persone{Lie}}

\def\Poincare{\persone{PoincarŽ}}

\def\Ampere{\persone{Ampre}}
\def\Lorentz{\persone{Lorentz}}

\def\Henry{\persone{Henry}}
\def\Faraday{\persone{Faraday}}
\def\Maxwell{\persone{Maxwell}}
\def\ClerkMaxwell{\persone{Clerk-Maxwell}}
\def\Stokes{\persone{Stokes}}
\def\Poincare{\persone{PoincarŽ}}
\def\Helmholtz{\persone{Helmholtz}}
\def\Gauss{\persone{Gauss}}
\def\Feynman{\persone{Feynman}}
\def\Cartan{\persone{Cartan}}
\def\Palais{\persone{Palais}}
\def\Galilei{\persone{Galilei}}

\def\Fubini{\persone{Fubini}}
\def\Reynolds{\persone{Reynolds}}
\def\Euclid{\persone{Euclid}}
\def\Hertz{\persone{Hertz}}

\def\Einstein{\persone{Einstein}}
\def\Thomson{\persone{Thomson}}
\def\Kelvin{\persone{Kelvin}}
\def\Heaviside{\persone{Heaviside}}
\def\Minkowski{\persone{Minkowski}}
\def\Volterra{\persone{Volterra}}
\def\Brouwer{\persone{Brouwer}}
\def\Ostrogradski{\persone{Ostrogradski}}
\def\Hamel{\persone{Hamel}}
\def\deRham{\persone{de Rham}}
\def\Weyl{\persone{Weyl}}
\def\Poynting{\persone{Poynting}}
\def\moto{{\Bvarphi}}
\def\conf{{\BOmega}}
\def\confo{{\BOmega_{o}}}
\newtheorem{remark}{Remark}
\newtheorem{proposition}{Proposition}

\newtheorem{lemma}{Lemma}
\newtheorem{definition}{Definition}
\newtheorem{axiom}{Axiom}
\numberwithin{theorem}{section}
\numberwithin{proposition}{section}
\numberwithin{lemma}{section}
\numberwithin{remark}{section}
\numberwithin{definition}{section}
\begin{document}
\begin{frontmatter}



\title{\textcolor{BrickRed}{On the Laws of Electromagnetic Induction}}

% \author[label2]{}

\author[label1]{\textcolor{BrickRed}{Giovanni Romano}}

\address[label1]{\textcolor{BrickRed}{
Department of Structural Engineering, University of Naples Federico II, 
via Claudio 21, 80125 - Naples, Italy\\
e-mail: romano@unina.it}}

\begin{abstract}

The \Faraday-\Ampere\ laws of electro-magnetic induction are
formulated in terms of even and odd differential forms,
taking in due account the body motion
in terms of \Lie\ time-derivatives.
It is shown that covariance of \Lie\ derivatives with respect to arbitrary relative motions,
and \Galilei\ invariance of the electro-magnetic fields, 
imply \Galilei\ invariance of the induction laws,
contrary to most claims in literature.
A noteworthy outcome of the theory is that
the \emph{so called} \Lorentz\ \emph{force} on a charged particle
is not an additional law of electromagnetism, but rather,
when corrected by a factor one-half,
a non-invariant contribution to the electric field evaluated, according to \Faraday\ law,
by an observer performing tests on a translating charged body crossing a region of uniform magnetic field.
The formulation of the laws of electromagnetism in the four dimensional classical space-time,
by stating the observer-dependent splitting for bodies in motions, 
provides a proof of \Galilei\ invariance of electric and magnetic fields.

\end{abstract}

\end{frontmatter}
% main text

\section{Introduction}
\label{int: Intro}

A geometric approach to the laws of electromagnetism
reveals the need for considering arbitrarily moving material circuits in the integral formulations,
so that every-day engineering applications can be investigated by the theory and
well-posedness and observer-invariance properties can be correctly deduced.
This revisitation shows that the laws of electromagnetic induction,
when correctly formulated, are in fact \Galilei\ invariant,
contrary to most claims in literature.
In the light of the proposed formulation, it is further shown that
the \emph{so called} \Lorentz\ \emph{force} acting a charged particle 
is rather an expression of the electric field evaluated,
according to \Faraday\ law of induction,
by an observer which tests a body in translational motion 
across a region of uniform and time-independent magnetic \emph{vortex}
(an alternative name for the \emph{magnetic induction} which underlines 
that it is a \emph{even} two-form, or equivalently a \emph{odd} vector field).

A critical discussion of previous treatments is performed and 
some important issues of classical electromagnetism are reconsidered in the new perspective.
In particular \Galilei\ invariance provides a simple direct answer to the troubles
concerning the induction effects due to the relative motion of a magnet and a conductor,
as expressed by \citet{Einstein1905} and still lasting in literature, see e.g. \citep[p. 477]{Griffiths1999}.

Some basic issues of integration on manifolds and of exterior differential calculus
are preliminarily summarized for the reader's convenience.
Integration of forms on inner oriented submanifolds and of odd forms on outer oriented submanifolds
in an oriented ambient manifold are illustrated in detail as basic tools for the
development of the theory.

The connection between the exterior calculus and the more usual vector calculus
is recalled and 
the basics of classical electromagnetism are reformulated according to both formats.
This treatment is propaedeutic to the main sections dealing with
\Galilei\ invariance and with the electromagnetics
of moving bodies, where the exterior differential calculus format is adopted,
being basic for a treatment of induction laws independent of metric properties of the ambient space.

A careful attention to the roles played by inner and outer 
orientations in the integration 
over surfaces and along their boundary cycles
leads naturally to propose a new terminology.
The \emph{electric field} one-form and 
the \emph{magnetic vortex} two-form are involved in \Faraday\ law of induction,
where an inner orientation of the involved surface and of its boundary circuit is considered.
The \emph{electric displacement flux} and \emph{electric current} \emph{odd} two-form, 
and the \emph{magnetic winding} \emph{odd} one-form,
are involved in \Ampere\ law of induction,
where an outer orientation of the involved surface and of its boundary circuit
is adopted. 
The former choice provides a clear physical interpretation of the \emf\ as circulation.
The latter provides a better physical description, as a flux rule, of the induction law
and of the equivalent condition of charge balance.

The formulation of electromagnetism in classical space-time,
an affine four-dimensional manifold, 
provides an impressively simple expression 
of balance laws for electric and magnetic charges as closedness conditions of three-forms. 
Induction laws are expressed as exactness conditions of the same forms.
Charge balance and induction laws are thus simply represented by equivalent
integrability conditions for exterior forms  and exactness conditions in terms of potential forms,
according to \Poincare\ Lemma.
The observer-dependent splitting into space and time components, is here extended to bodies in motion, 
and shows that the electric and magnetic spatial fields involved in the theory are all
\Galilei\ invariant.

A final discussion points out the innovative features of the present approach to
the laws of electromagnetic induction and suggests corrections to common misstatements.

\section{Calculus on manifolds}
\label{sec: calmany}

\subsection{Push, pull and Lie derivatives}
\label{sec: Pushandpull}

Let $\,\confo\,$ and $\,\conf\,$ 
be two submanifolds embedded in
a container \Riemann\ manifold $\,\coppia\EU{\metric}\,$
with metric tensor field
$\,\metric\,$.
The tangent map associated  with a
diffeomorphism $\,\srif\in\cont^1\di{\confo\sp\conf}\,$
between two manifolds $\,\confo\,$ and $\,\conf\,$
relates the velocity of a curve through a point $\,\Bx\in\confo\,$
in the domain to the corresponding velocity of the curve
at the point $\,\srif\di\Bx\in\conf\,$ in the codomain.
The tangent map to the diffeomorphism, and and its dual, are denoted by
%%%%%%%%%%%%%%%%%%%%%%%%%%%%%%%%%
$$\vcenter{\halign{
\hfil$#$&$#$\hfil&$#$\hfil&$#$\hfil\cr
\TT_\Bx\srif&\,
\in\Linmap{\TANG_{\Bx}\confo\sp\TANG_{\srif\di\Bx}\conf}\,,
\vspace{8pt}\cr
\CT_\Bx\srif=(\TT_\Bx\srif)^*&\,
\in\Linmap{\CTANG_{\srif\di\Bx}\conf\sp\CTANG_{\Bx}\confo}
\,,\cr}}$$
%%%%%%%%%%%%%%%%%%%%%%%%%%%%%%%%%
and, for every $\,\Baa_\Bx\in\CTANG_{\srif\di\Bx}\conf\,$, fulfill the identity
%%%%%%%%%%%%%%%%%%%%%%%%%%%%%%%%%%%
$$\,\scalar{\Baa_\Bx}{\TT_\Bx\srif\punto\Bb_\Bx}
=\scalar{\CT_\Bx\srif\punto\Baa_\Bx}{\Bb_\Bx}\,.$$
%%%%%%%%%%%%%%%%%%%%%%%%%%%%%%%%%%%
We have that
$\,\TT\inv\srif=\inv{(\TT\srif)}\,$
and $\,\CT\inv\srif=\inv{(\CT\srif)}\,$.
%%%%%%%%%%%%%%%%%%%%%%%%%%%%%%%%%
The push-forward of a scalar field $\,\ff\in\cont^1\di{\confo\sp\Re}\,$
is a change of its base points:
%%%%%%%%%%%%%%%%%%%%%%%%%%%%%%%%%%%
$$\,(\srif\push\ff)_{\srif\di\Bx}
\equaldef\ff\di{\Bx}\in\Re\,.$$
%%%%%%%%%%%%%%%%%%%%%%%%%%%%%%%%%%%
The push-forward of a tangent vector $\,\Bv_\Bx\in\TANG_\Bx\confo\,$
is the tangent vector defined by
%%%%%%%%%%%%%%%%%%%%%%%%%%%%%%%%%%%
$$\,\srif\push\Bv_\Bx
\equaldef\TT_\Bx\srif\punto\Bv_\Bx\in\TANG_{\srif\di\Bx}\conf\,.$$
%%%%%%%%%%%%%%%%%%%%%%%%%%%%%%%%%%%
The pull-back is the push induced by the inverse diffeomorphism.
The push of a covector $\,\dual\Bv_\Bx\in\CTANG_\Bx\confo\,$
is defined by invariance
%%%%%%%%%%%%%%%%%%%%%%%%%%%%%%%%%%%
$$\,\scalar{\srif\push\dual\Bv_\Bx}{\srif\push\Bv_\Bx}
=\srif\push\scalar{\dual\Bv_\Bx}{\Bv_\Bx}\,,$$
%%%%%%%%%%%%%%%%%%%%%%%%%%%%%%%%%%%
equivalent to
%%%%%%%%%%%%%%%%%%%%%%%%%%%%%%%%%%%
$\,\scalar{\CT_\Bx\srif\punto\srif\push\dual\Bv_\Bx}{\Bv_\Bx}
=\scalar{\dual\Bv_\Bx}{\Bv_\Bx}\in\CTANG_{\Bx}\conf\,$,
%%%%%%%%%%%%%%%%%%%%%%%%%%%%%%%%%%%
so that 
%%%%%%%%%%%%%%%%%%%%%%%%%%%%%%%%%%%
$\,\srif\push\dual\Bv_\Bx=\CT_\Bx\inv\srif\punto\dual\Bv_\Bx\,$.
%%%%%%%%%%%%%%%%%%%%%%%%%%%%%%%%%%%.
Pushes of tensors are also defined by invariance.

The \Lie\ derivative of a vector field 
$\,\Bv\in\cont^1\di{\EU\sp\TEU}\,$
along a flow $\,\moto\in\cont^1\di{\EU\times\II\sp\EU}\,$
with velocity field $\,\Bv_\moto\equaldef\parder\lambda0\moto_\lambda\,$,
is defined by:
%%%%%%%%%%%%%%%%%%%%%%%%%%%%%%%%%%%
$$\,\Lieder_{\Bv_\moto}\,\Bv\equaldef
\parder{\lambda}{0}\moto_{\lambda}\pull\Bv\,,$$
%%%%%%%%%%%%%%%%%%%%%%%%%%%%%%%%%%%
The time-derivation yields a vector since all vectors
$\,\moto_{\lambda}\pull\Bv_{\moto_\lambda\di\Bx}\,$ belong to the linear space $\,\TANG_{\Bx}\EU\,$.
The \Lie\ derivatives of tensor fields are analogously defined in terms of
the appropriate pull-back. 
A simple basic property of the \Lie\ derivative is the following
\citep{AbrahamMarsden2002,RomanoDiff2007}:
%%%%%%%%%%%%%%%%%%%%%%%%%%%%%%%%%%%
$$\,\moto_\lambda\pull(\Lieder_{\Bv_\moto}\,\Bv)=
\parder\mu\lambda(\moto_{\mu}\pull\Bv)\,.$$
%%%%%%%%%%%%%%%%%%%%%%%%%%%%%%%%%%%

\subsection{Parallel transport and parallel derivatives}
\label{sec: covder}

A linear connection in a manifold $\,\EU\,$ is expressed
by a derivation $\,\nabla\,$, called the \textit{parallel derivation} 
(also called \emph{covariant derivation}) fulfilling the properties:
%%%%%%%%%%%%%%%%%%%%%%%%%%%%%%%%%
$$\vcenter{\halign{
\hfil$#$&$#$\hfil&$#$\hfil&$#$\hfil\cr
\nabla_{\aa_1\Bv_1+\aa_2\Bv_2}\Bu&\,=\aa_1\nabla_{\Bv_1}\Bu+\aa_2\nabla_{\Bv_2}\Bu\,,
\vspace{8pt}\cr
\nabla_{\Bv}(\aa_1\Bu_1+\aa_2\Bu_2)&\,=\aa_1\nabla_{\Bv}\Bu+\aa_2\nabla_{\Bv}\Bu\,,
\vspace{8pt}\cr
\nabla_{\Bv}(\ff\Bu)&\,=\ff\,\nabla_{\Bv}\Bu+(\nabla_\Bv\ff)\Bu
\,.\cr}}$$
%%%%%%%%%%%%%%%%%%%%%%%%%%%%%%%%%
For a scalar field $\,\ff\in\cont^1\di{\EU\sp\Re}\,$, and more in general
for fields taking values in a linear space, the
parallel derivation is identical to the usual derivation.
The parallel derivation of a covector field is defined by
a formal application of the \Leibniz\ rule:
%%%%%%%%%%%%%%%%%%%%%%%%%%%%%%%%%%%
$$\,\scalar{\nabla_{\Bv_\Bx}\Bus}{\delta\Bv_\Bx}
=\nabla_{\Bv_\Bx}\scalar{\Bu}{\delta\Bv_\Bx}
-\scalar{\Bus}{\nabla_{\Bv_\Bx}\delta\Bv_\Bx}\,,
\perogni\delta\Bv_\Bx\in\TANG_\Bx\EU\,,$$
%%%%%%%%%%%%%%%%%%%%%%%%%%%%%%%%%%%
where $\,\delta\Bv_\Bx\in\TANG_\Bx\EU\,$ can be arbitrarily extended to a vector field
in a neighborhood of $\,\Bx\in\EU\,$ to perform the derivations.
Analogously, the  parallel derivation of a $\,2$-covariant tensor field is defined by:
%%%%%%%%%%%%%%%%%%%%%%%%%%%%%%%%%%%
$$\nabla_{\Bv_\Bx}\Bss\di{\Bu_1,\Bu_2}=
\nabla_{\Bv_\Bx}(\Bss\di{\Bu_1,\Bu_2})
-\Bss\di{\nabla_{\Bv_\Bx}\Bu_1,\Bu_2}
-\Bss\di{\Bu_1,\nabla_{\Bv_\Bx}\Bu_2}\,.$$
%%%%%%%%%%%%%%%%%%%%%%%%%%%%%%%%%%%
The integrated counterpart of the parallel derivation is provided by the notion of
\textit{parallel transport }
$\,\Bc_{\tau,\ttt}\forw\in\cont^1\di{\TANG_{\Bc\di\ttt}\EU\sp\TANG_{\Bc\di\tau}\EU}\,$
along a curve $\,\Bc\in\cont^1\di{\II\sp\EU}\,$.
Setting $\,\Bx=\Bc\di\ttt\,$ and $\,\Bv_\Bx=\parder\tau\ttt\Bc\di\tau\,$,
we have the formula:
%%%%%%%%%%%%%%%%%%%%%%%%%%%%%%%%%%%
$$\,\nabla_{\Bv_\Bx}\Bu=\parder\tau\ttt\Bc_{\ttt,\tau}\forw\Bu\di{\Bc\di\tau}\in\TANG_{\Bc\di\ttt}\EU\,.$$
%%%%%%%%%%%%%%%%%%%%%%%%%%%%%%%%%%% 
The derivation yields a vector since all vectors
$\,\Bc_{\ttt,\tau}\forw\Bu\di{\Bc\di\tau}\,$ belong to the linear space $\,\TANG_{\Bc\di\ttt}\EU\,$.
The parallel transport of tensor fields is defined by invariance.
The parallel transport of a vector field 
$\,\Bv\in\cont^1\di{\EU\times\II\sp\TM}\,$
along a flow $\,\moto\in\cont^1\di{\EU\times\II\times\II\sp\EU}\,$,
from time $\,\tau\,$ to time $\,\ttt\,$,
will accordingly be denoted by 
$\,\moto_{\ttt,\tau}\forw\Bv_\tau=\moto_{\tau,\ttt}\back\Bv_\tau\,$.
%%%%%%%%%%%%%%%%%%%%%%%%%%%%%%%%

For any $\,\Bv\in\cont^1\di{\EU\sp\TEU}\,$,
$\,\TORS\di{\Bv}\in\cont^1\di{\EU\sp\MIX\di{\EU}}\,$
is the mixed tensor field defined, at $\,\Bx\in\EU\,$, by
%%%%%%%%%%%%%%%%%%%%%%%%%%%%%%%%%
$$\TORS\di{\Bv_\Bx}\punto{\Bu_\Bx}
=\torstensor\di{\Bv_\Bx,\Bu_\Bx}\,,
\perogni\Bu_\Bx\in\TANG_\Bx\EU\,,$$
%%%%%%%%%%%%%%%%%%%%%%%%%%%%%%%%%
and $\,\torstensor\di{\Bv,\Bu}\equaldef
\nabla_{\Bv}\Bu-\nabla_{\Bu}\Bv-\brack{\Bv}{\Bu}\,$
is the torsion of the linear connection.
The \Lie\ bracket is given by:
$\,\brack{\Bv}{\Bu}\,\ff= \nabla_{\Bv}\nabla_{\Bu}\,\ff-\nabla_{\Bu}\nabla_{\Bv}\,\ff\,$,
for any scalar valued function $\,\ff\in\cont^2\di{\EU\sp\Re}\,$,
with $\,\brack{\Bv}{\Bu}=\Lieder_{\Bv}\Bu\,$,
\citep{AbrahamMarsden2002}.
%%%%%%%%%%%%%%%%%%%%%%%%%%%%%%%%

\subsection{Exterior forms and integrals}
\label{sec: manifolds}

Let $\,\MMM^n\,$ be an $\,n$-dimensional manifold,
$\,\surf^k\,$ a $\,k$-dimensional manifold ($\,k\le{n}$)
and $\,\moto\in\cont^1\di{\surf^k\sp\MMM^n}\,$
a diffeomorphism onto $\,\moto\di{\surf^k}\subset\MMM^n\,$.
We will denote by $\,\Boo^k_\moto\in\EXForms^k\di{\TANG\MMM^n\sp\Re}\,$
a material $\,k$-form, i.e.
a smooth tensor field of $\,k$-linear alternating maps ($\,k$-covectors)
defined on $\,\surf^k\,$ along $\,\moto\in\cont^1\di{\surf^k\sp\MMM^n}\,$,
according to the commutative diagram:
%%%%%%%%%%%%%%%%%%%%%%%%%%%%%%%%%s
$$\begin{aligned}
\xymatrix{ 
&\ALT^k\di{\TANG\MMM^n}
\ar@<-.7ex>[d]_{\projb}\\ 
{\surf} 
\ar[ur]^{\Boo^k_{\moto}} 
\ar[r]^{\moto} 
&{\moto\di{\surf^k}}
}
\end{aligned}
\equi
\vcenter{\halign{
\hfil$#$&$#$\hfil&$#$\hfil\cr
%\vspace{1pt}\cr
&\,\projb\circ\Boo^k_{\moto}=\moto\,.\cr}}
$$
%%%%%%%%%%%%%%%%%%%%%%%%%%%%%%%%%
In a time-interval $\,\II\,$, we consider a motion
$\,\moto\in\cont^1\di{\surf^k\times\II\sp\MMM^n}\,$
and the corresponding displacement
$\,\moto_{\tau,\ttt}\in\cont^1\di{\moto_\ttt\di{\surf^k}\sp\moto_\tau\di{\surf^k}}\,$,
defined by $\,\moto_{\tau,\ttt}\equaldef\moto_{\tau}\circ\inv\moto_{\ttt}\,$.
The push transformation $\,\moto_{\tau,\ttt}\push\,$ of $\,\Boo^k_{\moto,\ttt}\,$ 
to a $\,k$-form on $\,\moto_\tau\di{\surf^k}\,$ according to the relation
(taking $\,k=1\,$):
%%%%%%%%%%%%%%%%%%%%%%%%%%%%%%%%%%%
$$\,(\moto_{\tau,\ttt}\push\Boo^k_{\moto,\ttt})\punto(\moto_{\tau,\ttt}\push\Ba_{\moto,\ttt})
\equaldef\moto_{\tau,\ttt}\push(\Boo^k_{\moto,\ttt}\punto\Ba_{\moto,\ttt})\,$$
%%%%%%%%%%%%%%%%%%%%%%%%%%%%%%%%%%%
where the push of the tangent vector 
$\,\Ba_{\moto,\ttt}\di\Bx\in\TANG_{\moto_\ttt\di\Bx}\moto_\ttt\di{\surf^k}\,$
is performed by the tangent map as:
%%%%%%%%%%%%%%%%%%%%%%%%%%%%%%%%%%%
$$\,\moto_{\tau,\ttt}\push\Ba_{\moto,\ttt}
\equaldef\TT_{\moto_\ttt\di\Bx}\moto_{\tau,\ttt}\punto\Ba_{\moto_\ttt\di\Bx}
\in\TANG_{\moto_\tau\di\Bx}\moto_\tau\di{\surf^k}\,.$$
%%%%%%%%%%%%%%%%%%%%%%%%%%%%%%%%%%%
and the push of the scalar 
$\,\Boo^k_{\moto,\ttt}\punto\Ba_{\moto,\ttt}\,$ 
at $\,\moto_\tau\di\Bx\,$ is defined by invariance:
%%%%%%%%%%%%%%%%%%%%%%%%%%%%%%%%%%%
$$\,(\moto_{\tau,\ttt}\push(\Boo^k_{\moto,\ttt}\punto\Ba_{\moto,\ttt}))
_{\moto_\tau\di\Bx}\equaldef(\Boo^k_{\moto,\ttt}\punto\Ba_{\moto,\ttt})_{\moto_\ttt\di\Bx}\,.$$
%%%%%%%%%%%%%%%%%%%%%%%%%%%%%%%%%%%
Main tools of calculus on manifolds are the following
\citep{AbrahamMarsden2002,Bossavit2005,RomanoDiff2007}.
The formulae for change of integration domain (\CID):
%%%%%%%%%%%%%%%%%%%%%%%%%%%%%%%%%
$$\integrale{\moto\di{\surf^k}}{}\moto\push\Boo^k=\integrale{\surf^k}{}\Boo^k\,,\qquad
\integrale{\moto\di{\surf^k}}{}\Boo^k=\integrale{\surf^k}{}\moto\pull\Boo^k\,,$$
%%%%%%%%%%%%%%%%%%%%%%%%%%%%%%%%%
are a direct consequence of the definition of push.
By considering a motion $\,\moto\in\cont^1\di{\surf^k\times\II\sp\MMM}\,$,
leads to \Reynolds\  transport formula:
%%%%%%%%%%%%%%%%%%%%%%%%%%%%%%%%%
$$\parder\tau\ttt\integrale{\moto_{\tau}\di{\surf^k}}{}\Boo^k_\tau
=\integrale{\moto_{\ttt}\di{\surf^k}}{}\parder\tau\ttt\moto_{\tau,\ttt}\pull\Boo^k
=\integrale{\moto_{\ttt}\di{\surf^k}}{}\Lieder_{\Bv_\moto,\ttt}\,\Boo^k\,,$$
%%%%%%%%%%%%%%%%%%%%%%%%%%%%%%%%%
\Stokes\ formula:
%%%%%%%%%%%%%%%%%%%%%%%%%%%%%%%%%
$$\integrale{\surf^k}{}\der\Boo^{k-1}=\ointegrale{\partial\surf^k}{}\Boo^{k-1}\,,$$
%%%%%%%%%%%%%%%%%%%%%%%%%%%%%%%%%
introduces the \textit{exterior derivative} $\,\der\,$
by a generalization of the fundamental formula of integral calculus 
to manifolds of finite dimension higher than one.
As quoted in \citep{deRham1955}, according to \citet{Segre1951},
this general integral transformation was considered by 
\citet{Volterra1889,Poincare1895,Brouwer1906}.
It includes as special cases the classical formulae due to
\Gauss, \Green, \Ostrogradski\ and to \Ampere, \Kelvin, \Hamel,
that were taught by \Stokes\ at Cambridge.
\Stokes\ formula, in its modern general formulation,
might therefore at right be renamed \Volterra-\Poincare-\Brouwer\ (\VPB) formula.
Since boundaryless surfaces are said to be closed,
differential forms such that $\,\der\Boo^k=0\,$ are called \emph{closed} forms
due to the duality between the exterior differentiation $\,\der^{k-1}\,$
operating on $\,\EXForms^{k-1}\di{\TANG\MMM^n\sp\Re}\,$
and the boundary operator $\,\partial^k\,$,
operating on $\,k$-chains, resulting by rewriting \Stokes\ formula as follows:
%%%%%%%%%%%%%%%%%%%%%%%%%%%%%%%%%%%
$$\,\scalar{\surf^k}{\der^{k-1}\Boo^{k-1}}=\scalar{\partial^k\surf^k}{\Boo^{k-1}}\,.$$
%%\%%%%%%%%%%%%%%%%%%%%%%%%%%%%%%%%%
In general, $\,\surf^k\,$ is a chain and $\,\partial^k\,$
is the boundary operator.
Hence $\,\Boo^{k-1}\,$ is called a co-chain
and $\,\der^{k-1}\,$ is the co-boundary operator.
The relevant theory, first developed by 
\persone{Georges} \deRham\ in his famous $\,1931\,$ thesis,
is exposed in \citep{deRham1931,deRham1955}.
The basic results are expressed by the following annihilation relations
which extend to chain and co-chains well-known formulae 
for dual operators in linear algebra:
%%%%%%%%%%%%%%%%%%%%%%%%%%%%%%%%
$$\left\{\vcenter{\halign{
$\displaystyle#$\hfil&$\displaystyle#\qquad$\hfil\cr
\Ker\partial^k&\,=(\Im\der^{k-1})^0\,,
\vspace{8pt}\cr
\Ker\der^{k}&\,=(\Im\partial^{k+1})^0\,,
\cr}}\right.
\qquad
\left\{\vcenter{\halign{
$\displaystyle#$\hfil&$\displaystyle#\qquad$\hfil\cr
\Im\partial^{k+1}&\,=(\Ker\der^{k})^0\,,
\vspace{8pt}\cr
\Im\der^{k-1}&\,=(\Ker\partial^{k})^0\,,
\cr}}\right.$$
%%%%%%%%%%%%%%%%%%%%%%%%%%%%%%%%
where the annihilators are defined as exemplified by:
%%%%%%%%%%%%%%%%%%%%%%%%%%%%%%%%%%%
$$\,(\Im\partial^{k})^0\equaldef\set{\Bomega^{k-1}\in\EXForms^{k-1}\di{\MMM^n\sp\Re}\talechesia
\scalar{\Bomega^{k-1}}{\partial^{k}\surf^{k}}=0 \perogni\surf^k}\,.$$
%%%%%%%%%%%%%%%%%%%%%%%%%%%%%%%%%%%
The duality between \emph{homologies} and \emph{cohomologies}
of degree $\,k\,$, i.e. the quotient spaces:
%%%%%%%%%%%%%%%%%%%%%%%%%%%%%%%%%%%
$$\,\HH_k\di\MMM\equaldef\Ker\partial^{k}/\Im\partial^{k+1}
\quad\mathrm{and}\quad
\HH^k\di\MMM\equaldef\Ker\der^{k}/\Im\der^{k-1}\,,$$
%%%%%%%%%%%%%%%%%%%%%%%%%%%%%%%%%%%
is expressed by the \emph{period}, which is the integral of a cocycle (closed cochain) over a cycle (closed chain).
A direct application of \VPB\ formula provides the basic invariance property:
%%%%%%%%%%%%%%%%%%%%%%%%%%%%%%%%%%%
$$\,\ointegrale{\Bc^k}{}\Boo^k=\ointegrale{\Bc^k+\Bl^{k}}{}\Boo^k+\Baa^{k}\,,$$
%%%%%%%%%%%%%%%%%%%%%%%%%%%%%%%%%%%
with $\,\Bc^k\in\Ker\partial^k\,$ and $\,\Boo^k\in\Ker\der^k\,$, for all
$\,\Bl^{k}\in\Im\partial^{k+1}\,$ and $\,\Baa^k\in\Im\der^{k-1}\,$.
The \deRham\ annihilations reveal that
the duality provided by the \emph{period} is separating and this ensures the existence of an isomorphism
between the spaces of homologies and cohomologies of degree $\,k\,$.
Accordingly these will have  the same finite dimension, the $\,k$-dimensional \persone{Betti}'s number of $\,\MMM\,$.
The \emph{currents} introduced by  \deRham\ are the k-dimensional extension of
\emph{scalar distributions} of \persone{Laurent Schwartz}. 
Currents are linear functionals on the linear space of smooth exterior forms
with compact support on a manifold.
These topological notions are gaining a rapidly increasing attention 
in theoretical and computational aspects of electromagnetics,
\citep{Bossavit1991,Bossavit2004,Bossavit2005},
\citep{Tonti1995,Tonti2002},
\citep{GrossKotiuga2004}.

The following commutation property will be referred to in the sequel.
%%%%%%%%%%%%%%%%%%%%%%%%%%%%%%%%%
\begin{lemma}[Exterior derivatives and pushes]
\label{lm: extderpushes}
The pull-back of a form
by an injective immersion $\,\moto\in\cont^1\di{\MMM\sp\NNN}\,$
and the exterior derivative of differential forms commute:
%%%%%%%%%%%%%%%%%%%%%%%%%%%%%%%%%
$$\der_\MMM\circ\moto\pull=\moto\pull\circ\der_\NNN\,.$$
%%%%%%%%%%%%%%%%%%%%%%%%%%%%%%%%%
\end{lemma}
\proof
For any $\,k$-form $\,\Bomega^{k}\in\EXForms^k\di{\NNN\sp\Re}\,$ 
we have that $\,\moto\pull\Bomega^{k}\in\EXForms^k\di{\MMM\sp\Re}\,$
and the image of any $\,(k+1)$-dimensional chain $\,\surf^{k+1}\subset\MMM\,$
by the injective immersion $\,\moto\in\cont^1\di{\MMM\sp\NNN}\,$
is still a $\,(k+1)$-dimensional chain $\,\moto\di{\surf^k}\subset\NNN\,$.
Then, by \VPB\ and \CID\ formulas, we have the equality:
%%%%%%%%%%%%%%%%%%%%%%%%%%%%%%%%%
$$\vcenter{\halign{
\hfil$\displaystyle#$&$\displaystyle#$ \hfil &$\displaystyle#$\hfil\cr
\integrale{\surf^{k+1}}{}\der_\MMM(\moto\pull\Bomega^{k})
=&\,\ointegrale{\partial \surf^{k+1}}{}\moto\pull\Bomega^{k}
=\ointegrale{\moto\di{\partial \surf^{k+1}}}{}\Bomega^{k}
\vspace{6pt}\cr
=&\,\ointegrale{\partial\moto\di{\surf^{k+1}}}{}\Bomega^{k}
=\integrale{\moto\di {\surf^{k+1}}}{}\der_\NNN\Bomega^{k}=
\integrale{\surf^{k+1}}{}\moto\pull(\der_\NNN\Bomega^{k})\,,
\cr}}$$
%%%%%%%%%%%%%%%%%%%%%%%%%%%%%%%%
which yields the result.
\endprova
%%%%%%%%%%%%%%%%%%%%%%%%%%%%%%%%%
\Fubini's formula states that
the rate of variation of the integral of a volume-form $\,\Boo^{k+1}_\moto\,$,
on the $\,k+1$-dimensional flow tube
$\,\JJ_\Bv\di{\Path,\ttt}\,$,
traced by a $\,k$-dimensional submanifold $\,\moto_\tau\di\Path\,$,
with $\,\tau\in[0,\ttt]\,$, 
is equal to the $\,\Boo^{k+1}_\moto$-flux of the velocity field $\,\Bv\,$ 
of the flow through the tracing manifold $\,\moto_\ttt\di\Path\,$:
%%%%%%%%%%%%%%%%%%%%%%%%%%%%%%%%%
$$\parder\tau\ttt\integrale{\JJ_\Bv\di{\Path,\tau}}{}\Boo^{k+1}_\moto=
\integrale{\moto_\ttt\di\Path}{}(\Boo^{k+1}_\moto\punto\Bv)\,.$$
%%%%%%%%%%%%%%%%%%%%%%%%%%%%%%%%%
\Stokes\ and \Fubini\ formulae lead to the integral \textit{extrusion formula}:
%%%%%%%%%%%%%%%%%%%%%%%%%%%%%%%%
$$\vcenter{\halign{
$\displaystyle#$\hfil&$\displaystyle#\qquad$\hfil&$\displaystyle#$\hfil\cr
\parder\tau\ttt\,
\integrale{\moto_\tau\di{\surf^k}}{}\Boo^k
&\,=\integrale{\moto_\ttt\di{\surf^k}}{}(\der\Boo^k_{\moto,\ttt})\punto\Bv_{\moto,\ttt}
+\integrale{\partial\moto_\ttt\di{\surf^k}}{}\,\Boo^k_{\moto,\ttt}\punto\Bv_{\moto,\ttt}\,,
\cr}}$$
%%%%%%%%%%%%%%%%%%%%%%%%%%%%%%%%
and the related \persone{Henri} \Cartan\  \emph{magic formula} (or \emph{homotopy} formula)
provides the expression the \Lie\ derivative
in terms of the exterior derivative:
%%%%%%%%%%%%%%%%%%%%%%%%%%%%%%%%
$$\Lieder_{\Bv_\moto}\,\Boo^k\equaldef\parder\lambda0\,\moto_\lambda\pull\Boo^k
=(\der\Boo^{k})\punto\Bv_\moto+\der(\Boo^{k}\punto\Bv_\moto)\,,$$
%%%%%%%%%%%%%%%%%%%%%%%%%%%%%%%%
where $\,\surf^k\,$ is a $\,k$-chain with boundary $\,\partial\surf^k\,$,
$\,\Boo^{k}\,$ is a $\,k$-form 
and $\,\Boo^{k}\punto\Bv_\moto\,$ is the
contraction performed by inserting $\,\Bv_\moto\,$ as first argument of
$\,\Boo^{k}\,$.

The homotopy formula generalizes, in terms of forms and exterior derivatives,
the formulae for convective time-derivatives contributed by
\Maxwell\ and \Helmholtz\
in \citep{Maxwell1861,Helmholtz1874,Helmholtz1892}, see \citep[p. 406]{Darrigol2000}.

The homotopy formula may be readily inverted to get
\Palais\ formula for the exterior derivative. 
Indeed, by \Leibniz\ rule for the \Lie\ derivative, 
for any $\,1$-form $\,\Boo^{1}\in\cont^1\di{\MMM\sp\TM^*}\,$ and vector fields
$\,\Bv,\Bw\in\cont^1\di{\MMM\sp\TM}\,$ we have:
%%%%%%%%%%%%%%%%%%%%%%%%%%%%%%%%
$$\vcenter{\halign{
$\displaystyle#$\hfil&$\displaystyle#\qquad$\hfil&$\displaystyle#$\hfil\cr
\der\Boo^{1}\punto\Bv\punto\Bw
&\,=(\Lieder_\Bv\,\Boo^1)\punto\Bw-\der(\Boo^{1}\punto\Bv)\punto\Bw
\vspace{6pt}\cr
&\,=\der_\Bv\,(\Boo^1\punto\Bw)
-\Boo^1\punto\lineare{\Bv,\Bw}
-\der_\Bw\,(\Boo^{1}\punto\Bv)\,.
\cr}}$$
%%%%%%%%%%%%%%%%%%%%%%%%%%%%%%%%
The exterior derivative of a differential 
$\,1$-form is a two-form which is well-defined by \Palais\ formula
because the expression at the r.h.s. fulfills the tensoriality criterion.
The value of the exterior derivative at a point is independent of
the extension of argument vectors to vector fields,
extension needed to compute the 
involved directional and \Lie\ derivative.

An $\,n$-dimensional manifold $\,\MMM\,$ is a \textit{star-shaped manifold}
if there exists a point $\,\Bx_0\in\MMM\,$ and a \emph{homotopy}
$\,\homotopy_\ttt\in\cont^1\di{\MMM\sp\MMM}\,$,
continuous in $\,\ttt\in[0,1]\,$,
such that $\,\homotopy_1\,$ is the identity map, i.e.
$\, \homotopy_1\di\Bx=\Bx\,$ for all $\,\Bx\in\MMM\,$,
and $\,\homotopy_0\,$ is the constant map
$\,\homotopy_0\di\Bx=\Bx_0\,$ for all $\,\Bx\in\MMM\,$.
This homotopy is called a \textit{contraction} to $\,\Bx_0\in\MMM\,$.
Denoting by 
$\,\Bvt=\parder{\tau}{\ttt}\homotopy_\tau\circ\inv\homotopy_\ttt\in\cont^1\di{\MMM\sp\TM}\,$ 
the velocity of the homotopy,
we have the formula
%%%%%%%%%%%%%%%%%%%%%%%%%%%%%%%%
$$\Boo^k=\der\Baa^{(k-1)}+\Bbeta^k\,,$$
%%%%%%%%%%%%%%%%%%%%%%%%%%%%%%%%
with
%%%%%%%%%%%%%%%%%%%%%%%%%%%%%%%%
$$\Baa^{(k-1)}=
\integrale{0}{1}{\homotopy_\ttt}\pull(\Boo^k\punto\Bv)\inde\ttt\,,\qquad
\Bbeta^k=
\integrale{0}{1}\homotopy_\ttt\pull(\der\Boo^k\punto\Bv)\inde\ttt
\,.$$
%%%%%%%%%%%%%%%%%%%%%%%%%%%%%%%%
If $\,\der\Boo^k=0\,$ the form $\,\Boo^k\,$ is exact being $\,\Boo^k=\der\Baa^{(k-1)}\,$.
This is \Poincare\ Lemma:
in a star-shaped manifold any closed form is exact.

\subsection{Classical integral transformations}
\label{sec: integraltrans}

Let $\,\coppia{\MMM^n}{\volform^n}\,$ be a $\,n$-dimensional volume manifold.
The \textit{divergence} of a vector field
$\,\Bv\in\cont^1\di{\MMM\sp\TM}\,$
is defined as the constant of proportionality
between the \Lie\ derivative of the volume form 
along the flow of the vector field 
and the volume form itself:\index{divergence}
%%%%%%%%%%%%%%%%%%%%%%%%%%%%%%%%%
$$\Lieder_\Bv\volform\,=(\diverg\Bv)\,\volform\,.$$
%%%%%%%%%%%%%%%%%%%%%%%%%%%%%%%%%
The divergence may be equivalently defined 
in terms of the exterior derivative by the relation
%%%%%%%%%%%%%%%%%%%%%%%%%%%%%%%%
$$\der(\Bmu\Bv)=(\diverg\Bv)\,\Bmu\,.$$
%%%%%%%%%%%%%%%%%%%%%%%%%%%%%%%%
Indeed, $\,\der\Bmu=0\,$ identically as $\,\der\Bmu\,$ is an $\,(n+1)$-form in an 
$\,n$-dimensional manifold, so that by the homotopy formula:
%%%%%%%%%%%%%%%%%%%%%%%%%%%%%%%%
$$\Lieder_\Bv\,\Bmu=(\der\Bmu)\Bv+\der(\Bmu\Bv)=\der(\Bmu\Bv)\,.$$
%%%%%%%%%%%%%%%%%%%%%%%%%%%%%%%%
From the \VPB\ formula, introduced in section \ref{sec: manifolds}, we may derive all classical
integral transformation formulas, as special cases. Indeed being: 
%%%%%%%%%%%%%%%%%%%%%%%%%%%%%%%%
$$\vcenter{\halign{
\textrm{#}\qquad\hfil&$\displaystyle#$\hfil&$\displaystyle#$\hfil&$#$\hfil\cr
%\vspace{2pt}\cr
gradient:&\der\,\ff\,&=\metric\punto\nabla\ff\,,\qquad&\dim\MMM=n
\vspace{8pt}\cr
curl:&\der(\metric\Bv)\,&=(\rotor\Bv)\,\volform\,,\qquad&\dim\MMM=2
\vspace{8pt}\cr
curl:&\der(\metric\Bv)\,&=\volform\punto(\rotor\Bv)\,,\qquad&\dim\MMM=3
\vspace{8pt}\cr
divergence:&\der(\volform\Bv)\,&=(\diverg\Bv)\volform\,,\qquad&\dim\MMM=n\cr
}}$$
%%%%%%%%%%%%%%%%%%%%%%%%%%%%%%%%
we get the following statements:
%%%%%%%%%%%%%%%%%%%%%%%%%%%%%%%%
\begin{itemize}\item
$\,\surf^1\subset\MMM^n:\,$ 
the \textit{gradient formula}:
\index{gradient theorem}
%%%%%%%%%%%%%%%%%%%%%%%%%%%%%%%%
$$\vcenter{\halign{
$\displaystyle#$\hfil&$\displaystyle#\qquad$\hfil&$\displaystyle#$\hfil\cr
&\integrale{\surf^1}{}\der\ff=
\integrale{\surf^1}{}\metric\nabla\ff=
\integrale{\surf^1}{}\metric\di{\nabla\ff,\Bt}\,\,(\metric\,\Bt)=
\integrale{\partial\surf^1}{}\ff=
\ff\di\BBb-\ff\di\BBa\,,\phantom{aaaaaaaa}
\vspace{9pt}\cr}}$$
%%%%%%%%%%%%%%%%%%%%%%%%%%%%%%%%
with $\,\BBa\,$, $\,\BBb\,$ end points of the curve $\,\surf^1\,$
oriented from $\,\BBa\,$ to $\,\BBb\,$
and $\,\metric\,\Bt\,$ volume form (the signed-length) induced along the curve $\,\surf\,$.
\end{itemize}
%%%%%%%%%%%%%%%%%%%%%%%%%%%%%%%%
\begin{itemize}\item
$\, \surf^2\subset\MMM^3:\,$ 
the \textit{curl formula}:\index{curl formula}
\end{itemize}
%%%%%%%%%%%%%%%%%%%%%%%%%%%%%%%%
$$\vcenter{\halign{
$\displaystyle#$\hfil&$\displaystyle#\qquad$\hfil&$\displaystyle#$\hfil\cr
\vspace{3pt}\cr
\integrale{\surf}{}\der(\metric\Bv)&\,=
\integrale{\surf}{}\volform(\rotor\Bv)=
\integrale{\surf}{}\metric\di{\rotor\Bv,\Bn}\,(\volform\Bn)
=\integrale{\partial\surf}{}\metric\Bv=
\integrale{\partial\surf}{}\metric\di{\Bv,\Bt}\,(\metric\,\Bt)\,,
\vspace{3pt}\cr
\cr}}$$
%%%%%%%%%%%%%%%%%%%%%%%%%%%%%%%%
with $\,\Bn\,$ piecewise smooth field of unit normals to the surface $\,\surf\,$ and
$\,\Bt\,$ unit tangent to the boundary of the surface.
For $\,\dim\MMM=3\,,\quad\dim\surf=2\,$ the \textit{curl theorem} writes:
%%%%%%%%%%%%%%%%%%%%%%%%%%%%%%%%
$$\vcenter{\halign{
$\displaystyle#$\hfil&$\displaystyle#\qquad$\hfil&$\displaystyle#$\hfil\cr
\integrale{\surf}{}\volform\punto(\rotor\Bv)=
\integrale{\surf}{}\der(\metric\punto\Bv)
=\integrale{\partial\surf}{}\metric\punto\Bv\,.
\vspace{3pt}\cr
\cr}}$$
%%%%%%%%%%%%%%%%%%%%%%%%%%%%%%%%
It is evident that the \emph{curl} vector or scalar fields in the formulas above 
are orientation dependent.
%%%%%%%%%%%%%%%%%%%%%%%%%%%%%%%%
\begin{itemize}\item
$\,\dim\MMM=n\,,\dim\surf=k\le{n}\,$ the \textit{divergence formula}:
\index{divergence theorem}
%%%%%%%%%%%%%%%%%%%%%%%%%%%%%%%%
$$\vcenter{\halign{
$\displaystyle#$\hfil&$\displaystyle#\qquad$\hfil&$\displaystyle#$\hfil\cr
\integrale{\surf}{}(\diverg\Bv)\,\volform
=\integrale{\surf}{}\der(\volform\punto\Bv)
=\integrale{\partial \surf}{}\volform\punto\Bv
=\integrale{\partial \surf}{}\metric\di{\Bv,\Bn}\,(\volform\punto\Bn)\,,
\cr}}$$
%%%%%%%%%%%%%%%%%%%%%%%%%%%%%%%%
with $\,\Bn\,$ unit normal to the boundary $\,\partial \surf\,$.
\end{itemize}
%%%%%%%%%%%%%%%%%%%%%%%%%%%%%%%%
\begin{remark}
The definition of gradient, curl and divergence in $\,\Re^3\,$ given above
are based on the following algebraic results \citep{RomanoDiff2007}.
\begin{itemize}\item
To any one-form $\,\der\ff\,$ on $\,\Re^n\,$ there correspond a unique vector 
$\,\nabla\ff\,$ in $\,\Re^n\,$ such that $\,\der\ff=\metric\punto\nabla\ff\,$.
\item
To any two-form $\,\Bomega^2\,$ on $\,\Re^3\,$ there correspond a unique vector 
$\,\Bw\,$ in $\,\Re^3\,$ such that $\,\Bomega^2=\Bmu\punto\Bw\,$, with $\,\Bmu\,$
a given volume form.
\item
All volume forms $\,\Bmu\,$ on $\,\Re^n\,$ are proportional one another.
\end{itemize}
%%%%%%%%%%%%%%%%%%%%%%%%%%%%%%%%
\end{remark}
%%%%%%%%%%%%%%%%%%%%%%%%%%%%%%%%
A noteworthy formula, 
due to \persone{Hermann von} \Helmholtz,
is also a direct consequence of the homotopy formula, see \citep{Deschamps1970,Deschamps1981}.

To see this, given a time-dependent tangent vector field $\,\Bu_\ttt\in\cont^2\di{\MMM\sp\TM}\,$,
we set $\,\Boo^2_\ttt=\volform\punto\Bu_\ttt\,$.
To evaluate the flux of the field $\,\Bu_\ttt\in\cont^2\di{\MMM\sp\TM}\,$
through a surface $\,\surf^2\,$
drifted by a flow $\,\moto_{\tau,\ttt}\in\cont^2\di{\MMM\sp\MMM}\,$, we 
set $\,\dot\Boo^2_\ttt\equaldef\parder\tau\ttt\Boo^2_\tau\,$
and apply the homotopy formula to get:
%%%%%%%%%%%%%%%%%%%%%%%%%%%%%%%%%%%
$$\vcenter{\halign{
\hfil$\displaystyle#$&$\displaystyle#$\hfil&$#$\hfil&$#$\hfil\cr
\parder\tau\ttt\integrale{\moto_{\tau,\ttt}\di{\surf^2}}{}\Boo^2_\tau
&\,=\integrale{\surf^2}{}\dot\Boo^2_\ttt+\Lieder_{\Bv_\moto,\ttt}\,\Boo^2
=\integrale{\surf^2}{}\dot\Boo^2_\ttt+\der(\Boo^2_\ttt\punto\Bv_{\moto,\ttt})+(\der\Boo^2_\ttt)\punto\Bv_{\moto,\ttt}
\,.\cr}}$$
%%%%%%%%%%%%%%%%%%%%%%%%%%%%%%%%%%%
Translating  into the language of vector analysis, recalling that
%%%%%%%%%%%%%%%%%%%%%%%%%%%%%%%%%%%
$$\vcenter{\halign{
\hfil$#$&$#$\hfil&$#$\hfil&$#$\hfil\cr
\volform\punto\Bu_\ttt\punto\Bv_{\moto,\ttt}&\,=\metric\punto(\Bu_\ttt\times\Bv_{\moto,\ttt})\,,
\vspace{8pt}\cr
\der(\metric\punto(\Bu_\ttt\times\Bv_{\moto,\ttt}))&\,=\volform\punto(\rotor(\Bu_\ttt\times\Bv_{\moto,\ttt}))
\,,\cr}}$$
%%%%%%%%%%%%%%%%%%%%%%%%%%%%%%%%%%%
we have:
%%%%%%%%%%%%%%%%%%%%%%%%%%%%%%%%%%%
$$\vcenter{\halign{
\hfil$#$&$#$\hfil&$#$\hfil&$#$\hfil\cr
\der(\Boo^2_\ttt\punto\Bv_{\moto,\ttt})
&\,=\der(\volform\punto\Bu_\ttt\punto\Bv_{\moto,\ttt})
=\volform\punto(\rotor(\Bu_\ttt\times\Bv_{\moto,\ttt}))\,,
\vspace{8pt}\cr
(\der\Boo^2_\ttt)\punto\Bv_{\moto,\ttt}&\,=\der(\volform\punto\Bu_\ttt)\punto\Bv_{\moto,\ttt}
=(\diverg\Bu_\ttt)\volform\punto\Bv_{\moto,\ttt}
\,.\cr}}$$
%%%%%%%%%%%%%%%%%%%%%%%%%%%%%%%%%%%
Substituting into the first expression, we get \Helmholtz's formula:
%%%%%%%%%%%%%%%%%%%%%%%%%%%%%%%%%%%
$$\vcenter{\halign{
\hfil$\displaystyle#$&$\displaystyle#$\hfil&$#$\hfil&$#$\hfil\cr
\parder\tau\ttt\integrale{\moto_{\tau,\ttt}\di{\surf^2}}{}\Boo^2_\tau
&\,=\integrale{\surf^2}{}
\volform\punto(\dot\Bu_\ttt+\rotor(\Bu_\ttt\times\Bv_{\moto,\ttt}))+(\diverg\Bu_\ttt)\volform\punto\Bv_{\moto,\ttt}
\,.\cr}}$$
%%%%%%%%%%%%%%%%%%%%%%%%%%%%%%%%%%%

\section{Tensor bundles and tensor fields}
\label{sec: TBTF}

In a mathematical field theory in physics the geometrical notion of fibre bundle plays a basic role.
A comprehensive exposition may be found in \citep{Saunders1989} and an application oriented treatment
is available in \citep{RomanoDiff2007}.
Here we only give the intuitive idea that a fibre bundle (the \emph{total space}) consists of a manifold
(the \emph{base}) with diffeomorphic manifolds (the \emph{fibers}) attached at each of its points.

Given a point in the fibre bundle the key property is that it is possible to detect univocally
the fiber at which it belongs. This information is provided by a map,
the \emph{projection} from the \emph{total space} onto the \emph{base},
which is differentiable with an injective tangent map.
The fibers endow the manifold with a geometric structure and in applications
are most commonly linear tensor spaces.
The physical concept of tensor fields is geometrically described as follows.

A generic tensor bundle will be denoted by
$\,\terna{\TENS\di\MMM}{\projb_\TENS}\MMM\,$
or often, shortly, by $\,\TENS\di\MMM\,$.
The projection $\,\projb_\TENS\in\cont^1\di{\TENS\di\MMM\sp\MMM}\,$
is surjective with surjective tangent maps at all points (a surjective submersion).
A section $\,\Bs\in\cont^1\di{\MMM\sp\TENS\di\MMM}\,$,
of a fiber bundle $\,\projb_\TENS\in\cont^1\di{\TENS\di\MMM\sp\MMM}\,$,
is a map such that $\,\projb_\TENS\circ\Bs=\ID\MMM\,$, the identity map on $\,\MMM\,$.

On a manifold $\,\MMM\,$ with $\,\dim\MMM=n\,$ we consider the bundles
listed below with the corresponding sections:
%%%%%%%%%%%%%%%%%%%%%%%%%%%%%%%%%
$$\vcenter{\halign{
#\hfil&#\hfil\cr
$\,\terna{\FUN\di\MMM}{\projb_\FUN}\MMM\,$
&$\,\equi\,$ scalar fields,
\vspace{8pt}\cr
$\,\terna{\TANG\MMM}{\projb_\TAN}\MMM\,$
&$\,\equi\,$ tangent vector fields,
\vspace{8pt}\cr
$\,\terna{\CTANG\MMM}{\projb_\CTAN}\MMM\,$
&$\,\equi\,$ cotangent vector fields,
\vspace{8pt}\cr
$\,\terna{\COV\di\MMM}{\projb_\COV}\MMM\,$
&$\,\equi\,$ covariant tensor fields,
\vspace{8pt}\cr
$\,\terna{\CON\di\MMM}{\projb_\CON}\MMM\,$
&$\,\equi\,$ contravariant tensor fields.
%\cr}}$$
%%%%%%%%%%%%%%%%%%%%%%%%%%%%%%%%%
%%%%%%%%%%%%%%%%%%%%%%%%%%%%%%%%%
%$$\vcenter{\halign{
%#\hfil&#\hfil\cr
\vspace{8pt}\cr
$\,\terna{\MIX\di\MMM}{\projb_\MIX}\MMM\,$
&$\,\equi\,$ mixed tensor fields.
\vspace{8pt}\cr
$\,\terna{\ALT^k\di\MMM}{\projb_\ALT}\MMM\,$
&$\,\equi\,$ alternating $\,k$-tensor fields ($\,k$-forms).
\vspace{8pt}\cr
$\,\terna{\VOL\di\TM}{\projb_\VOL}\MMM\,$
&$\,\equi\,$ nowhere null $\,n$-forms.
\cr}}$$
%%%%%%%%%%%%%%%%%%%%%%%%%%%%%%%%%

\section{Inner and outer orientations, odd forms}
\label{sec: orientation}

The reader interested in the issues of orientation of manifolds and integration over compact manifolds,
whether orientable or not, is addressed to the mathematical treatment given in 
\citep{AbrahamMarsden2002}.
A presentation of basic aspects and a discussion with applications to electromagnetism is provided in
\citep{Bossavit1991,Bossavit2004}, \citep{Tonti1995,Tonti2002} and references therein.

A treatment of \emph{odd} and \emph{even} (or \emph{plain} and \emph{twisted}) forms in oriented affine manifolds,
with emphasis on formulation of \Maxwell\ equations
in the $\,4$D space-time and in \Minkowski\ relativistic space-time,
has been provided in 
\citep{HehlObukhov2003} and revisited with a punctual analysis in \citep{Marmo2005,MarmoTulczyjew2006}.
Due to orientability of space-time, the relevance of odd forms in physics has been
questioned in a recent article by \cite{daRocha2010},
with an ongoing controversy 
\citep{Itin2010,daRochabis2010}.

In fact, the notion of
\emph{even and odd $\,k$-covectors}
and of \emph{even and odd $\,k$-forms},
introduced in \citep{deRham1931,deRham1955, Schouten1951},
is required not only to perform integration over non-orientable manifolds, but also
to define the flux of a field across a surface or the winding of a field around a cycle,
in such a way that the result depends only on the outer orientation of the integration manifold,
but neither on the inner orientation of the manifold nor on the orientation of the ambient manifold.

In the context of electromagnetic induction theory, integration over non-orientable
manifold is required, for instance, to evaluate the global electric charge
on a \persone{M\"obius} strip or on a \persone{Kline} bottle.
On the other hand, integration over outer oriented manifold and on its boundary is
required to properly formulate the \Ampere\ law of induction,
see Section \ref{sec: MatAMP}.

Let us preliminarily provide the definition of immersed manifold.
%%%%%%%%%%%%%%%%%%%%%%%%%%%%%%%%%%%
\begin{definition}[Immersion]\label{def: immersion}
A smooth map $\,\uu\in\cont^1\di{\surf^k\sp\MMM^n}\,$
is called an immersion, of the $\,k$-manifold $\,\surf^k\,$
into the $\,n$-manifold $\,\MMM^n\,$ with $\,k\le{n}\,$,
if for any $\,\Bx\in\surf^k\,$ the tangent map
$\,\TT_\Bx\uu\in\cont^1\di{\TANG_\Bx\surf^k\sp\TANG_{\uu\di\Bx}\MMM^n}\,$
is injective.
\end{definition}
%%%%%%%%%%%%%%%%%%%%%%%%%%%%%%%%%%%
The range of an injective immersion $\,\uu\in\cont^1\di{\surf^k\sp\MMM^n}\,$,
of a compact and connected $\,k$-dimensional manifold $\,\surf^k\,$ 
with boundary into an $\,n$-dimensional manifold $\,\MMM^n\,$ without boundary,
is a connected $\,k$-dimensional submanifold $\,\uu\di{\surf^k}\,$ of $\,\MMM^n\,$.
Denoting by $\,\set{\partial_1,\ldots,\partial_k}\,$ the standard basis of $\,\Re^k\,$,
let us consider a tesselation of $\,\surf^k\,$ whose  simplicial map 
$\,\simp\in\cont^1\di{\Simp^k\sp\MMM}\,$ at $\,\Bx=\simp\di{0^k}\in\surf^k\,$
has domain is the reference simplex
%%%%%%%%%%%%%%%%%%%%%%%%%%%%%%%%%%%
$$\Simp^k=\set{\xx\in\Re^n\talechesia\xx_\ii\ge0\perogni\ii\,,\,\sum_{\ii=1,k}\xx_i\le1}\,,$$
%%%%%%%%%%%%%%%%%%%%%%%%%%%%%%%%%%%
and maps the basis of $\,\Re^k\,$ in the basis
$\,\set{\Be_{1},\ldots,\Be_k}_\Bx\,$ of $\,\TANG_\Bx\surf^k\,$ with:
%%%%%%%%%%%%%%%%%%%%%%%%%%%%%%%%%%%
$\,\Be_i=\simp\di{\partial_i}\,$.
%%%%%%%%%%%%%%%%%%%%%%%%%%%%%%%%%%%
\begin{definition}[Volumes and point-orientations]\label{def: orientation}
In a $\,n$-dimensional manifold $\,\MMM^n\,$,
a volume $\,\volform^n\di\Bx\in\ALT^n\di{\TANG_\Bx\MMM^n}\,$
 is a non-null $\,n$-covector at $\,\Bx\in\MMM^n\,$.
Being the linear space of $\,n$-covectors at $\,\Bx\in\MMM^n\,$ one dimensional,
the equivalence relation of positive proportionality defines,
at $\,\Bx\in\MMM^n\,$,  two disjoint classes of volumes $\,\orientpairx\,$,
named point-orientations.
\end{definition}
%%%%%%%%%%%%%%%%%%%%%%%%%%%%%%%%%%%
%%%%%%%%%%%%%%%%%%%%%%%%%%%%%%%%%%%
\begin{definition}[Inner orientation, volume manifolds]\label{def: orientable}
A manifold $\,\MMM^n\,$ endowed with a smooth volume form, 
viz. with a nowhere vanishing section 
$\,\volformn\in\cont^1\di{\MMM^n\sp\VOL\di{\TM^n}}\,$
of the bundle $\,\terna{\VOL\di{\TM^n}}{\projb_\VOL}{\MMM^n}\,$
is said to be inner oriented.
The pair $\,\coppia{\MMM^n}{\volform^n}\,$ is called a smooth volume manifold.
\end{definition}
%%%%%%%%%%%%%%%%%%%%%%%%%%%%%%%%%%%
%%%%%%%%%%%%%%%%%%%%%%%%%%%%%%%%%%%
\begin{figure}[h]
\centering
%\setlength{\fboxsep}{0mm}
%\fbox{			
\includegraphics[width=.5\textwidth]{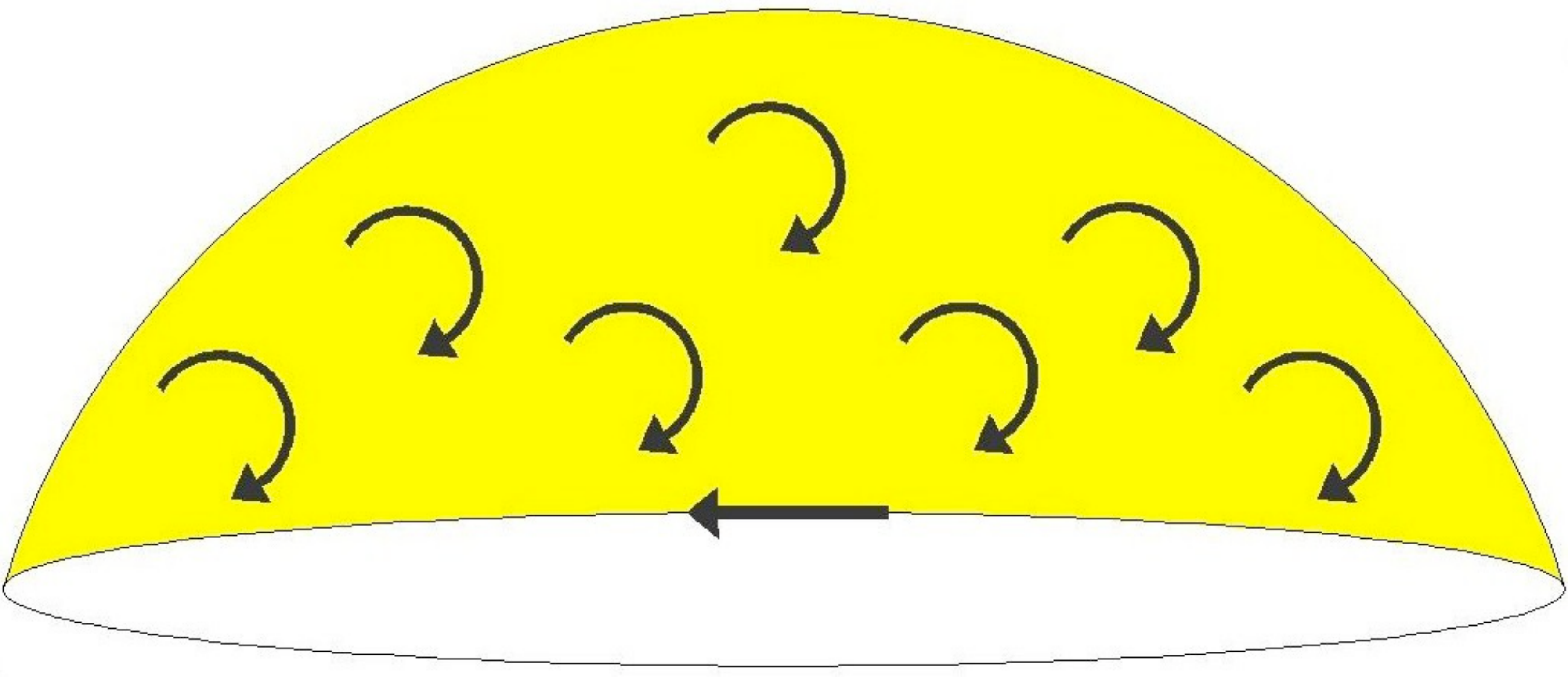}
%}
\caption{inner-oriented surface and boundary}
\label{fig: inner}
\end{figure}
%%%%%%%%%%%%%%%%%%%%%%%%%%%%%%%%%%%
Let us adopt the redundant terminology of \emph{even} (or \emph{plain}) form,
to contrast \emph{odd} (or \emph{twisted}) form.
%%%%%%%%%%%%%%%%%%%%%%%%%%%%%%%%%%%
\begin{definition}[Even and odd covectors]\label{def: evenodd}
In a $\,n$-dimensional manifold $\,\MMM^n\,$,
$\,k$-covectors $\,\Boo^k\di\Bx\in\ALT^k\di{\TANG_\Bx\MMM^n}\,$,
with $\,k\le{n}\,$
are assumed to be function of the orientation of the manifold.
\emph{Even} covectors are invariant with respect to the orientation,
while \emph{odd} covectors change sign as the orientation changes.
\end{definition}
%%%%%%%%%%%%%%%%%%%%%%%%%%%%%%%%%%%
%%%%%%%%%%%%%%%%%%%%%%%%%%%%%%%%%%%
\begin{definition}[Integral over inner oriented submanifolds]\label{def: innerint}
Given a even $\,k$-form $\,\Boo^k:{\MMM^n}\mapsto\ALT^k\di{\TM^n}\,$
in an $\,n$-manifold $\,\MMM^n\,$,
the integral, over an inner oriented $\,k$-manifold
$\,\coppia{\surf^k_\inner}{\volform^k_\inner}\,$
with immersion $\,\uu\in\cont^1\di{\surf^k\sp\MMM^n}\,$,
of the pull-back $\,k$-form $\,\uu\pull\Boo^k:{\surf^k}\mapsto\ALT^k\di{\TANG\surf^k}\,$,
is denoted by:
%%%%%%%%%%%%%%%%%%%%%%%%%%%%%%%%%%%
$$\,\integrale{\surf^k_\inner}{}\uu\pull\Boo^k\,$$
%%%%%%%%%%%%%%%%%%%%%%%%%%%%%%%%%%%
and is defined, \emph{ˆ la \Riemann}, as the inductive limit,
along a family of simplicial tesselations directed by refinement,
of finite sums of scalar terms:
%%%%%%%%%%%%%%%%%%%%%%%%%%%%%%%%%%%
$$\vcenter{\halign{
\hfil$#$&$#$\hfil&$#$\hfil&$#$\hfil\cr
\signum\di{\volform^k_\inner\di{\Be_1,\ldots,\Be_k}_\Bx}\,\inv{(k!)}\,
&\Boo^k\di{\uu\push\Be_1,\ldots,\uu\push\Be_k}_{\uu\di\Bx}
\,,\qquad\Bx\in\surf^k\,.\cr}}$$
%%%%%%%%%%%%%%%%%%%%%%%%%%%%%%%%%%%
\end{definition}
%%%%%%%%%%%%%%%%%%%%%%%%%%%%%%%%%%%
The sign of the integral, as defined above, is independent of permutations of the
basis $\,\set{\Be_{1},\ldots,\Be_k}_\Bx\,$ in $\,\TANG_\Bx\surf^k\,$,
the significant property being the following:
%%%%%%%%%%%%%%%%%%%%%%%%%%%%%%%%
\begin{itemize}\item
Changing the inner orientation results in changing the integral of a form into its opposite.
\end{itemize}
%%%%%%%%%%%%%%%%%%%%%%%%%%%%%%%%
This definition is suitable to compare the value of a global vortex on an inner oriented 
surface in the \Euclid $\,3$-space,
with the corresponding value of the global circulation around its inner oriented boundary circuit,
see fig. \ref{fig: inner}.
%%%%%%%%%%%%%%%%%%%%%%%%%%%%%%%%%%%
\begin{definition}[Volume manifolds, induced measures and densities]\label{def: volume}
A \emph{volume form} in a $\,n$-manifold $\,\MMM^n\,$ is a field of volumes
$\,\volform^n\in\EXForms^n\di{\TANG\MMM^n\sp\Re}\,$.
The pair $\,\coppia{\MMM^n}{\volform^n}\,$ is called a volume manifold.
The induced measure is defined by the map:
%%%%%%%%%%%%%%%%%%%%%%%%%%%%%%%%%%%
$$\,\meas\di{\volform^n}\equaldef\signum\di{\volform^n}\circ\volform^n\,.$$
%%%%%%%%%%%%%%%%%%%%%%%%%%%%%%%%%%%
The density associated with a scalar field $\,\rho:\MMM^n\mapsto\Re\,$ 
and a volume form $\,\volform^n\in\EXForms^n\di{\TANG\MMM^n\sp\Re}\,$ 
is the product
$\,\rho\,\meas\di{\volform^n}\,$
\end{definition}
%%%%%%%%%%%%%%%%%%%%%%%%%%%%%%%%%%%
\begin{definition}[Integral of a density]\label{def: densintegral}
Let us consider in a compact $\,n$-manifold $\,\MMM^n\,$
a density $\,\rho\,\meas\di{\volform^n}\in\EXForms^n\di{\TANG\MMM^n\sp\Re}\,$.
Then, its integral over a manifold $\,\surf^k\,$
with immersion $\,\uu\in\cont^1\di{\surf^k\sp\MMM^n}\,$:
%%%%%%%%%%%%%%%%%%%%%%%%%%%%%%%%%%%
$$\,\integrale{\surf^k}{}\,\uu\pull(\rho\,\meas\di{\volform^k})\,$$
%%%%%%%%%%%%%%%%%%%%%%%%%%%%%%%%%%%
is defined, \emph{ˆ la \Riemann}, as the inductive limit of finite sums of scalar terms:
%%%%%%%%%%%%%%%%%%%%%%%%%%%%%%%%%%%
$$\vcenter{\halign{
\hfil$#$&$#$\hfil&$#$\hfil&$#$\hfil\cr
\inv{(n!)}\,
&\rho\di{\uu\di\Bx}\,\meas\di{\volform^k}\punto\di{\uu\push\Be_1,\ldots,\uu\push\Be_k}_{\uu\di\Bx}
\,,\quad\Bx\in\surf^k\,,\cr}}$$
%%%%%%%%%%%%%%%%%%%%%%%%%%%%%%%%%%%
along a family of simplicial tesselations directed by refinement.
The integral is then independent of permutations of the basis vectors 
$\,\uu\push\Be_i, i=1,\ldots,k\,$.
\end{definition}
%%%%%%%%%%%%%%%%%%%%%%%%%%%%%%%%%%%
Densities can be integrated over even non-orientable manifolds,
since arbitrary changes of point-orientations do not affect the integral.
The next notion provides a generalization of densities to exterior forms of lower order.
%%%%%%%%%%%%%%%%%%%%%%%%%%%%%%%%%%%
\begin{definition}[Twisted forms]\label{def: twistform}
In a volume manifold $\,\coppia{\MMM^n}{\volform^n}\,$,
a map assigning, to a point $\,\Bx\in\MMM\,$ and to a point-orientation 
$\,\orient_\Bx\in\orientpairx\,$ of the
tangent space $\,\TANG_\Bx\MMM\,$,
a $\,k$-covector $\,\Boo^k_\Bx\in\ALT^k\di{\TANG_\Bx\MMM^n}\,$, $\,k\le{n}\,$,
or its opposite depending on whether $\,\orient_\Bx=\orient^+_\Bx\,$ or $\,\orient=\orient^-_\Bx\,$,
is called a odd $\,k$-form, and is written as:
%%%%%%%%%%%%%%%%%%%%%%%%%%%%%%%%%%%
$$\,\signum\di{\volform^n}\circ\Boo^k\,.$$
%%%%%%%%%%%%%%%%%%%%%%%%%%%%%%%%%%%
\end{definition}
%%%%%%%%%%%%%%%%%%%%%%%%%%%%%%%%%%%
Accordingly, densities are \emph{odd} volume forms.
In an analogous way, the notion of \emph{odd} vector fields may be introduced as follows.
%%%%%%%%%%%%%%%%%%%%%%%%%%%%%%%%%%%
\begin{definition}[Odd vector fields]\label{def: twistvec}
A \emph{odd} vector field on a $\,n$-manifold $\,\MMM^n\,$
is a map which assigns to a point $\,\Bx\in\MMM\,$ and to a point-orientation 
$\,\orient_\Bx\in\orientpairx\,$ of the
tangent space $\,\TANG_\Bx\MMM\,$ a
vector $\,\Bv_\Bx\in\TANG_\Bx\MMM^n\,$
or its opposite depending on whether $\,\orient_\Bx=\orient^+_\Bx\,$ or $\,\orient=\orient^-_\Bx\,$,
and may then be written as: $\,\signum\di{\volform^n}\circ\Bv\,$.
\end{definition}
%%%%%%%%%%%%%%%%%%%%%%%%%%%%%%%%%%%
%%%%%%%%%%%%%%%%%%%%%%%%%%%%%%%%%%%
\begin{definition}[Outer orientability]\label{def: outorientation}
In a volume manifold $\,\coppia{\MMM^n}{\volform^n}\,$, 
a $\,k$-manifold $\,\surf^k\,$
with immersion $\,\uu\in\cont^1\di{\surf^k\sp\MMM^n}\,$
is outer orientable if there exists a $\,(n-k)$-tuple of
linearly independent smooth vector fields
$\,\Bn_i\in\cont^1\di{\surf^k\sp\TANG_{\uu\di{\surf^k}}\MMM^n}\,$
along $\,\uu\in\cont^1\di{\surf^k\sp\MMM^n}\,$,
that is vector fields fulfilling the commutative diagram: $\,\,$
%%%%%%%%%%%%%%%%%%%%%%%%%%%%%%%%%s
$$\begin{aligned}
\xymatrix{ 
&\TM
\ar[d]^{\projb_\TAN}\\ 
{\surf^k} 
\ar[ur]^{\Bn_i} 
\ar[r]^{\uu} 
&{\MMM^n}
}
\end{aligned}
\equi
\projb_\TAN\circ\Bn_i=\uu\,,
$$
%%%%%%%%%%%%%%%%%%%%%%%%%%%%%%%%%
whose values $\,\Bn_i\di\Bx\in\TANG_{(\uu\di\Bx)}\MMM^n\,$, at each point $\,\Bx\in\surf^k\,$,
are transversal to $\,\TANG_{\uu\di\Bx}\uu\di{\surf^k}=\uu\push(\TANG_\Bx\surf^k)\,$, i.e.
are such that $\,\TANG_{(\uu\di\Bx)}\MMM^n=\TANG_{\uu\di\Bx}\uu\di{\surf^k}\oplus\BN_\Bx\,$,
where $\,\BN_\Bx\,$ is the linear span of the vectors $\,\Bn_i\di\Bx\,$.
\end{definition}
%%%%%%%%%%%%%%%%%%%%%%%%%%%%%%%%%%%
%%%%%%%%%%%%%%%%%%%%%%%%%%%%%%%%%%%
\begin{definition}[Global, inner and outer volume forms]\label{def: gioorientation}
Let $\,\coppia{\MMM^n}{\volform^n}\,$ be a smooth volume manifold 
and $\,\surf^k\,$, $\,k\le{n}\,$, an outer orientable immersed manifold
with immersion $\,\uu\in\cont^1\di{\surf^k\sp\MMM^n}\,$.
A smooth pair of related outer volume form:
%%%%%%%%%%%%%%%%%%%%%%%%%%%%%%%%%%%
$$\,\volform^{n-k}_\outer\in\cont^1\di{\surf^k\sp\ALT^{n-k}\di{\TANG_{\uu\di{\surf^k}}\MMM^n}}\,,$$
%%%%%%%%%%%%%%%%%%%%%%%%%%%%%%%%%%%
and inner volume form 
$\,\volform^k_\inner\in\cont^1\di{\surf^k\sp\VOL\di{\TANG\surf^k}}\,$
may be defined by setting, at each $\,\Bx\in\surf^k\,$:
%%%%%%%%%%%%%%%%%%%%%%%%%%%%%%%%%%%
$$\vcenter{\halign{
\hfil$#$&$#$\hfil&$#$\hfil&$#$\hfil\cr
\volform^k_\inner\,\di{\Be_1,\ldots,\Be_k}_\Bx\,&\,\volform^{n-k}_\outer\,\di{\Bn_1,\ldots,\Bn_{(n-k)}}_{\uu\di\Bx}
\vspace{8pt}\cr
&\,=\volform^n\,\di{\Bn_{1}\ldots,\Bn_{(n-k)},\uu\push\Be_1,\ldots,\uu\push\Be_k}_{\uu\di\Bx}
\,,\cr}}$$
%%%%%%%%%%%%%%%%%%%%%%%%%%%%%%%%%%%
for any basis $\,\set{\Be_{1},\ldots,\Be_k}_\Bx\,$ of $\,\TANG_\Bx\surf^k\,$
and for any list $\,\set{\Bn_{k+1}\ldots,\Bn_n}_{\uu\di\Bx}\,$
of $\,n-k\,$ vector fields fulfilling the requirement of
Definition \ref{def: outorientation}.
\end{definition}
%%%%%%%%%%%%%%%%%%%%%%%%%%%%%%%%%%%

%%%%%%%%%%%%%%%%%%%%%%%%%%%%%%%%%%%
\begin{definition}[Integral over outer oriented submanifolds]\label{def: outerint}
Let a vo\-lume $\,n$-manifold $\,\coppia{\MMM^n}{\volform^n}\,$
and a \emph{odd} $\,k$-form $\,\Boo^k:{\MMM^n}\mapsto\ALT^k\di{\MMM^n}\,$ be given.
The integral, over a connected outer oriented $\,k$-manifold 
$\,\coppia{\surf^k_\outer}{\volform^k_\outer}\,$
with immersion $\,\uu\in\cont^1\di{\surf^k\sp\MMM^n}\,$,
of the $\,k$-form 
$\,\uu\pull\Boo^k:{\surf^k}\mapsto\ALT^k\di{\TANG\surf^k}\,$
is denoted by:
%%%%%%%%%%%%%%%%%%%%%%%%%%%%%%%%%%%
$$\,\integrale{\surf^k_\outer}{}\uu\pull\Boo^k\,,$$
%%%%%%%%%%%%%%%%%%%%%%%%%%%%%%%%%%%
and is defined, ˆ la \Riemann, as the inductive limit,
along a family of simplicial tesselations directed by refinement,
of finite sums of scalar terms:
%%%%%%%%%%%%%%%%%%%%%%%%%%%%%%%%%%%
$$\vcenter{\halign{
\hfil$#$&$#$\hfil&$#$\hfil&$#$\hfil\cr
\signum\di{\volform^k_\inner\di{\Be_1,\ldots,\Be_k}_\Bx}\,\inv{(k!)}\,
&\Boo^k\di{\uu\push\Be_1,\ldots,\uu\push\Be_k}_{\uu\di\Bx}
\,,\qquad\Bx\in\surf^k_\outer\,,\cr}}$$
%%%%%%%%%%%%%%%%%%%%%%%%%%%%%%%%%%%
the volume form $\,\volform^k_\inner\in\cont^1\di{\surf^k\sp\VOL\di{\TANG\surf^k}}\,$
being the one induced by the forms:
%%%%%%%%%%%%%%%%%%%%%%%%%%%%%%%%%%%
$$\vcenter{\halign{
\hfil$#$&$#$\hfil&$#$\hfil&$#$\hfil\cr
\volform^{n-k}_\outer&\,\in\cont^1\di{\surf^k\sp\ALT^{n-k}\di{\TANG_{\uu\di{\surf^k}}\MMM^n}}\,,
\vspace{8pt}\cr
\volformn&\,\in\cont^1\di{\surf^k\sp\VOL^{n}\di{\TANG_{\uu\di{\surf^k}}\MMM^n}}
\,,\cr}}$$
%%%%%%%%%%%%%%%%%%%%%%%%%%%%%%%%%%%
according to Definition \ref{def: outorientation}.
\end{definition}
%%%%%%%%%%%%%%%%%%%%%%%%%%%%%%%%%%%
%%%%%%%%%%%%%%%%%%%%%%%%%%%%%%%%
\begin{figure}[h]
\centering
%\setlength{\fboxsep}{0mm}
%\fbox{			
\includegraphics[width=.5\textwidth]{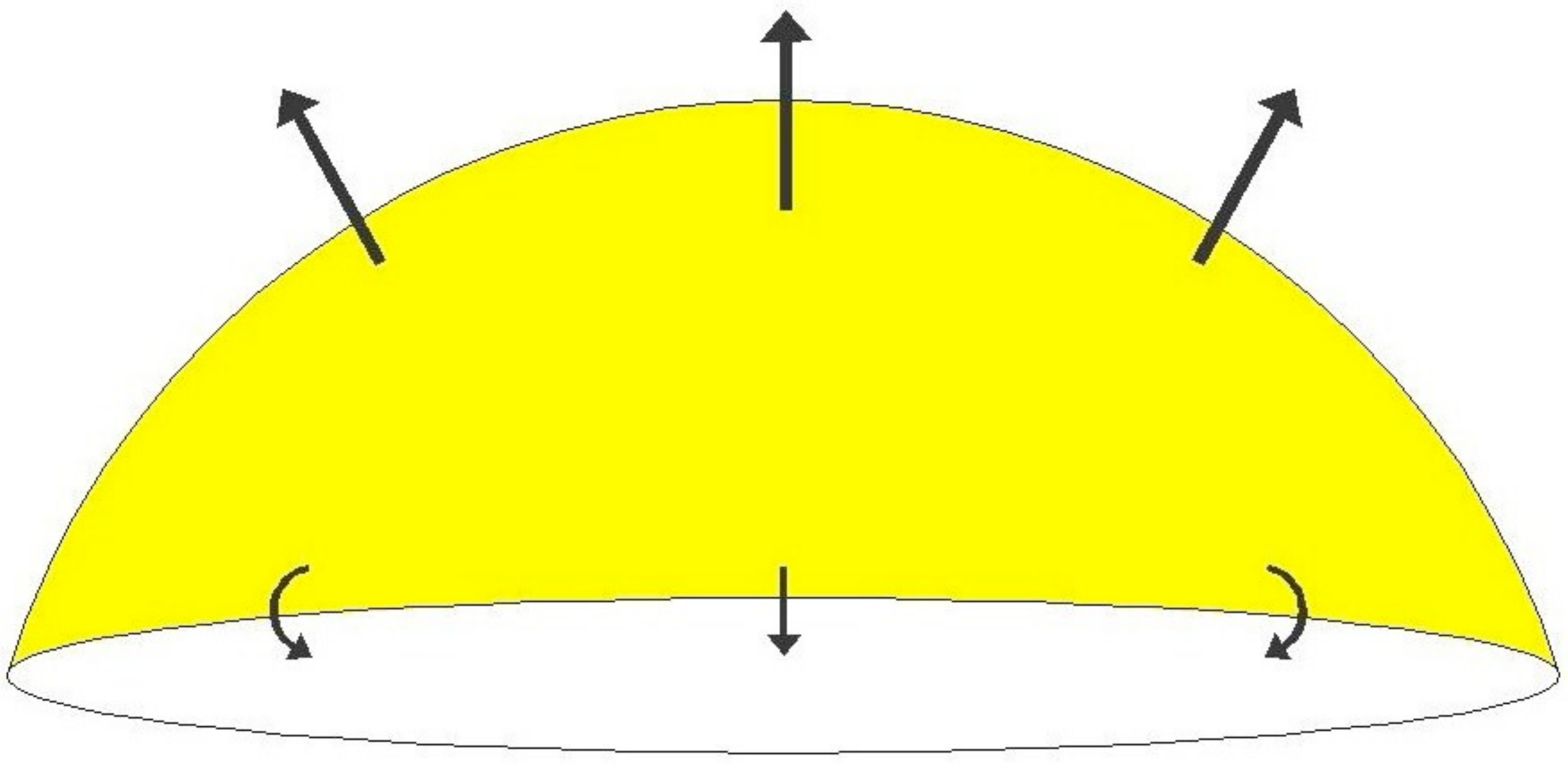}
%}
\caption{outer-oriented surface and boundary}
\label{fig: outer}
\end{figure}
%%%%%%%%%%%%%%%%%%%%%%%%%%%%%%%%

%%%%%%%%%%%%%%%%%%%%%%%%%%%%%%%%%%%
\begin{itemize}\item
Changing the orientation of the ambient manifold results in changing the induced
inner orientation of the integration manifold but not the value of the integral,
because the integrand is a \emph{odd} form which also changes sign.
\item
Changing the outer orientation of the integration manifold results in changing the 
inner orientation induced by the ambient orientation.
The integral is then changed into its opposite.
\end{itemize}
%%%%%%%%%%%%%%%%%%%%%%%%%%%%%%%%%%%

This definition is suitable to define the global flux across an outer oriented 
surface in the \Euclid $\,3$-space,
and, likewise, to define the global winding around its outer oriented boundary circuit,
see fig. \ref{fig: outer}.
In inner oriented volume manifolds,
the integral of a \emph{odd} form over an outer oriented hypersurface as the physical meaning of 
a flux across the hypersurface, because its sign depends only upon the surface outer orientation,
and the integral over the outer oriented boundary cycle has the meaning of winding
around the circuit, see fig. \ref{fig: outer}.
%%%%%%%%%%%%%%%%%%%%%%%%%%%%%%%%%%%

Let us now consider an orientable compact and connected 
$\,k$-manifold $\,\surf^k\,$ 
and the canonical immersion $\,\partial\uu\in\cont^1\di{\partial\surf^k\sp\surf^k}\,$
of its $\,(k-1)$-dimensional boundary manifold $\,\partial\surf^k\,$
into the $\,k$-manifold $\,\surf^k\,$.

For an inner oriented surface $\,\surf^k_\inner\,$,
see fig. \ref{fig: inner},
the \VPB\ formula of Section \ref{sec: manifolds} for a \emph{even} $\,k$-form $\,\Boo^k\,$ writes:
%%%%%%%%%%%%%%%%%%%%%%%%%%%%%%%%%
$$\integrale{\surf^k_\inner}{}\der(\uu\pull\Boo^{k})
=\ointegrale{\partial\uu\di{\partial\surf^k_\inner}}{}\uu\pull\Boo^{k}
=\ointegrale{\partial\surf^k_\inner}{}\partial\uu\pull(\uu\pull\Boo^{k})\,.$$
%%%%%%%%%%%%%%%%%%%%%%%%%%%%%%%%%
\begin{itemize}\item
Changing the inner orientation of the surface $\,\surf_\inner\,$, 
all integrals in the equality will change sign, so that the equality is still valid.
\end{itemize}
%%%%%%%%%%%%%%%%%%%%%%%%%%%%%%%%
By definition \ref{def: outerint},
the \VPB\ formula holds also for 
the integrals of \emph{odd} forms over outer oriented manifolds.
In fact, let an outer orientation across a $\,k$-manifold $\,\surf^k_\outer\,$,
see fig. \ref{fig: outer},
(a crossing direction for the flux)
and the induced outer orientation around the boundary circuit 
(a turning sense for the winding) be given.
The \VPB\ formula writes:
%%%%%%%%%%%%%%%%%%%%%%%%%%%%%%%%%
$$\integrale{\surf^k_\outer}{}\der(\uu\pull\Boo^{k})
=\ointegrale{\partial\uu\di{\partial\surf^k_\outer}}{}\uu\pull\Boo^{k}
=\ointegrale{\partial\surf^k_\outer}{}\partial\uu\pull(\uu\pull\Boo^{k})\,.$$
%%%%%%%%%%%%%%%%%%%%%%%%%%%%%%%%%
\begin{itemize}\item 
Changing the orientation of the ambient manifold results in changing the induced
inner orientations of the integration manifolds but not the value of the integrals
due to the sign change of the \emph{odd} integrand form.
\item 
Changing the outer orientation of the surface,
and the associated outer orientation on the boundary circuit, all integrals will change sign
and the equality still holds.
\end{itemize}
%%%%%%%%%%%%%%%%%%%%%%%%%%%%%%%%

\section{Motions and displacements}
\label{sec: MD}

The \emph{ambient space} $\,\coppia{\EU}{\metric}\,$
is a finite dimensional \Riemann\ manifold $\, \EU\,$ without boundary,
endowed with a metric tensor field $\,\metric\,$.
Points in the ambient space are denoted by $\,\Bx\in\EU\,$.
The usual ambient space is the flat \Euclid\ $\,3$-D space.

The \emph{material body} $\,\CORPO\,$
is a set of labels, the \emph{particles} $\,\partic\in\CORPO\,$,
which become available to physical experience in their  
\textit{spatial motion} $\,\moto:\CORPO\times\II\mapsto\EU\,$
through the space during an open observation time interval $\,\II\,$.

\emph{Spatial events} and
\emph{material events}  
are respectively the elements of the manifolds
$\,\EU\times\II\,$ and $\,\CORPO\times\II\,$.

To a spatial motion there correspond at each time $\,\ttt\in\II\,$
a \emph{material configu\-ration} map
$\,\moto_\ttt\in\cont^1\di{\CORPO\sp\conft}\,$
which is a diffeomorphisms of the body manifold $\,\CORPO\,$
onto the placement manifold $\,\conft\,$.

The \textit{material displacement}
from a source placement $\,\conf_\ttt=\moto_\ttt\di\CORPO\,$
to the target placement $\,\conf_\tau=\moto_\tau\di\CORPO\,$,
is the diffeomorphism:
%%%%%%%%%%%%%%%%%%%%%%%%%%%%%%%%%s
$$\,\moto_{\tau,\ttt}\equaldef\moto_\tau\circ\inv\moto_\ttt
\in\cont^1\di{\conf_\ttt\sp\conf_\tau}\,,$$
%%%%%%%%%%%%%%%%%%%%%%%%%%%%%%%%%s
providing the position in $\,\conf_\tau\,$ at time $\,\tau\in\II\,$
of the particle which occupies the given position in $\,\conf_\ttt\,$ at time $\,\ttt\in\II\,$.

To emphasize the distinction between the material body and the ambient space,
it is expedient to consider the inclusion map
$\,\Bi_{\moto_\ttt}\in\cont^1\di{\conft\sp\EU}\,$
and then define the
\emph{spatial configuration} map by
$\,\moto_\ttt=\Bi_{\moto_\ttt}\di{\moto_\ttt}\in\cont^1\di{\CORPO\sp\Bi_{\moto_\ttt}\di\conft}\,$
and the \textit{spatial displacement} map
$\,\moto_{\tau,\ttt}\in\cont^1\di{\Bi_{\moto_\ttt}\di{\conf_\ttt}\sp\Bi_{\moto,\tau}\di{\conf_\tau}}\,$
corresponding to material displacement map
$\,\moto_{\tau,\ttt}\in\cont^1\di{\conf_\ttt\sp\conf_\tau}\,$,
according to the commutative diagram:
%%%%%%%%%%%%%%%%%%%%%%%%%%%%%%%%%s
$$\begin{aligned}
\xymatrix{ 
\EU
\ar[r]^{\moto_{\tau,\ttt}}
&\EU\\ 
{\conft}
\ar[r]_{\moto_{\tau,\ttt}}
\ar[u]^{\Bi_{\moto_\ttt}} 
&{\conf_\tau}
\ar[u]_{\Bi_{\moto,\tau}} 
}
\end{aligned}
\equi
\moto_{\tau,\ttt}\circ\Bi_{\moto_\ttt}\equaldef\Bi_{\moto,\tau}\circ\moto_{\tau,\ttt}\,.
$$
%%%%%%%%%%%%%%%%%%%%%%%%%%%%%%%%%
Acting with the tangent functor, gives the commutative diagram:
%%%%%%%%%%%%%%%%%%%%%%%%%%%%%%%%%s
$$\begin{aligned}
\xymatrix{ 
\TEU
\ar[r]^{\TT\moto_{\tau,\ttt}}
&\TEU\\ 
{\TANG\conft}
\ar[r]_{\TT\moto_{\tau,\ttt}}
\ar[u]^{\TT\Bi_{\moto_\ttt}} 
&{\TANG\conf_\tau}
\ar[u]_{\TT\Bi_{\moto,\tau}} 
}
\end{aligned}
\equi
\TT\moto_{\tau,\ttt}\circ\TT\Bi_{\moto_\ttt}=\TT\Bi_{\moto,\tau}\circ\TT\moto_{\tau,\ttt}\,.
$$
%%%%%%%%%%%%%%%%%%%%%%%%%%%%%%%%%
The roles of the material and the spatial displacement maps may be illustrated by the following remark.

Let the body $\,\CORPO\,$ be two-dimensional (a membrane) and let
two coordinate line systems be drawn in the placements 
$\,\conf_\ttt\,$ and $\,\conf_\tau\,$.
Then:
%%%%%%%%%%%%%%%%%%%%%%%%%%%%%%%%%
\begin{itemize}\item
The \emph{material displacement} map is assigned by a rule which evaluates 
the pair of coordinates of the target point in $\,\conf_\tau\,$ as a function of 
the pair of coordinates of a source point in $\,\conf_\ttt\,$.
If the coordinated system is convected by the motion, the pair of coordinates
in $\,\conf_\ttt\,$ and in $\,\conf_\tau\,$ are the same.
\item
The \emph{spatial displacement} map in the three-dimensional ambient space is
instead assigned by the rule which evaluates 
the triplet of coordinates of the target point in 
$\,\Bi_{\moto,\tau}\di{\conf_\tau}\subset\EU\,$ 
as a function of the triplet of coordinates of a source point in
$\,\Bi_{\moto,\ttt}\di{\conf_\ttt}\subset\EU\,$ in a spatial coordinate system.
\end{itemize}
%%%%%%%%%%%%%%%%%%%%%%%%%%%%%%%%%

A body, in motion $\,\moto\in\cont^1\di{\CORPO\times\II\sp\EU}\,$
in the ambient space $\,\EU\,$, describes a \emph{spatial trajectory} which is the
codomain of the map describing the motion in the time interval $\,\II\,$:
%%%%%%%%%%%%%%%%%%%%%%%%%%%%%%%%%%%
$$\,\track\equaldef\moto\di{\CORPO\times\II}\subset\EU\,.$$
%%%%%%%%%%%%%%%%%%%%%%%%%%%%%%%%%%%
To any point $\,\Bx\in\track\,$ in the wake, there corresponds a 
nonempty set $\,\II_{\Bx}\subseteq\II\,$ of time instants
ensuring the existence of a (unique) particle $\,\partic\in\CORPO\,$
passing through $\,\Bx\in\track\,$ at time $\,t\in\II_{\Bx}\,$,
so that $\,\moto_\ttt\di\partic=\Bx\,$.

We denote by $\,\BPi_\II\in\cont^1\di{\CORPO\times\II\sp\II}\,$
the cartesian projection on the second component.

%%%%%%%%%%%%%%%%%%%%%%%%%%%%%%%%%
\begin{definition}[Motion]\label{def: eventsmap}
The  map$\,\coppia{\moto}{\BPi_\II}\in\cont^1\di{\CORPO\times\II\sp\EU\times\II}\,$
is called the motion
and its codomain
$\,\trajectory\equaldef\coppia{\moto}{\BPi_\II}\di{\CORPO\times\II}\,$
 is the trajectory manifold
which is included in the spatial events manifold
$\,\EU\times\II\,$.
\end{definition}
%%%%%%%%%%%%%%%%%%%%%%%%%%%%%%%%%
\begin{axiom}[Non-compenetration]\label{ax: compene}
The motion map is injective.
This mean that two distinct particles of the body cannot 
be at the same location at the same time instant.
\end{axiom}
%%%%%%%%%%%%%%%%%%%%%%%%%%%%%%%%%
The velocity of the spatial motion
$\,\moto\in\cont^1\di{\CORPO\times\II\sp\EU}\,$
at time $\,\ttt\in\II\,$ is the \emph{spatial-valued material vector field}
$\,{\Bv}_{\moto,\ttt}\in\cont^1\di{\CORPO\sp\TEU}\,$,
defined by:
%%%%%%%%%%%%%%%%%%%%%%%%%%%%%%%%%
$$\,{\Bv}_{\moto,\ttt}\di\partic
=\parder\tau\ttt\moto_\tau\di{\partic}
=\parder\tau\ttt(\Bi_{\moto,\tau}\circ\moto_\tau)\di{\partic}
\in\TANG_{(\Bi_{\moto_\ttt}\circ\moto_\ttt)\di\partic}\EU\,.$$
%%%%%%%%%%%%%%%%%%%%%%%%%%%%%%%%%
Its spatial description is evaluated by the time-derivative of the spatial displacement map,
according to the formula:
%%%%%%%%%%%%%%%%%%%%%%%%%%%%%%%%%
$$\vcenter{\halign{
\hfil$#$&$#$\hfil&$#$\hfil&$#$\hfil\cr
\Bv^\SPA_{\moto,\ttt}
&\,=\parder\tau\ttt(\moto_{\tau,\ttt}\circ\Bi_{\moto_\ttt})
=\parder\tau\ttt(\Bi_{\moto,\tau}\circ\moto_{\tau,\ttt})
\vspace{8pt}\cr
&\,=\parder\tau\ttt(\Bi_{\moto,\tau}\circ\moto_{\tau}\circ\inv\moto_\ttt)
={\Bv}_{\moto,\ttt}\circ\inv\moto_\ttt
\in\cont^1\di{\conf_\ttt\sp\TANG\EU}
\,.\cr}}$$
%%%%%%%%%%%%%%%%%%%%%%%%%%%%%%%%%

\section{Material and spatial fields}
\label{sec: MSTF}

In dealing with fundamentals of continuum mechanics or electrodynamics,
a distinction is to be made between 
\emph{spatial tensors} and \emph{material tensors}, and between
\emph{spatial fields} and \emph{material fields}.

A first, basic difference is between curves drawn in the ambient space
and curves drawn in the material body or in any of its placements.
The former are made of points in space, for instance subsequent positions 
of a particle in motion, while the latter are made of material particles in the body.
Vectors tangent to the former are spatial, while vectors tangent to the latter are material.

The physical distinction between spatial and material tangent vectors 
is often hidden by the coincidence of bodies and space dimensions, but
becomes geometrically apparent when lower dimensional bodies are considered
(for instance wires or membranes in the \Euclid\ $\,3$-space).

The main consequence of the distinction, between spatial curves and material 
curves and between the respective tangent vectors, is that appropriate transformations
should be envisaged for each of them, and precisely:
%%%%%%%%%%%%%%%%%%%%%%%%%%%%%%%%%%
\begin{itemize}\item[-]
Spatial tangent vectors transform according to
a given parallel transport defined along a path in the space manifold.
A special instance is the canonical translation in the \Euclid\ space
(which is a path independent parallel transport).
\item[-]
Material curves transform according to a displacement map
of the body in space.
By the chain rule of differential calculus, material tangent vectors
will transform according to the corresponding tangent map. 
The transformation is independent of any connection
(or parallel transport) chosen in the space manifold.
\end{itemize}
%%%%%%%%%%%%%%%%%%%%%%%%%%%%%%%%%%
%%%%%%%%%%%%%%%%%%%%%%%%%%%%%%%%%%
\begin{remark}\label{rem: matbody}
In defining constitutive properties of a material body,
only material tensor fields may be considered, and therefore the
simple rules stated above translate into prescriptions to be respected
in formulating constitutive relations.
Although seemingly evident and physically clear,
these rules have been often violated in continuum mechanics
and in magneto-electro-dynamics,
with highly undesirable consequences.
\end{remark}
%%%%%%%%%%%%%%%%%%%%%%%%%%%%%%%%%%
Let us now fix the nomenclature adopted hereafter.
%%%%%%%%%%%%%%%%%%%%%%%%%%%%%%%%%%
\begin{itemize}\item[-]
\emph{Spatial tensors} are bilinear maps over a tangent space
to the space manifold.
\item[-]
\emph{Material tensors} are bilinear maps that operate, at each time instant,
over a tangent space at a point of the body's placement along the motion.
\item[-]
\emph{Spatial fields} are defined at each point of the ambient space manifold 
and at any time, their values being spatial tensors based at that point,
independently of whether there is a body particle crossing it or not.
They are sections of the bundle $\,\terna{\TENS\di\EU}{\projb_\TENS}\EU\,$.
\item[-]
\emph{Material fields} are defined, at any given instant of time, 
at particles of the body manifold
and their values are material tensors based at the particle location evolving in the motion.
They are sections of the bundle $\,\terna{\TENS\di\conft}{\projb_\TENS}{\conft}\,$.
\item[-]
\emph{Spatial-valued material fields} are defined, at any given instant of time, 
an the body placement, their values being spatial tensors
based at the particle location evolving in the motion.
They are sections of the bundle $\,\terna{\TENS_{\conft}\di{\EU}}{\projb_\TENS}{\conft}\,$.
\end{itemize}
%%%%%%%%%%%%%%%%%%%%%%%%%%%%%%%%%%
Most fields of interest in continuum mechanic
are material fields, for instance,
stretch, stretching, stress, stressing, 
temperature, heat flow, entropy and thermodynamical potentials.
In continuum dynamics, velocity, acceleration, force and momentum
are spatial-valued material fields.

The injective \emph{inclusion map}
$\,\Bi_{\moto,\ttt}\in\cont^1\di{\conft\sp\EU}\,$,
helps in making the distinction, 
between material tangent vectors and spatial tangent vectors, more explicit.
The tangent maps 
$\,\TT_{\Bx}\Bi_{\moto,\ttt}\in\Linmap{\TANG_{\Bx}\conft\sp\TANG_{\Bi_{\moto,\ttt}\di\Bx}\EU}\,$
are injective, so that  the map 
$\,\Bi_{\moto,\ttt}\in\cont^1\di{\conft\sp\EU}\,$ is an \emph{immersion}
and the image $\,\TT_{\Bx}\Bi_{\moto,\ttt}\punto\Bv_{\Bx,\ttt}\in\TANG_{\Bi_{\moto,\ttt}\di\Bx}\EU\,$
of a tangent vector $\,\Bv_{\Bx,\ttt}\in\TANG_{\Bx}\conft\,$
is called an \emph{immersed} tangent vector.
%%%%%%%%%%%%%%%%%%%%%%%%%%%%%%%%%%
\begin{itemize}\item[-]
The \emph{material pull-back} 
$\,\Bs_{\moto,\ttt}\in\cont^1\di{\conft\sp\COV\di\conft}\,$
of a time-dependent covariant spatial field 
$\,\Bs^\SPA_{t}\in\cont^1\di{\EU\sp\COV\di\EU}\,$ at time $\,\ttt\in\II\,$,
is the material field defined by:
%%%%%%%%%%%%%%%%%%%%%%%%%%%%%%%%%%%
$$\,\Bs_{\moto,\ttt}\di{\Ba_{\moto,\ttt},\Bb_{\moto,\ttt}}\equaldef
\Bs_{t}\di{
\Bi_{\moto,\ttt}\push\Ba_{\moto,\ttt},
\Bi_{\moto,\ttt}\push\Bb_{\moto,\ttt}}\circ\Bi_{\moto,\ttt}\,,$$
%%%%%%%%%%%%%%%%%%%%%%%%%%%%%%%%%%%
for all material tangent fields
$\,\Ba_{\moto,\ttt},\Bb_{\moto,\ttt}\in\cont^1\di{\conft\sp\TANG\conft}\,$.
%%%%%%%%%%%%%%%%%%%%%%%%%%%%%%%%%%%
\item[-]
The \emph{spatial description} 
$\,\Bs^\SPA_{\trajectory,\ttt}\in\cont^1\di{\Bi_{\moto,\ttt}\di\conft\sp\COV\di\EU}\,$
of a spatial-valued material field 
$\,\Bs^\SPA_{\moto,\ttt}\in\cont^1\di{\conft\sp\COV\di\EU}\,$,
is defined along the body's trajectory in space-time, by
$$\,\Bs^\SPA_\moto=\Bs^\SPA_\trajectory\circ\coppia{\moto}{\BPi_\II}\,,$$ 
that is
$\,\Bs^\SPA_\trajectory\di{\Bx,\ttt}\equaldef\Bs^\SPA_{\moto}\di{\partic,\ttt}\,$
with $\,\Bx=\moto\di{\partic,\ttt}\,$.
%%%%%%%%%%%%%%%%%%%%%%%%%%%%%%%%%%%s
\end{itemize}
%%%%%%%%%%%%%%%%%%%%%%%%%%%%%%%%%

%%%%%%%%%%%%%%%%%%%%%%%%%%%%%%%%%%
\begin{remark}\label{rem: mettens}
The metric tensor field, being non singular, provides, between different kinds of tensors, 
a linear one-to-one correspondence (a linear isomorphism)
called an alteration, defined by:
Alteration of tensors is defined by the relations:
%%%%%%%%%%%%%%%%%%%%%%%%%%%%%%%%%
$$\,\Baa_\Bx^\MIX=\inv\metric_\Bx\circ\Baa_\Bx^\COV=\Baa_\Bx^\CON\circ\metric_\Bx\,,$$
%%%%%%%%%%%%%%%%%%%%%%%%%%%%%%%%%
which, in components form, correspond to lowering and rising of indexes.
\end{remark}
%%%%%%%%%%%%%%%%%%%%%%%%%%%%%%%%%%
The \Lie\ (convective) time-derivative of a material tensor field $\,\Baa_\moto\,$ along a motion
$\,\moto\in\cont^1\di{\CORPO\times\II\sp\EU}\,$
is defined by
%%%%%%%%%%%%%%%%%%%%%%%%%%%%%%%%%%%
$$\,\Lieder_{\moto,\ttt}\,\Baa_\moto\equaldef\parder\tau\ttt\moto_{\tau,\ttt}\pull\Baa_{\moto,\tau}\,.$$
%%%%%%%%%%%%%%%%%%%%%%%%%%%%%%%%%%%
If the material tensor field $\,\Baa_\moto\,$ admits a
regular spatial description $\,\Baa_{\trajectory}\,$, the relevant
convective time-derivative may be spit by \Leibniz\ rule:
%%%%%%%%%%%%%%%%%%%%%%%%%%%%%%%%%%%
$$\,\Lieder_{\moto,\ttt}\,\Baa_\trajectory
=\parder\tau\ttt\Baa_{\trajectory,\tau}
+\Lieder_{\Bv_{\trajectory,\ttt}}\,\Baa_{\trajectory,\ttt}\,,$$
%%%%%%%%%%%%%%%%%%%%%%%%%%%%%%%%%%%
as the sum of the partial time-derivative and of the \Lie\ derivative of the spatial field at frozen time.
The critical assumption here concerns the partial time-derivative 
$\,\parder\tau\ttt\Baa_{\trajectory,\tau}\,$
at a fixed spatial point.
Indeed, in general, the time set, in which particles cross that point, will not be an open
interval and may even consist of isolated time instants.

\section{Invariance under relative motions}
\label{sec: invariance}

Let $\,\moto\in\cont^1\di{\CORPO\times\II\sp\EU}\,$
be a motion of a body $\,\CORPO\,$ in the ambient space and
$\,\relat\in\cont^1\di{\EU\times\II\sp\EU}\,$
be a time-dependent automorphism of the ambient space onto itself
which will be called a \emph{relative motion},
as depicted in the diagram below.
%%%%%%%%%%%%%%%%%%%%%%%%%%%%%%%%%s
$$\begin{aligned}
\xymatrix{ 
\relat_\ttt\di\conft
\ar[r]^{(\relat\push\moto)_{\tau,\ttt}}
&\relat_\tau\di{\conf_\tau}
\\ 
{\conft}
\ar[r]^{\moto_{\tau,\ttt}} 
\ar[u]^{\relat_\ttt} 
&{\conf_\tau}
\ar[u]^{\relat_\tau} 
}
\end{aligned}
$$
%%%%%%%%%%%%%%%%%%%%%%%%%%%%%%%%%
The pushed motion 
$\,\relat\push\moto\in\cont^1\di{\EU\times\II\sp\EU}\,$
according to the relative motion $\,\relat\in\cont^1\di{\EU\times\II\sp\EU}\,$
is defined by the composition:
%%%%%%%%%%%%%%%%%%%%%%%%%%%%%%%%%%%
$$\,(\relat\push\moto)_\ttt\equaldef\relat_\ttt\circ\moto_\ttt\,,$$
%%%%%%%%%%%%%%%%%%%%%%%%%%%%%%%%%%%
and the corresponding displacement from time $\,\ttt\in\II\,$ to time $\,\tau\in\II\,$
along the pushed motion is given by:
%%%%%%%%%%%%%%%%%%%%%%%%%%%%%%%%%%%
$$\,(\relat\push\moto)_{\tau,\ttt}
=(\relat_\tau\circ\moto_\tau)\circ\inv{(\relat_\ttt\circ\moto_\ttt)}
=\relat_\tau\circ\moto_{\tau,\ttt}\circ\inv{\relat}_\ttt\,.$$
%%%%%%%%%%%%%%%%%%%%%%%%%%%%%%%%%%%
In terms of the velocity of the relative motion
%%%%%%%%%%%%%%%%%%%%%%%%%%%%%%%%%%%
$$\,\Bv_{\relat,\ttt}\equaldef\parder\tau\ttt\relat_\tau\circ\inv{\relat}_\ttt\,,$$
%%%%%%%%%%%%%%%%%%%%%%%%%%%%%%%%%%%
the velocity of the pushed motion is expressed by:
%%%%%%%%%%%%%%%%%%%%%%%%%%%%%%%%%%%
%%%%%%%%%%%%%%%%%%%%%%%%%%%%%%%%%%%
$$\vcenter{\halign{
\hfil$#$&$#$\hfil&$#$\hfil&$#$\hfil\cr
\Bv_{(\relat\push\moto),\ttt}\equaldef
\parder\tau\ttt(\relat\push\moto)_{\tau,\ttt}
&\,=\parder\tau\ttt\relat_\tau\circ\moto_{\tau,\ttt}\circ\inv{\relat}_\ttt
\vspace{8pt}\cr
&\,=\parder\tau\ttt\relat_\tau\circ\inv{\relat}_\ttt+(\TT\relat_\ttt\punto\Bv_{\moto,\ttt})\circ\inv{\relat}_\ttt
\vspace{8pt}\cr
&\,=\Bv_{\relat,\ttt}+\relat_\ttt\push\Bv_{\moto,\ttt}
\,.\cr}}$$
%%%%%%%%%%%%%%%%%%%%%%%%%%%%%%%%%%%
%%%%%%%%%%%%%%%%%%%%%%%%%%%%%%%%%%%
\begin{lemma}[Covariance of convective time-derivatives]
\label{lm: covariance}
The convective time-derivative of a material tensor field,
according to the body motion,
fulfills is covariant with respect to relative motions:
%%%%%%%%%%%%%%%%%%%%%%%%%%%%%%%%%%%
$$\vcenter{\halign{
\hfil$#$&$#$\hfil&$#$\hfil&$#$\hfil\cr
\Lieder_{(\relat\push\moto),\ttt}\,(\relat\push\Baa_\moto)
&\,=\relat_\ttt\push\Lieder_{\moto,\ttt}\,\Baa_\moto
\,.\cr}}$$
%%%%%%%%%%%%%%%%%%%%%%%%%%%%%%%%%%%
\end{lemma}
\proof
By \Leibniz\ rule:
%%%%%%%%%%%%%%%%%%%%%%%%%%%%%%%%%%%
$$\vcenter{\halign{
\hfil$#$&$#$\hfil&$#$\hfil&$#$\hfil\cr
(\relat\push\moto)_{\tau,\ttt}\pull(\relat_\tau\push\Baa_{\moto,\tau})
&\,=(\relat_\tau\circ\moto_{\tau,\ttt}\circ\inv{\relat}_\ttt)\pull(\relat_\tau\push\Baa_{\moto,\tau})
\vspace{8pt}\cr
&\,=\relat_\ttt\push(\moto_{\tau,\ttt}\pull\circ\relat_\tau\pull)(\relat_\tau\push\Baa_{\moto,\tau})
\vspace{8pt}\cr
&\,=\relat_\ttt\push(\moto_{\tau,\ttt}\pull\Baa_{\moto,\tau})
\,.\cr}}$$
%%%%%%%%%%%%%%%%%%%%%%%%%%%%%%%%%%%
Taking the time-derivative $\,\parder\tau\ttt\,$ the result follows by 
the definition of convective time-derivative:
%%%%%%%%%%%%%%%%%%%%%%%%%%%%%%%%%%%
$$\vcenter{\halign{
\hfil$#$&$#$\hfil&$#$\hfil&$#$\hfil\cr
\Lieder_{\moto,\ttt}\,\Baa_\moto
&\,\equaldef\parder\tau\ttt\moto_{\tau,\ttt}\pull\Baa_{\moto,\tau}\,,
\vspace{8pt}\cr
\Lieder_{(\relat\push\moto),\ttt}\,(\relat\push\Baa_\moto)
&\,\equaldef\parder\tau\ttt(\relat\push\moto)_{\tau,\ttt}\pull(\relat\push\Baa)_\tau
\,,\cr}}$$
%%%%%%%%%%%%%%%%%%%%%%%%%%%%%%%%%%%
and the fiberwise linearity of the map 
$\,\relat_\ttt\push\in\cont^0\di{\TEU\sp\TEU}\,$
which implies commutation between 
the time-derivative $\,\parder\tau\ttt\,$ and the push $\,\relat_\ttt\push\,$.
\endprova
%%%%%%%%%%%%%%%%%%%%%%%%%%%%%%%%%%%
%%%%%%%%%%%%%%%%%%%%%%%%%%%%%%%%%%%
\begin{definition}[Invariance of spatial tensor fields]\label{def: definvariance}
The invariance of a material tensor field 
$\,\Baa_\moto\,$ under a relative motion
$\,\relat\in\cont^1\di{\EU\times\II\sp\EU}\,$
is expressed by the drag condition:
%%%%%%%%%%%%%%%%%%%%%%%%%%%%%%%%%%%
$$\,\relat_\ttt\push\Baa_{\moto,\ttt}
=\Baa_{\moto,\ttt}\,,\perogni\ttt\in\II\,.$$
%%%%%%%%%%%%%%%%%%%%%%%%%%%%%%%%%%%
For twice covariant tensors, the invariance property is written, explicitly:
%%%%%%%%%%%%%%%%%%%%%%%%%%%%%%%%%%%
$$\,(\relat_\ttt\pull\Baa_{\moto,\ttt})\coppia\Ba\Bb
=\Baa_{\moto,\ttt}\coppia{\relat_{\ttt}\push\Ba}{\relat_{\ttt}\push\Bb}\circ\gamma_{\ttt}
= \Baa_{\moto,\ttt}\coppia\Ba\Bb\,,$$
%%%%%%%%%%%%%%%%%%%%%%%%%%%%%%%%%%%
for all $\,\tau\in\II\,$ and $\,\Ba,\Bb\in\TANG\EU\,$.
\end{definition}
%%%%%%%%%%%%%%%%%%%%%%%%%%%%%%%%%%%
The basic result concerning invariance is provided by the next Lemma.
%%%%%%%%%%%%%%%%%%%%%%%%%%%%%%%%%%%
\begin{lemma}[Invariance of convective time-derivatives]
\label{lm: invarianceconv}
Invariance of a material tensor field $\,\Baa_\moto\,$
with respect to a relative motion,
implies invariance of its convective time-derivative:
%%%%%%%%%%%%%%%%%%%%%%%%%%%%%%%%%%%
$$\vcenter{\halign{
\hfil$#$&$#$\hfil&$#$\hfil&$#$\hfil\cr
\relat_\ttt\push\Baa_{\moto,\ttt}=\Baa_{\moto,\ttt}
\implies
\Lieder_{(\relat\push\moto),\ttt}\,\Baa_\moto
&\,=\relat_\ttt\push\Lieder_{\moto,\ttt}\,\Baa_\moto
\,.\cr}}$$
%%%%%%%%%%%%%%%%%%%%%%%%%%%%%%%%%%%
\end{lemma}
\proof
The result follows directly from Lemma \ref{lm: covariance} and the
Definition \ref{def: definvariance} of invariance.
\endprova
%%%%%%%%%%%%%%%%%%%%%%%%%%%%%%%%%%%
%%%%%%%%%%%%%%%%%%%%%%%%%%%%%%%%%%%
\begin{lemma}[Time-derivatives and relative motions]\label{lm: timerelmot}
The partial time-derivatives, of a spatial tensor field 
$\,\Baa_{\moto,\ttt}\in\cont^1\di{\EU\sp\TENS\di\EU}\,$
and of its push according to a relative motion
$\,\relat_\ttt\in\cont^1\di{\EU\times\II\sp\EU}\,$,
are related by:
%%%%%%%%%%%%%%%%%%%%%%%%%%%%%%%%%%%
$$\,\parder\tau\ttt(\relat\push\Baa_\moto)_\tau
=\relat_\ttt\push(\parder\tau\ttt\Baa_{\moto,\tau})
-\Lieder_{\Bv_{\relat,\ttt}}\,(\relat\push\Baa_\moto)\,.$$
%%%%%%%%%%%%%%%%%%%%%%%%%%%%%%%%%%%
\end{lemma}
\proof
Being
$\,\Bv_{(\relat\push\moto),\ttt}
=\Bv_{\relat,\ttt}+\relat_\ttt\push\Bv_{\moto,\ttt}\,$
and
$\,\Lieder_{\moto,\ttt}\,\Baa
=\parder\tau\ttt\Baa_{\moto,\tau}
+\Lieder_{\Bv_{\moto},\ttt}\,\Baa_{\moto,\ttt}\,$ we have that:
%%%%%%%%%%%%%%%%%%%%%%%%%%%%%%%%%%%
$$\vcenter{\halign{
\hfil$#$&$#$\hfil&$#$\hfil&$#$\hfil\cr
\Lieder_{(\relat\push\moto),\ttt}\,(\relat\push\Baa)
&\,=\parder\tau\ttt(\relat\push\Baa)_\tau
+\Lieder_{\Bv_{(\relat\push\moto),\ttt}}\,(\relat\push\Baa)_\ttt
\vspace{8pt}\cr
&\,=\parder\tau\ttt(\relat\push\Baa)_\tau
+\Lieder_{\Bv_{\relat,\ttt}}\,(\relat\push\Baa)_\ttt
+\Lieder_{\relat_\ttt\push\Bv_{\moto,\ttt}}\,(\relat\push\Baa)_\ttt
\vspace{8pt}\cr
&\,=\parder\tau\ttt(\relat\push\Baa)_\tau
+\Lieder_{\Bv_{\relat,\ttt}}\,(\relat\push\Baa)_\ttt
+\relat_\ttt\push\Lieder_{\Bv_{\moto,\ttt}}\,\Baa_{\moto,\ttt}
\vspace{8pt}\cr
&\,=\parder\tau\ttt(\relat\push\Baa)_\tau
+\Lieder_{\Bv_{\relat,\ttt}}\,(\relat\push\Baa)_\ttt
+\relat_\ttt\push\Lieder_{{\moto,\ttt}}\,\Baa_{\moto,\ttt}
-\relat_\ttt\push(\parder\tau\ttt\Baa_{\moto,\tau})
\,.\cr}}$$
%%%%%%%%%%%%%%%%%%%%%%%%%%%%%%%%%%%
By Lemma \ref{lm: covariance}:
$\, \Lieder_{(\relat\push\moto),\ttt}\,(\relat\push\Baa)
=\relat_\ttt\push\Lieder_{\moto,\ttt}\,\Baa\,$
and hence the result.
\endprova
%%%%%%%%%%%%%%%%%%%%%%%%%%%%%%%%%%%
Invariance of a time-dependent spatial tensor field $\,\Baa\,$
with respect to a relative motion, implies that:
%%%%%%%%%%%%%%%%%%%%%%%%%%%%%%%%%%%
$$\,\parder\tau\ttt\Baa_{\moto,\tau}
=\relat_\ttt\push(\parder\tau\ttt\Baa_{\moto,\tau})
-\Lieder_{\Bv_{\relat,\ttt}}\,\Baa_{\moto,\ttt}\,,$$
%%%%%%%%%%%%%%%%%%%%%%%%%%%%%%%%%%%
which is equivalent to
%%%%%%%%%%%%%%%%%%%%%%%%%%%%%%%%%%%
$$\, \Lieder_{\relat,\ttt}\,\Baa=\relat_\ttt\push(\parder\tau\ttt\Baa_{\moto,\tau})\,.$$
%%%%%%%%%%%%%%%%%%%%%%%%%%%%%%%%%%%
We may then conclude that the partial time-derivative of an invariant
time-dependent spatial tensor field $\,\Baa\,$ is not invariant,
unless its \Lie\ derivative along the relative motion vanishes identically in time.

From the expression of the \Lie\ derivative in terms of parallel derivative,
performed according to a torsion-free connection,
\citep{RomanoDiff2007}:
%%%%%%%%%%%%%%%%%%%%%%%%%%%%%%%%
$$\Lieder_{\Bv_{\relat,\ttt}}\,\Baa_{\moto,\ttt}=
\nabla_{\Bv_{\relat,\ttt}}\,\Baa_{\moto,\ttt}
+\Baa_{\moto,\ttt}\circ\nabla{\Bv_{\relat,\ttt}}
+\dual\nabla{\Bv_{\relat,\ttt}}\circ\Baa_{\moto,\ttt}\,,$$
%%%%%%%%%%%%%%%%%%%%%%%%%%%%%%%%
we infer that the vanishing of the parallel derivative
$\,\nabla{\Bv_{\relat,\ttt}}=0\,$
of the velocity field of the relative motion, implies that
$\,\Lieder_{\Bv_{\relat,\ttt}}\,\Baa_{\moto,\ttt}=
\nabla_{\Bv_{\relat,\ttt}}\,\Baa_{\moto,\ttt}\,$ and hence also that
%%%%%%%%%%%%%%%%%%%%%%%%%%%%%%%%%%%
$$\, \nabla_{\relat,\ttt}\,\Baa
=\parder\tau\ttt\Baa_{\moto,\tau}
+ \nabla_{\Bv_{\relat,\ttt}}\,\Baa_{\moto,\ttt}
= \Lieder_{\relat,\ttt}\,\Baa\,.$$
%%%%%%%%%%%%%%%%%%%%%%%%%%%%%%%%%%%

\section{Galilei invariance}
\label{sec: Galilei}

%%%%%%%%%%%%%%%%%%%%%%%%%%%%%%%%%%%
\begin{definition}[Translational relative motion]
A relative motion in the ambient space $\,\relat\in\cont^1\di{\EU\times\II\sp\EU}\,$ 
is translational at time $\,\ttt\in\II\,$, according to a spatial connection,
if the relevant spatial velocity field, frozen at time $\,\ttt\in\II\,$,
has a vanishing parallel derivative, viz.:
%%%%%%%%%%%%%%%%%%%%%%%%%%%%%%%%%%%
$$\,\nabla\Bv_{\relat,\ttt}=0\,.$$
%%%%%%%%%%%%%%%%%%%%%%%%%%%%%%%%%%%
\end{definition}
%%%%%%%%%%%%%%%%%%%%%%%%%%%%%%%%%%%
%%%%%%%%%%%%%%%%%%%%%%%%%%%%%%%%%%%
\begin{definition}[Stationary relative motion]
A relative motion in the ambient  space $\,\relat\in\cont^1\di{\EU\times\II\sp\EU}\,$ 
is stationary if the partial time-derivative
of the relevant spatial velocity field vanishes, viz.:
%%%%%%%%%%%%%%%%%%%%%%%%%%%%%%%%%%%
$$\,\parder{\tau}{\ttt}\Bv_{\relat,\tau}=0\,.$$
%%%%%%%%%%%%%%%%%%%%%%%%%%%%%%%%%%%
\end{definition}
%%%%%%%%%%%%%%%%%%%%%%%%%%%%%%%%%%%
The acceleration field:
%%%%%%%%%%%%%%%%%%%%%%%%%%%%%%%%%%%
$$\,\nabla_{\relat,\ttt}\,\Bv_{\relat}
\equaldef\parder{\tau}{\ttt}\relat_{\tau,\ttt}\back\Bv_{\relat,\tau}
=\parder\tau\ttt\Bv_{\relat,\tau}
+\nabla_{\Bv_{\moto,\ttt}}\,\Bv_{\relat,\ttt}\,,$$
%%%%%%%%%%%%%%%%%%%%%%%%%%%%%%%%%%% 
of a motion which is stationary and translational at all times, vanishes identically.
By definition, the relative spatial motion
$\,\relat\in\cont^1\di{\EU\times\II\sp\EU}\,$
between two \Galilei\ observers is stationary and translational,
and parallel transport and push along relative motions are coincident.
Moreover the standard connection is path-independent, so that
all definitions and results of Section \ref{sec: invariance} 
may be applied to this special circumstance.

A \Galilei\ transformation is metric-preserving and then also
volume-preserving so that $\,\relat_\ttt\pull\volform=\volform\,$.
Setting $\, \volform\punto\BF_\ttt=\Boo^2_{\BF,\ttt}\,$
we then have:
%%%%%%%%%%%%%%%%%%%%%%%%%%%%%%%%%%%
$$\,\relat_\ttt\pull(\volform \punto \BF_\ttt)
=(\relat_\ttt\pull\volform)\punto(\relat_\ttt\pull \BF_\ttt)
=\volform\punto(\relat_\ttt\pull \BF_\ttt)\,.$$
%%%%%%%%%%%%%%%%%%%%%%%%%%%%%%%%%%%
so that the two-form $\,\Boo^2_{\BF,\ttt}\,$
is \Galilei\ invariant iff the vector field $\, \BF_\ttt\,$ is such.
Moreover, taking the time derivative and applying the \Leibniz\ rule:
%%%%%%%%%%%%%%%%%%%%%%%%%%%%%%%%%%%
$$\,\Lieder_{\relat,\ttt}\,(\volform \punto \BF_\ttt)
=(\Lieder_{\relat,\ttt}\, \volform) \punto \BF_\ttt
+ \volform\punto(\Lieder_{\relat,\ttt}\, \BF_\ttt)\,.$$
%%%%%%%%%%%%%%%%%%%%%%%%%%%%%%%%%%%
Being $\,\Lieder_{\relat,\ttt}\, \volform=0\,$. It follows that:
%%%%%%%%%%%%%%%%%%%%%%%%%%%%%%%%%%%
$$\,\Lieder_{\relat,\ttt}\, (\volform \punto \BF_\ttt)
=\volform \punto(\Lieder_{\relat,\ttt}\, \BF_\ttt)\,,$$
%%%%%%%%%%%%%%%%%%%%%%%%%%%%%%%%%%%
and we may conclude that the convective time-derivativive of the two-form field
$\,\Boo^2_{\BF,\ttt}\,$
is \Galilei\ invariant iff the convective time-derivativive of the vector field $\, \BF_\ttt\,$ is such.
%%%%%%%%%%%%%%%%%%%%%%%%%%%%%%%%%%%
\begin{definition}[Galilei time-independence]
The \Galilei\ time-indepen\-dence, of a spatial tensor field 
$\,\Baa_{\moto,\ttt}\,$
is expressed by the requirement that there exists a \Galilei\ observer
which sees a time-independent $\,k$-form at all times:
%%%%%%%%%%%%%%%%%%%%%%%%%%%%%%%%%%%
$$\,\parder\tau\ttt\Baa_{\moto,\tau}=0\,,\perogni\ttt\in\II\,.$$
%%%%%%%%%%%%%%%%%%%%%%%%%%%%%%%%%%%
\end{definition}
%%%%%%%%%%%%%%%%%%%%%%%%%%%%%%%%%%%

\section{Electromagnetic induction: standard treatment}
\label{sec: EM}

A noteworthy physical application of the theory of integration
on manifolds is to the laws of Electromagnetism, see e.g.
\citep{Deschamps1970}.

We will denote by $\,\pair\EU\metric\,$
the \Euclid\ ambient $\,3$-D manifold without boundary,
endowed with the standard metric tensor field $\,\metric\,$.
Customarily, the $\,3$-form $\,\volform\,$ is the volume form induced in
$\,\EU\,$ by the metric tensor field.
The geometric objects involved in electrodynamics are 
the following \emph{odd} and \emph{even} exterior forms.
They are related to the vectorial or scalar
representations by the linear isomorphisms
generated by the metric tensor (for one-forms)
and by the volume form (for two-forms and three-forms),
as explicitly illustrated in the following lists, pertaining to:
%%%%%%%%%%%%%%%%%%%%%%%%%%%%%%%%%
\begin{itemize}\item
\Faraday\ law:
\end{itemize}
%%%%%%%%%%%%%%%%%%%%%%%%%%%%%%%%%
$$\vcenter{\halign{
\hfil$#$&$#$\hfil&$\quad$#\hfil&$\,$#\hfil&$\,$#\hfil\cr
\elefieldttt\,&=\metric \punto\elevector_\ttt&electric circulation &(one-form\,,\,vector field)\,,
\vspace{8pt}\cr
\magcurlttt\,&=\volform \punto\magindttt& magnetic vortex &(two-form\,,\,\emph{odd} vector field)\,,
%\vspace{8pt}\cr
\cr}}$$
%%%%%%%%%%%%%%%%%%%%%%%%%%%%%%%%%
\begin{itemize}\item
\Ampere\ law:
\end{itemize}
%%%%%%%%%%%%%%%%%%%%%%%%%%%%%%%%%
$$\vcenter{\halign{
\hfil$#$&$#$\hfil&$\quad$#\hfil&$\,$#\hfil&$\,$#\hfil\cr
\magwindttt\,&=\metric \punto\magvector_\ttt&magnetic winding &(\emph{odd} one-form\,,\,\emph{odd} vector field)\,,
\vspace{8pt}\cr
\elecurrformttt\,&=\volform\punto\elecurr_\ttt& current flux &(\emph{odd} two-form\,,\,vector field)\,,
\vspace{8pt}\cr
\elefluxttt\,&=\volform \punto\eledispttt&electric flux &(\emph{odd} two-form\,,\,vector field)\,,
\vspace{8pt}\cr
\elechargeformttt\,&=\chargescalarttt\,\volform&electric charge &(\emph{odd} three-form\,,\,scalar field)\,.
\cr}}$$
%%%%%%%%%%%%%%%%%%%%%%%%%%%%%%%%%
%%%%%%%%%%%%%%%%%%%%%%%%%%%%%%%%%
In engineering and physics literature, it is customary to
express the laws of electromagnetic induction
in terms of the spatial description of the vector fields
$\, \elevector_\ttt,\magindttt,\magvector_\ttt,\elecurr_\ttt,\eledispttt\in\cont^1\di{\EU\sp\TEU}\,$
and of the scalar field
$\,\chargescalarttt\in\cont^1\di{\EU\sp\Re}\,$, electric charge density per unit volume,
by the integral relations:
%%%%%%%%%%%%%%%%%%%%%%%%%%%%%%%%
$$\vcenter{\halign{
$\hfil\displaystyle#$&$\displaystyle#$\hfil&$\quad\textrm{#}$\hfil\cr
\ointegrale{\partial\surfin}{}\metric \punto\elevector_\ttt
&\,=-\integrale{\surfin}{}\volform\punto(\parder\tau\ttt\magind_\tau)
&\Henry-\Faraday(1831)\cr
\vspace{6pt}\cr
\ointegrale{\partial\conf^\outer_\ttt}{}\volform \punto\magindttt
&\,=0
&\Gauss(1831)\cr
%\cr}}$$
%%%%%%%%%%%%%%%%%%%%%%%%%%%%%%%%
%%%%%%%%%%%%%%%%%%%%%%%%%%%%%%%%
%$$\vcenter{\halign{
%$\hfil\displaystyle#$&$\displaystyle#$\hfil&$\quad\textrm{#}$\hfil\cr
\vspace{6pt}\cr
\ointegrale{\partial\surfout}{}\metric\punto\magvector_\ttt
&\,=\integrale{\surfout}{}\volform\punto(\parder\tau\ttt\eledisp_\tau+\elecurrttt)
&\Maxwell(1861)-\Ampere(1826)\cr
\vspace{6pt}\cr
\ointegrale{\partial\conf^\outer_\ttt}{}\volform \punto\eledispttt
&\,=\integrale{\conf^\outer_\ttt}{} \chargescalarttt\,\volform
&\Gauss(1835)
\cr}}$$
%%%%%%%%%%%%%%%%%%%%%%%%%%%%%%%%
with $\,\parder\tau\ttt\eledisp_\tau\,$ and 
$\,\parder\tau\ttt\magind_\tau\,$ 
partial time-derivatives at a fixed point in the ambient space,
as seen by a \Galilei\ observer,
$\,\surf\,$ a bounded connected surface
and $\,\conft\,$ bounded connected domain in $\,\EU\,$.
Applying \Ampere\ law to the closed surface $\,\surft =\partial\conft\,$, we infer that:
%%%%%%%%%%%%%%%%%%%%%%%%%%%%%%%%%%%
$$\,\parder\tau\ttt\rho_{\BE,\tau}+\diverg\elecurrttt=0\,$$
%%%%%%%%%%%%%%%%%%%%%%%%%%%%%%%%%%%
which expresses the so called \emph{equation of continuity}.
In all the equations above, the change in time of the surface $\,\surft\,$
and of the domain $\,\conft\,$, as seen by a \Galilei\ observer, are not considered,
neither the effect of a change of observer is taken into acount.
A critically discussion about these equations, 
which are customary in literature, will be performed in the sequel,
with the equation of continuity discussed in Remark \ref{rem: eqcontinuity}.

\section{Electromagnetic induction in continuous bodies in motion}
\label{sec: EMMB}

The standard formulation of the laws of electromagnetic induction 
is introduced hereafter, with innovative features: 
material and spatial fields are carefully distinguished,
\Galilei\ invariant laws are formulated and their well-posedness is discussed, 
leading to correct expressions of electric and magnetic charge conservation.
The laws are first introduced as integral balance laws,
over arbitrarily drawn two-dimensional submanifolds,
and then translated into the equivalent differential form.
While the integral form provides a direct tool for the evaluation of electromotive 
or magnetomotive forces along circuits,
the differential form opens the way for the introduction of potential fields
(respectively one-forms and zero-forms)
and for their evaluation.
Metric independent formulations of electromagnetic induction were introduced by
\citet{Murnaghan1921,Kottler1922,Cartan1924,Dantzig1934}.

\section{Electromotive induction by magnetic vortex rate}
\label{sec: EMMF}

\subsection{Integral Faraday law}
\label{sec: MatFAR}

The magnetic vortex $\,\magcurlttt\,$ is
a \Galilei\ invariant, material \emph{even} two-form.
In \citep[p. 284]{Tonti1995} it is said:
\emph{Therefore, the magnetic flux is associated with a surface 
element and inner orientation, i.e. with a prescribed direction along its boundary.}
The name \emph{flux} is however not appropriate for an inner orientation dependent
extensive quantity, no crossing direction across the surface being specified.
So we prefer to adopt the name \emph{magnetic vortex},
suggested by the sketch in  fig.\ref{fig: inner}, instead of 
\emph{magnetic flux}, more apt to describe extensive quantities
related to outer oriented surface, as depicted in fig.\ref{fig: outer},
which will be considered with reference to electrical induction, in Section \ref{sec: MMEF}.

To formulate \Faraday\ law of induction,
let us consider in the body placement $\,\conft\,$ at time $\,\ttt\in\II\,$
an inner oriented surface $\,\surfin\,$, with the induced inner orientation on its boundary
$\,\partial\surfin\,$
(see fig. \ref{fig: inner}).
The law of magnetic induction, named after \persone{Michael} \Faraday\ 
who discovered it in $\,1821\,$,
is expressed in material formulation as:
%%%%%%%%%%%%%%%%%%%%%%%%%%%%%%%%
$$\framebox{$
\vcenter{\halign{
$\hfil\displaystyle#$&$\displaystyle#$\hfil&$\displaystyle#$\hfil\cr
-\ointegrale{\partial\surfin}{}\elefieldttt
&\,=\parder\tau\ttt
\integrale{\moto_{\tau,\ttt}\di{\surfin}}{}\magcurltau
%\vspace{8pt}\cr
&\,=\integrale{\surfin}{}\Lieder_{\moto,\ttt}\,\magcurlttt\,.
\cr}}
$}$$
%%%%%%%%%%%%%%%%%%%%%%%%%%%%%%%%
Here $\,\elefieldttt\,$ is the \emph{even} electric field one-form and
$\,\Lieder_{\moto,\ttt}\,\magcurl\,$ is the 
convective time-derivative of the \emph{even} magnetic vortex two-form
along the motion $\,\moto\in\cont^1\di{\CORPO\times\II\sp\EU}\,$.
By \VPB\ formula and localization,
\Faraday\ law may be expressed by the differential condition: 
%%%%%%%%%%%%%%%%%%%%%%%%%%%%%%%%%%%
$$\framebox{$
\,-\der\elefieldttt
=\Lieder_{\moto,\ttt}\,\magcurl
\,.$}$$
%%%%%%%%%%%%%%%%%%%%%%%%%%%%%%%%%%%
Defined by $\,\elefieldttt=\metric\punto\elevector_\ttt\,$,
the electric field $\,\elevector_\ttt\,$ is an \emph{even} vector field,
while, being defined by
$\,\magcurlttt=\volform \punto\magindttt\,$,
the magnetic vector field $\,\magindttt\,$ is a \emph{odd} vector field.

\subsection{Well-posedness of Faraday law}
\label{sec: wellFaraday}

In order that the integral \Faraday\ formula be meaningful, its r.h.s. should be proven 
to be independent of the choice of the surface $\,\surfin\,$,
for a given boundary $\,\partial\surfin\,$, and independent of the
motion of the surface $\,\surfin\,$ for a given motion of the boundary $\,\partial\surfin\,$.
This condition may be formalized by requiring that the 
time derivatives of the integrals:
%%%%%%%%%%%%%%%%%%%%%%%%%%%%%%%%
$$\,\integrale{\moto^1_{\tau,\ttt}\di{\surfin^1}}{}
\Lieder_{\moto,\ttt}\,\magcurl\,,
\qquad\integrale{\moto^2_{\tau,\ttt}\di{\surfin^2}}{}
\Lieder_{\moto,\ttt}\,\magcurl\,,$$
%%%%%%%%%%%%%%%%%%%%%%%%%%%%%%%%
be the same for any motions such that
%%%%%%%%%%%%%%%%%%%%%%%%%%%%%%%%
$$\,\partial(\moto^1_{\tau,\ttt}\di{\surfin^1})=\partial(\moto^2_{\tau,\ttt}\di{\surfin^2})\,,$$
%%%%%%%%%%%%%%%%%%%%%%%%%%%%%%%%
which is equivalent to require that, for any control-window $\,\controlt\,$:
%%%%%%%%%%%%%%%%%%%%%%%%%%%%%%%%%%%
$$\,\parder\tau\ttt
\integrale{\moto_{\tau,\ttt}\di{\partial\controlt}}{}\magcurlttt
=\integrale{\partial\controlt}{}\Lieder_{\moto,\ttt}\,\magcurl
=\integrale{\controlt}{} \der(\Lieder_{\moto,\ttt}\,\magcurl)=0\,.$$
%%%%%%%%%%%%%%%%%%%%%%%%%%%%%%%%%%%
By localizing and recalling the commutation property in Lemma \ref{lm: extderpushes},
this is equivalent to:
%%%%%%%%%%%%%%%%%%%%%%%%%%%%%%%%%%%
$$\,\der(\Lieder_{\moto,\ttt}\,\magcurl)
=\Lieder_{\moto,\ttt}\,(\der\magcurl)=0\,,$$
%%%%%%%%%%%%%%%%%%%%%%%%%%%%%%%%%%%
a condition assured by \Gauss\ law for the magnetic vortex:
$\,\der\magcurl=0\,$.
By \Poincare\ Lemma, the closedness condition
$\,\der(\Lieder_{\moto,\ttt}\,\magcurl)=0\,$
assures the existence of a one-form 
$\,\elefield\,$ electric field, fulfilling the differential \Faraday\ law:
$\,-\der\elefield=\Lieder_{\moto,\ttt}\,\magcurl\,$.

\subsection{Differential form of Faraday law}
\label{sec: diffFaraday}

The theory of electromotive induction is based on the assumption that
the body, in which the electric field $\,\elefieldttt\,$
and the magnetic vortex $\,\magcurl\,$ are defined, 
is spread over the whole ambient space,
being either a material body or the \emph{empty space} (or \ether).

The \ether\ is assumed to be homogeneous, isotropic and mass-free,
so that no motion of it can be detected.
As a consequence the induction law in the \ether\ is written in terms of partial time derivatives
by any observer since the \ether\ appears as fixed, to any observer.

A careful attention must be devoted to
singularities in the time dependence of spatial descriptions of the fields at a point,
during the transit of body particles,
at those time instants when sudden changes of material properties occur,
as tested by an observer.

To recover a standard form of \Faraday\ law, we consider
the spatial description of the magnetic vortex $\, \magcurlttt\,$
at a fixed space-point and at a time instant 
in whose neighbor it has a smooth time-dependence.
Then the convective time-derivative can be split as sum of 
the partial time-derivative and of the \Lie\ derivative 
along the flow of the velocity field $\,\Bv_{\moto,\ttt}\,$ at frozen time:
%%%%%%%%%%%%%%%%%%%%%%%%%%%%%%%%%%%
$$\,\Lieder_{\moto,\ttt}\, \magcurl
=\parder\tau\ttt\magcurltau
+\Lieder_{\Bv_{\moto,\ttt}}\, \magcurlttt\,,$$
%%%%%%%%%%%%%%%%%%%%%%%%%%%%%%%%%%%
so that the spatial description of \Faraday\ differential law writes:
%%%%%%%%%%%%%%%%%%%%%%%%%%%%%%%%%%%
$$\framebox{$
\,-\der\elefield
=\parder\tau\ttt\magcurltau
+\Lieder_{\Bv_{\moto,\ttt}}\, \magcurlttt
\,.$}$$
%%%%%%%%%%%%%%%%%%%%%%%%%%%%%%%%%%%
Expressing the \Lie\ derivative of the magnetic vortex $\,\magcurlttt\,$ 
by the homotopy formula, and recalling that $\,\der\magcurlttt=0\,$, we get:
%%%%%%%%%%%%%%%%%%%%%%%%%%%%%%%%%%%
$$\,\Lieder_{\Bv_{\moto,\ttt}}\, \magcurl
=\der(\magcurlttt\punto\Bv_{\moto,\ttt})
+(\der\magcurlttt)\punto\Bv_{\moto,\ttt}
=\der(\magcurlttt\punto\Bv_{\moto,\ttt})\,,$$
%%%%%%%%%%%%%%%%%%%%%%%%%%%%%%%%%%%
and the \Faraday\ law may be rewritten in integral form as:
%%%%%%%%%%%%%%%%%%%%%%%%%%%%%%%%
$$\framebox{$
\vcenter{\halign{
$\hfil\displaystyle#$&$\displaystyle#$\hfil&$\quad\textrm{#}$\hfil\cr
-\ointegrale{\partial\surfin}{}\elefield
&\,=\integrale{\surfin}{}(\parder\tau\ttt\magcurltau
+\Lieder_{\Bv_{\moto,\ttt}}\, \magcurlttt)
\vspace{8pt}\cr
&\,=\integrale{\surfin}{}\parder\tau\ttt\magcurltau
+\ointegrale{\partial\surfin}{}\magcurlttt\punto\Bv_{\moto,\ttt}\,,
\cr}}$}$$
%%%%%%%%%%%%%%%%%%%%%%%%%%%%%%%%
where $\,\surfin\,$ is a surface in the body placement $\,\conft\,$.
The first integral at the r.h.s. is wrongly omitted in the formula proved in \citep[p.240]{Greiner1998}. 
The differential expression in $\, \conft\,$ is:
%%%%%%%%%%%%%%%%%%%%%%%%%%%%%%%%%%%
$$\framebox{$
\vcenter{\halign{
$\hfil\displaystyle#$&$\displaystyle#$\hfil&$\quad\textrm{#}$\hfil\cr
-\der\elefield
&\,=\Lieder_{\moto,\ttt}\, \magcurl
\vspace{8pt}\cr
&\,=\parder\tau\ttt\magcurltau
+\Lieder_{\Bv_{\moto,\ttt}}\, \magcurlttt
\vspace{8pt}\cr
&\,=\parder\tau\ttt\magcurltau
+\der(\magcurlttt\punto\Bv_{\moto,\ttt})\,,\cr}}$}$$
%%%%%%%%%%%%%%%%%%%%%%%%%%%%%%%%%%%
which in vectorial notation becomes:
%%%%%%%%%%%%%%%%%%%%%%%%%%%%%%%%%%%
$$\,-\rotor\elevector_\ttt
=\parder\tau\ttt\magind_\tau
+\rotor(\magindttt\times\Bv_{\moto,\ttt})\,.$$
%%%%%%%%%%%%%%%%%%%%%%%%%%%%%%%%%%%
We emphasize that, being the cross product 
between \emph{odd} and \emph{even} vector fields,
the vector field $\,\magindttt\times\Bv_{\moto,\ttt}\,$ is \emph{even} and that
$\,\rotor(\magindttt\times\Bv_{\moto,\ttt})\,$, the rotor of a \emph{even} vector field,
is a \emph{odd} vector field.

The expression above is usually reported, see e.g. \cite[eq.$\,9.16\,$]{Sadiku2010},
as the result of the sum of two distinct contributions
to the electromotive force (\emf) in a circuit:
transformer \emf\ and motional \emf.
The former is derived from the usual expression of \Faraday's law
for a fixed body and the latter is deduced from the so called \Lorentz\ force law. 
We have instead shown that the whole
expression is a direct consequence of \Faraday's law.
The single addend fields are not \Galilei\ invariant and hence
do not have the physical meaning of a force, as discussed in 
Remarks \ref{rem: lorfor} and \ref{rem: thomson}.

\subsection{Galilei invariance of Faraday law}
\label{sec: GalileiFaraday}

The \Galilei\ invariance of 
$\,\elefieldttt\,$ and
$\,\magcurlttt\,$ is expressed by
$\,\gamma_\ttt\push\elefieldttt=\elefieldttt\,$
and
$\,\gamma_\ttt\push\magcurlttt=\magcurlttt\,$.
By Lemma \ref{lm: invarianceconv} these invariance properties
assure the \Galilei\ invariance of \Faraday\ law:
%%%%%%%%%%%%%%%%%%%%%%%%%%%%%%%%
$$\framebox{$
\vcenter{\halign{
$\hfil\displaystyle#$&$\displaystyle#$\hfil&$\quad\textrm{#}$\hfil\cr
\ointegrale{\partial\surfin}{}-\elefield
&\,=\integrale{\surfin}{}\Lieder_{\moto,\ttt}\,\magcurl
\equi
\ointegrale{\partial\relat_\ttt\di{\surfin}}{}-\elefield
=\integrale{\relat_\ttt\di{\surfin}}{}\Lieder_{(\relat\push\moto),\ttt}\,\magcurl\,.
\cr}}
$}$$
%%%%%%%%%%%%%%%%%%%%%%%%%%%%%%%%
Indeed it is:
%%%%%%%%%%%%%%%%%%%%%%%%%%%%%%%%
$$\vcenter{\halign{
$\hfil\displaystyle#$&$\displaystyle#$\hfil&$\quad\textrm{#}$\hfil\cr
\ointegrale{\partial\surfin}{}\elefieldttt
=\ointegrale{\partial\relat_\ttt\di{\surfin}}{}\relat_\ttt\push\elefieldttt
&\,=\ointegrale{\partial\relat_\ttt\di{\surfin}}{}\elefieldttt\,,
\cr}}$$
%%%%%%%%%%%%%%%%%%%%%%%%%%%%%%%%
and
%%%%%%%%%%%%%%%%%%%%%%%%%%%%%%%%
$$\vcenter{\halign{
$\hfil\displaystyle#$&$\displaystyle#$\hfil&$\quad\textrm{#}$\hfil\cr
\integrale{\surfin}{}\Lieder_{\moto,\ttt}\,\magcurl
=\integrale{\relat_\ttt\di{\surfin}}{}\relat_\ttt\push\Lieder_{\moto,\ttt}\,\magcurl
&\,=\integrale{\relat_\ttt\di{\surfin}}{}\Lieder_{(\relat\push\moto),\ttt}\,\magcurl\,.
\cr}}$$
%%%%%%%%%%%%%%%%%%%%%%%%%%%%%%%%

\subsection{Faraday potential one-form}
\label{sec: elepot}

An explicit expression for the electric field $\,\elefield\,$
can be got by observing that, being $\,\der\magcurl=0\,$, \Poincare\ lemma 
ensures that the closed form of magnetic vortex $\,\magcurl\,$
admits a potential $\,\faradpot\,$,
the \emph{even} \Faraday\ one-form, so that we may set:
%%%%%%%%%%%%%%%%%%%%%%%%%%%%%%%%%%%
$$\,\magcurlttt=\der\faradpotttt
\equi\magindttt=\rotor\potmag_\ttt\,,$$
%%%%%%%%%%%%%%%%%%%%%%%%%%%%%%%%%%%
where $\,\potmag_\ttt\,$ is the \emph{even} vector magnetic potential.
By relying on the commutation property in Lemma \ref{lm: extderpushes},
\Faraday\ differential law may then be written as:
%%%%%%%%%%%%%%%%%%%%%%%%%%%%%%%%%%%
$$-\der\elefieldttt
=\Lieder_{\moto,\ttt}\,\magcurl
=\Lieder_{\moto,\ttt}\,\der\,\faradpot
=\der\,\Lieder_{\moto,\ttt}\,\faradpot\,,$$
%%%%%%%%%%%%%%%%%%%%%%%%%%%%%%%%%%%
and leads to the following formula,
in terms of the \emph{even} scalar electric potential 
$\,\VV_{\elevector,\ttt}\in\cont^1\di{\EU\sp\FUN\di{\TEU}}\,$
%%%%%%%%%%%%%%%%%%%%%%%%%%%%%%%%%%%
$$\framebox{$
-\elefieldttt
=\Lieder_{\moto,\ttt}\,\faradpot
+\der\VV_{\elevector,\ttt}\,.
$}$$
%%%%%%%%%%%%%%%%%%%%%%%%%%%%%%%%%%%
To get a \Galilei\ invariant electric field, the \Faraday\ one-form $\,\faradpotttt\,$
and electric zero-form $\,\VV_{\elevector,\ttt}\,$ are assumed to be \Galilei-invariant.
Splitting the spatial description according to \Leibniz\ rule and resorting to the homotopy formula, we get:
%%%%%%%%%%%%%%%%%%%%%%%%%%%%%%%%%%%
$$\framebox{$\vcenter{\halign{
\hfil$#$&$#$\hfil&$#$\hfil&$#$\hfil\cr
-\elefieldttt
&\,=\parder\tau\ttt\faradpottau
+\Lieder_{\Bv_{\moto,\ttt}}\,\faradpotttt
+\der \VV_{\elevector,\ttt}
\vspace{8pt}\cr
&\,=\parder\tau\ttt\faradpottau
+\der(\faradpotttt\punto\Bv_{\moto,\ttt})
+(\der\faradpotttt)\punto\Bv_{\moto,\ttt}
+\der \VV_{\elevector,\ttt}
\vspace{8pt}\cr
&\,=\parder\tau\ttt\faradpottau
+\der(\faradpotttt\punto\Bv_{\moto,\ttt})
+\magcurlttt\punto\Bv_{\moto,\ttt}
+\der \VV_{\elevector,\ttt}
\,,\cr}}$}$$
%%%%%%%%%%%%%%%%%%%%%%%%%%%%%%%%%%%
and in vector notation:
%%%%%%%%%%%%%%%%%%%%%%%%%%%%%%%%%%%
$$\framebox{$\vcenter{\halign{
\hfil$#$&$#$\hfil&$#$\hfil&$#$\hfil\cr
-\elevector_\ttt
&\,=\parder\tau\ttt\potmag_\tau
+\der(\,\metric\di{\potmag_\ttt,\Bv_{\moto,\ttt}})
-\Bv_{\moto,\ttt}\times\magindttt
+\der \VV_{\elevector,\ttt}
\,,\cr}}$}$$
%%%%%%%%%%%%%%%%%%%%%%%%%%%%%%%%%%%
This expression should be compared with the standard,
not \Galilei\ invariant, formula, see e.g. \cite[eq.$\,9.45\,$]{Sadiku2010},
which may be obtained by dropping the convective derivative:
%%%%%%%%%%%%%%%%%%%%%%%%%%%%%%%%%%%
$$\,-\elefieldttt
=\parder\tau\ttt\faradpottau
+\der\VV_{\elevector,\ttt}
\equi
-\elevector_\ttt=\parder\tau\ttt\potmag_\tau+\grad\VV_{\elevector,\ttt}\,.$$
%%%%%%%%%%%%%%%%%%%%%%%%%%%%%%%%%%%
%%%%%%%%%%%%%%%%%%%%%%%%%%%%%%%%%%%
\begin{remark}\label{rem: twoobs}
The \Galilei\ invariant formula for the electric field,
 in terms of the \Lie\ time-derivative along the motion of the \Faraday\ one-form $\,\faradpotttt\,$,
should give up with the claim about the fact that two \Galilei\ observers,
one fixed to the magnets and the other drifted by a relative translational motion,
should evaluate the electric field induced by the magnetic vortex
by resorting to different laws of electrodynamics
(see e.g. \citep[p. 477]{Griffiths1999}), 
an issue yet discussed by
\persone{Albert} \Einstein\ 
at the very beginning of his celebrated paper 
on the electrodynamics of moving bodies \citep{Einstein1905}.
\end{remark}
%%%%%%%%%%%%%%%%%%%%%%%%%%%%%%%%%%%
\begin{remark}\label{rem: Thomson}
After having independently developed the present treatment,
in reading the original paper of \cite{Thomson1893},
the author became aware of the fact that the 
very same \Galilei\ invariant formula for the electric field, expressed in cartesian coordinates, 
was there reported in ch. VII, p. 534, as depicted in fig. \ref{fig: SC} below.
%%%%%%%%%%%%%%%%%%%%%%%%%%%%%%%%
\begin{figure}[h]
\centering
%\setlength{\fboxsep}{0mm}
%\fbox{			
\includegraphics[width=1.05\textwidth]{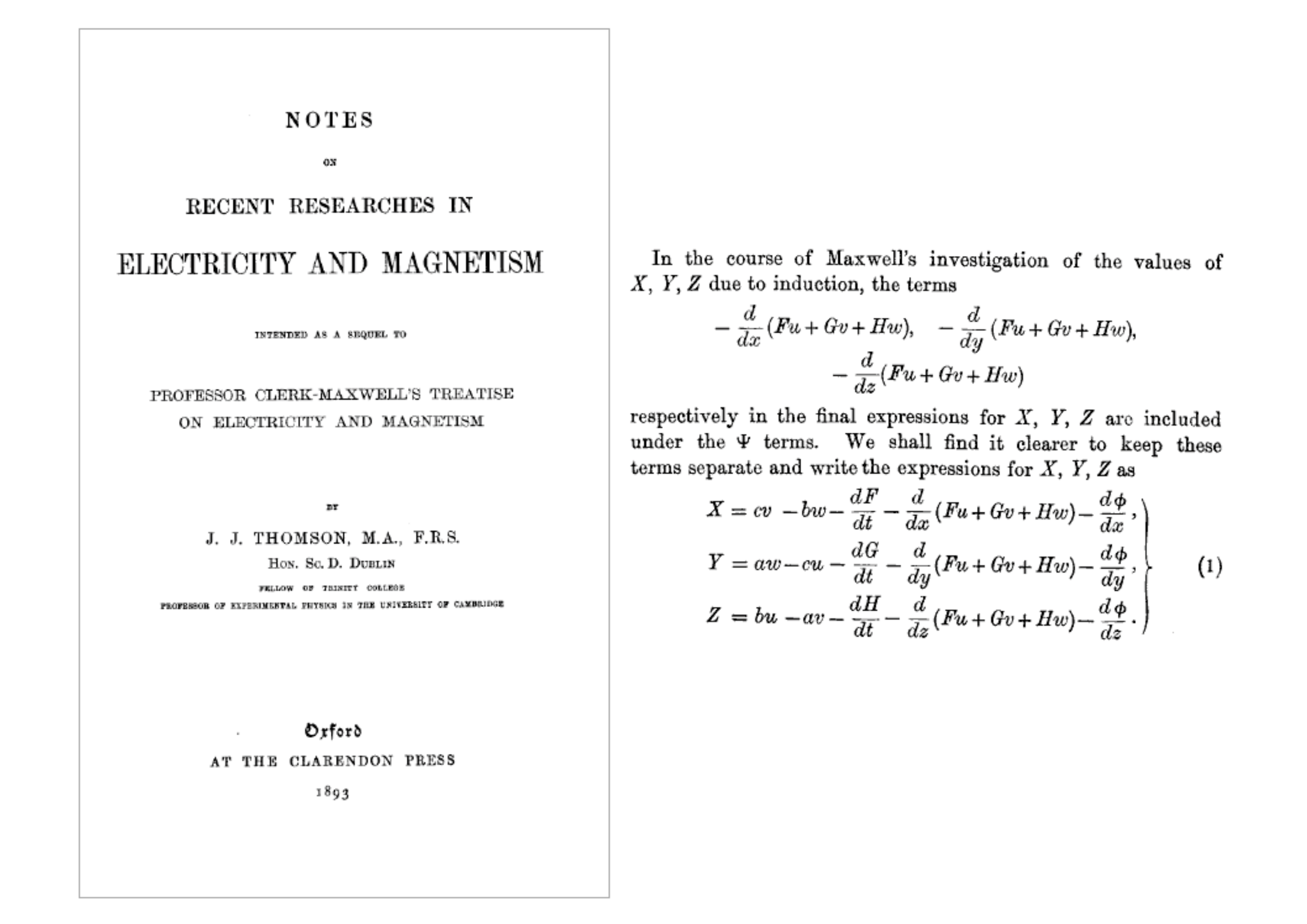}
%}
\caption{J.J. Thomson formulation}
\label{fig: SC}
\end{figure}
%%%%%%%%%%%%%%%%%%%%%%%%%%%%%%%%
It easy to check that the formula in fig. \ref{fig: SC}, when written in our notations, becomes:
%%%%%%%%%%%%%%%%%%%%%%%%%%%%%%%%%%%
$$\,\elefieldttt=-\magcurlttt\punto\Bv_{\moto,\ttt}
-\parder\tau\ttt\faradpottau
-\der(\faradpotttt\punto\Bv_{\moto,\ttt})
-\der \VV_{\elevector,\ttt}\,,$$
%%%%%%%%%%%%%%%%%%%%%%%%%%%%%%%%%%%
or in vector notation:
%%%%%%%%%%%%%%%%%%%%%%%%%%%%%%%%%%%
$\,\elevector_\ttt=\Bv_{\moto,\ttt}\times\magindttt
-\parder\tau\ttt\potmag_\tau
-\der\,\metric\di{\potmag_\ttt,\Bv_{\moto,\ttt}}
-\der\VV_{\elevector,\ttt}\,$.
%%%%%%%%%%%%%%%%%%%%%%%%%%%%%%%%%%%
As \persone{J.J.} \Thomson\ says, he got this expression by
a modification of the original formula by \cite{Maxwell1873} who, setting
%%%%%%%%%%%%%%%%%%%%%%%%%%%%%%%%%%%
$\,\UU_{\elevector,\ttt}=
\VV_{\elevector,\ttt}+\faradpotttt\punto\Bv_{\moto,\ttt}\,$,
%%%%%%%%%%%%%%%%%%%%%%%%%%%%%%%%%%%
wrote instead the electric field as:
%%%%%%%%%%%%%%%%%%%%%%%%%%%%%%%%%%%
$$\vcenter{\halign{
\hfil$#$&$#$\hfil&$#$\hfil&$#$\hfil\cr
\elefieldttt
&\,=-\parder\tau\ttt\faradpottau
-\magcurlttt\punto\Bv_{\moto,\ttt}
-\der\UU_{\elevector,\ttt}
\,,\cr}}$$
%%%%%%%%%%%%%%%%%%%%%%%%%%%%%%%%%%%
or in vector form
%%%%%%%%%%%%%%%%%%%%%%%%%%%%%%%%%%%
$\,\elevector_\ttt=-\parder\tau\ttt\potmag_\tau
+\Bv_{\moto,\ttt}\times\magindttt
-\nabla\UU_{\elevector,\ttt}\,$.
%%%%%%%%%%%%%%%%%%%%%%%%%%%%%%%%%%%
This expression was in fact originarily introduced in \citep[eq. 77]{Maxwell1861}
as the one of his equations
which includes the magnetic induction of the electric field,
but without explicit connection with \Faraday\ flux rule.
It is really surprising that engineers and physicists,
having had at hand the \Galilei\ invariant expression of the electric field
as formulated by
\persone{James} \ClerkMaxwell\ and
\persone{Joseph John} \Thomson,
have instead adopted, and still do, a non-invariant expression.
The reason may probably be found in that the wave equation in \emph{empty space}
is readily obtained from the expression without the convective term.
In our opinion, the two seemingly contradictory requirements,
i.e. \Galilei\ invariance and recovery of the wave equation in empty space,
may be reconciled by observing that the vanishing of the velocity field is a consequence
of isotropy and homogeneity of the electromagnetic constitutive properties
of the mass-free empty space, which make any motion of it to be undetectable.
%%%%%%%%%%%%%%%%%%%%%%%%%%%%%%%%%%%
\end{remark}
%%%%%%%%%%%%%%%%%%%%%%%%%%%%%%%%%%%
%%%%%%%%%%%%%%%%%%%%%%%%%%%%%%%%%%%
\begin{remark}\label{rem: lorfor}
In literature, the term $\,\Bv_{\moto,\ttt}\times\magindttt\,$ is 
referred to as the magnetic \Lorentz\ force per unit electric charge on a body in motion
\citep{Lorentz1899}, and most often introduced as a 
fundamental rule to be assumed in addition to the
law of magnetic induction,
see e.g. 
\citep[p.88]{Barut1980},
\cite[II.17-2]{Feynman1964},
\citep[p.238]{Greiner1998}
\citep[p.3]{Jackson1999},
\cite[ch.5.1.2]{Griffiths1999},
\citep[sec.15]{Kovetz2000}, 
\cite[ch.9.3B]{Sadiku2010}
\cite[6.1.2, p.344]{Lehner2010}.
The physical significance of a not \Galilei-invariant force is however highly questionable.
\end{remark}
%%%%%%%%%%%%%%%%%%%%%%%%%%%%%%%%%%%
It is important to underline that the term $\,\faradpotttt\punto\Bv_{\moto,\ttt}\,$
is spatially differentiable only under a regularity assumption for the 
velocity field which is likely to be violated in applications
(for instance when considering the motion of a transverse conductive bar 
sliding on a pair of parallel rails).
In these situations singular terms due to jumps in the velocity field must be
properly taken into account, see Sect. \ref{sec: sliding}, \ref{sec: FaradPar}.

\section{Magnetomotive induction by electric flux rate}
\label{sec: MMEF}

\subsection{Ampre law}
\label{sec: MatAMP}

The discovery by \persone{Hans Christian} \Oersted\ (1820)
that a magnetic field was induced by an electric current,
was immediately followed by a mathematical formulation of the law of electric induction, 
due to \persone{AndrŽ-Marie} \Ampere\ (1820), 
subsequently modified by \persone{James} \ClerkMaxwell\
who envisaged the basic additional term concerning the electric displacement
\citep{Maxwell1861}.
According to the point of view exposed in this paper,
\Ampere\ law is expressed by:
%%%%%%%%%%%%%%%%%%%%%%%%%%%%%%%%
$$\framebox{$
\vcenter{\halign{
$\hfil\displaystyle#$&$\displaystyle#$\hfil&$\quad\textrm{#}$\hfil\cr
\ointegrale{\partial\surfout}{}\magwindttt
&\,=\parder\tau\ttt
\integrale{\moto_{\tau,\ttt}\di{\surfout}}{}\elefluxtau
+\integrale{\surfout}{}\elecurrformttt
=\integrale{\surfout}{}\Lieder_{\moto,\ttt}\,\eleflux+\elecurrformttt\,,
\cr}}$}$$
%%%%%%%%%%%%%%%%%%%%%%%%%%%%%%%%
for any outer oriented circuit $\,\partial\surfout\,$ bounding a
correspondingly outer oriented surface 
$\,\surfout\,$, in the body placement at time $\,\ttt\in\II\,$,
see fig. \ref{fig: outer}.
The magnetic winding $\,\magwindttt\,$ is a \emph{odd} one-form,
the electric displacement flux
and the conduction current flux $\, \elefluxttt,\elecurrformttt\,$ 
are \emph{odd} two-forms.

Hence, the electric displacement $\,\eledispttt\,$,
defined by $\,\elefluxttt=\volform\eledispttt\,$, is an \emph{even} vector field,
and the magnetic field $\,\magvector_\ttt\,$,
defined by $\,\magwindttt=\metric\magvector_\ttt\,$,
is a \emph{odd} vector field.
The electric current field $\,\elecurrttt\,$,
defined by $\, \elecurrformttt=\volform\elecurrttt\,$
is an \emph{even} vector field.

\subsection{Well-posedness of Ampre law}
\label{sec: wellAmpere}

In order that \Ampere\ law be meaninful, it is to be
proven that the r.h.s. is independent of the choice of surface $\,\surf\,$,
for a given circuit $\,\partial\surf\,$, and independent of the
motion of surface $\,\surf\,$ for a given motion of circuit $\,\partial\surf\,$.
This condition may be formalized by requiring that the 
time derivatives of the integrals:
%%%%%%%%%%%%%%%%%%%%%%%%%%%%%%%%
$$\,\integrale{\moto^1_{\tau,\ttt}\di{\surfout^1}}{}
\Lieder_{\moto,\ttt}\,\eleflux+\elecurrformttt\,,
\qquad\integrale{\moto^2_{\tau,\ttt}\di{\surfout^2}}{}
\Lieder_{\moto,\ttt}\,\eleflux+\elecurrformttt\,,$$
%%%%%%%%%%%%%%%%%%%%%%%%%%%%%%%%
be the same for motions such that
%%%%%%%%%%%%%%%%%%%%%%%%%%%%%%%%
$$\,\partial(\moto^1_{\tau,\ttt}\di{\surfout^1})=\partial(\moto^2_{\tau,\ttt}\di{\surfout^2})\,.$$
%%%%%%%%%%%%%%%%%%%%%%%%%%%%%%%%
The chain $\,\moto^1_{\tau,\ttt}\di{\surfout^1}-\moto^2_{\tau,\ttt}\di{\surfout^2}\,$
is then a closed surface for any $\,\tau\in\II\,$.
It follows that,
for any outer-oriented control-window $\,\controltout\,$,
the flux across its boundary surface $\,\partial\controltout\,$ should vanish:
%%%%%%%%%%%%%%%%%%%%%%%%%%%%%%%%%%%
$$\,\ointegrale{\partial\controltout}{}(\Lieder_{\moto,\ttt}\,\eleflux+\elecurrformttt)=
\integrale{\controltout}{} \der(\Lieder_{\moto,\ttt}\,\eleflux+\elecurrformttt)=0\,.$$
%%%%%%%%%%%%%%%%%%%%%%%%%%%%%%%%%%%
Localizing, we get the equivalent closedness condition:
%%%%%%%%%%%%%%%%%%%%%%%%%%%%%%%%%%%
$$\,\der(\Lieder_{\moto,\ttt}\,\eleflux+\elecurrformttt)=0\,,$$
%%%%%%%%%%%%%%%%%%%%%%%%%%%%%%%%%%%
which, recalling the commutative property:
%%%%%%%%%%%%%%%%%%%%%%%%%%%%%%%%%%%
$$\,\der(\Lieder_{\moto,\ttt}\,\eleflux)
=\Lieder_{\moto,\ttt}\,(\der\elefluxttt)\,,$$
%%%%%%%%%%%%%%%%%%%%%%%%%%%%%%%%%%%
and resorting to \Gauss\ law for the electric displacement flux,
$\,\der\eleflux=\elechargeform\,$, is readily shown to be
equivalent to the \emph{electric charge balance law}:
%%%%%%%%%%%%%%%%%%%%%%%%%%%%%%%%%%%
$$\framebox{$\Lieder_{\moto,\ttt}\,\elechargeform+\der\elecurrformttt=0\,,$}$$
%%%%%%%%%%%%%%%%%%%%%%%%%%%%%%%%%%%
and, in integral form:
%%%%%%%%%%%%%%%%%%%%%%%%%%%%%%%%
$$\vcenter{\halign{
$\displaystyle#$\hfil&$\displaystyle#$\hfil&$\displaystyle#$\hfil&$\displaystyle#$\hfil\cr
&\,\integrale{\controltout}{}\Lieder_{\moto,\ttt}\,\elechargeform+\der\elecurrformttt
=\parder\tau\ttt
\integrale{\moto_{\tau,\ttt}\di{\controltout}}{}\elechargeformtau
+\integrale{\controltout}{} \der\elecurrformttt
\vspace{8pt}\cr
=&\,\ointegrale{\partial\controltout}{}\Lieder_{\moto,\ttt}\,\eleflux+\elecurrformttt
%\vspace{8pt}\cr
=\parder\tau\ttt
\ointegrale{\moto_{\tau,\ttt}\di{\partial\controltout}}{}\elefluxtau
+\ointegrale{\partial\controltout}{}\elecurrformttt
=0\,.
\cr}}$$
%%%%%%%%%%%%%%%%%%%%%%%%%%%%%%%%
The electric charge $\, \elechargeform\,$ is a \emph{odd} three-form 
which may be integrated over even non-orientable
manifolds to evaluate the total charge.

Observing that the outer orientations of open $\,3$D manifolds in \Euclid\ space, 
\emph{spring} and \emph{sink}, respectively correspond to
outer orientations \emph{outward} and \emph{inward} for its boundary $\,2$D manifold,
the electric charge balance law has to be read as:
%%%%%%%%%%%%%%%%%%%%%%%%%%%%%%%%
\begin{itemize}\item[-]
The time-rate of increase of the total electric charge,
in a traveling control-window, 
is equal to the inward flux of electric conduction current
through the window boundary.
\end{itemize}
%%%%%%%%%%%%%%%%%%%%%%%%%%%%%%%%
We emphasize that the assumption of absence of bulk sources of electric charge
plays a basic role in ensuring well-posedness of \Ampere\ law.
%%%%%%%%%%%%%%%%%%%%%%%%%%%%%%%%
\begin{remark}\label{rem: eqcontinuity}
In literature, the electric charge balance law is usually written,
in terms of spatial description of the involved fields, as:
$$\,\parder\tau\ttt\elechargeformtau+\der\elecurrformttt=0\,,$$
and is called the equation of continuity, see e.g.
\citep[p.90]{Barut1980},
\citep[p.127]{Purcell1985},
\citep[II.18-1]{Feynman1964},
\citep[p.9]{Schwinger1998},
\citep[p.251]{Greiner1998},
\citep[p.238]{Jackson1999},
\citep[p.345]{Griffiths1999},
\citep[p.50]{Wegner2003},
\citep[p.10]{Thide2010},
\citep[p.385]{Sadiku2010}.
The expression:
$\,\Lieder_{\moto,\ttt}\,\elechargeform+\der\elecurrform=0\,$
introduced above,
reduces to the usual one by assuming a translating body
and a \Galilei\ observer sitting on it.
We remark that, according to Lemma \ref{lm: invarianceconv}, the term 
$\,\Lieder_{\moto,\ttt}\,\elechargeform\,$
is  \Galilei\ invariant since such is the electric charge form $\, \elechargeformttt\,$.
If the formulation of the equation of continuity in terms of partial time derivative
of the electric charge is assumed to be (as usually made in literature) 
a general physical law, \Ampere\ law of induction
would be well-posed only for \Galilei\ observers testing
time-invariant material circuits, clearly a completely unsatisfactory conclusion.
\end{remark}
%%%%%%%%%%%%%%%%%%%%%%%%%%%%%%%%
When the spatial description of the material tensor $\,\elechargeform\,$
has a regular time-dependence, we may write:
%%%%%%%%%%%%%%%%%%%%%%%%%%%%%%%%%%%
$$\,\Lieder_{\moto,\ttt}\,\elechargeform=
\parder\tau\ttt\elechargeformtau+\Lieder_{\Bv_{\moto,\ttt}}\,\elechargeformttt\,.$$
%%%%%%%%%%%%%%%%%%%%%%%%%%%%%%%%%%%
Then, by the homotopy formula, being $\,\der\elechargeformttt=0\,$, we infer that:
%%%%%%%%%%%%%%%%%%%%%%%%%%%%%%%%%%%
$$\,\Lieder_{\Bv_{\moto,\ttt}}\,\elechargeformttt
=\der(\elechargeformttt\punto\Bv_{\moto,\ttt})+(\der\elechargeformttt)\punto\Bv_{\moto,\ttt}
=\der(\elechargeformttt\punto\Bv_{\moto,\ttt})\,,$$
%%%%%%%%%%%%%%%%%%%%%%%%%%%%%%%%%%%
and the spatial description of electric charge balance law may be written,
in terms of exterior derivatives, as:
%%%%%%%%%%%%%%%%%%%%%%%%%%%%%%%%%%%
$$\,\parder\tau\ttt\elechargeformtau
+\der(\elechargeformttt\Bv_{\moto,\ttt})
+\der\elecurrformttt=0\,.$$
%%%%%%%%%%%%%%%%%%%%%%%%%%%%%%%%%%%
In vector notations we recover the well-known \Helmholtz\ equation:
%%%%%%%%%%%%%%%%%%%%%%%%%%%%%%%%%%%
$$\,\parder\tau\ttt\chargescalartau
+\diverg(\chargescalarttt\,\Bv_{\moto,\ttt})
+\diverg\elecurrttt=0\,,$$
%%%%%%%%%%%%%%%%%%%%%%%%%%%%%%%%%%%
as quoted in \citep{Darrigol2000} who refers to \citep{Helmholtz1870}.

\subsection{Differential form of Ampre law}
\label{sec: diffAmpere}

Upon localization, \Ampere's law may be formulated in differential terms as
according to the equivalent notations:
%%%%%%%%%%%%%%%%%%%%%%%%%%%%%%%%%%%
$$\framebox{$\der\magwindttt
=\Lieder_{\moto,\ttt}\,\eleflux+\elecurrformttt\,.$}$$
%%%%%%%%%%%%%%%%%%%%%%%%%%%%%%%%%%%
%%%%%%%%%%%%%%%%%%%%%%%%%%%%%%%%%%%
$$\,\der(\metric\magvector_\ttt)
=\Lieder_{\moto,\ttt}\,(\volform\eledisp)+ \volform\elecurrttt\,,$$
%%%%%%%%%%%%%%%%%%%%%%%%%%%%%%%%%%%
%%%%%%%%%%%%%%%%%%%%%%%%%%%%%%%%%%%
$$\,\volform\punto(\rotor\magvector_\ttt)
=\Lieder_{\moto,\ttt}\,(\volform\eledisp)+\volform\elecurrttt\,,$$
%%%%%%%%%%%%%%%%%%%%%%%%%%%%%%%%%%%
%%%%%%%%%%%%%%%%%%%%%%%%%%%%%%%%%%%
$$\,\rotor\magvector_\ttt
=\Lieder_{\moto,\ttt}\,\eledispttt+(\diverg\Bv_{\moto,\ttt})\,\eledispttt+\elecurrttt\,.$$
%%%%%%%%%%%%%%%%%%%%%%%%%%%%%%%%%%%
Setting:
%%%%%%%%%%%%%%%%%%%%%%%%%%%%%%%%%%%
$\,\elefluxttt=\elechargedue+\Boo^2_{\nocharge,\ttt}\,$
%%%%%%%%%%%%%%%%%%%%%%%%%%%%%%%%%%%
with
%%%%%%%%%%%%%%%%%%%%%%%%%%%%%%%%%%%
$\,\der\elechargedue=\elechargeformttt\,$ and
$\,\der\Boo^2_{\nocharge,\ttt}=0\,$,
%%%%%%%%%%%%%%%%%%%%%%%%%%%%%%%%%%%
we may introduce the \Ampere\ electric potential one-form $\,\amperpotttt\,$ such that:
%%%%%%%%%%%%%%%%%%%%%%%%%%%%%%%%%%%
$$\,\Boo^2_{\nocharge,\ttt} =\der\amperpotttt\,,$$
%%%%%%%%%%%%%%%%%%%%%%%%%%%%%%%%%%%
and the differential form of \Ampere\ law may be written as:
%%%%%%%%%%%%%%%%%%%%%%%%%%%%%%%%%%%
$$\der\magwindttt
=\Lieder_{\moto,\ttt}\,\elechargedue
+ \der\Lieder_{\moto,\ttt}\,\amperpotttt
+ \elecurrformttt\,.$$
%%%%%%%%%%%%%%%%%%%%%%%%%%%%%%%%%%%
Being $\,\der(\Lieder_{\moto,\ttt}\,\elechargedue+\elecurrformttt)=0\,$
we may set
$\,\Lieder_{\moto,\ttt}\,\elechargedue+ \elecurrformttt
=\der\Boo^1_{\moto,\elechargeform,\elecurr,\ttt}\,$
and write:
%%%%%%%%%%%%%%%%%%%%%%%%%%%%%%%%%%%
$$\vcenter{\halign{
\hfil$#$&$#$\hfil&$#$\hfil&$#$\hfil\cr
\magwindttt
&\,=\Lieder_{\moto,\ttt}\,\amperpotttt
+\Boo^1_{\moto,\elechargeform,\elecurr,\ttt}
+\der\VV_{\magvector,\ttt}
\vspace{8pt}\cr
&\,=\parder\tau\ttt\,\amperpotttt
+\der(\amperpotttt\punto\Bv_{\moto,\ttt})
+\elefluxttt\punto\Bv_{\moto,\ttt}
+\Boo^1_{\moto,\elechargeform,\elecurr,\ttt}
+\der\VV_{\magvector,\ttt}
\,.\cr}}$$
%%%%%%%%%%%%%%%%%%%%%%%%%%%%%%%%%%%
all the terms, at the r.h.s. of the first equality, being \Galilei\ invariant.

The field theory of electromagnetic induction is based on the assumption that
the electric displacement flux $\,\elefluxttt\,$, the electric current $\,\elecurrformttt\,$
and the magnetic winding $\,\magwindttt\,$
are spread throughout the ambient space.
Then, the spatial description of \Gauss\ law should be formulated as
$\,\der\elefluxttt=\elechargeformttt\,$ with $\,\elechargeformttt=0\,$
in free space, i.e. at spatial events $\,\coppia\Bx\ttt\in\EU\times\II\,$
such that no charged material particle is passing through $\,\Bx\in\EU\,$
at time $\,\ttt\in\II\,$.
\Ampere\ law is accordingly written in spatial description as:
%%%%%%%%%%%%%%%%%%%%%%%%%%%%%%%%
$$\framebox{$
\vcenter{\halign{
$\hfil\displaystyle#$&$\displaystyle#$\hfil&$\quad\textrm{#}$\hfil\cr
\ointegrale{\partial\surfout}{}
\magwindttt
&\,=\integrale{\surfout}{}
(\elecurrformttt
+\parder\tau\ttt\elefluxtau)
+\ointegrale{\partial\surfout}{}
\elefluxttt\punto\Bv_{\moto,\ttt}
+\integrale{\surfout}{}\elechargeformttt\punto\Bv_{\moto,\ttt}\,,
\cr}}$}$$
%%%%%%%%%%%%%%%%%%%%%%%%%%%%%%%%
and in differential form:
%%%%%%%%%%%%%%%%%%%%%%%%%%%%%%%%%%%
$$\framebox{$
\der\magwindttt
=\parder\tau\ttt\elefluxtau
+\elecurrformttt
+\der(\elefluxttt\punto\Bv_{\moto,\ttt})
+\elechargeformttt\punto\Bv_{\moto,\ttt}
\,,$}$$
%%%%%%%%%%%%%%%%%%%%%%%%%%%%%%%%%%%
or in vector analysis notation:
%%%%%%%%%%%%%%%%%%%%%%%%%%%%%%%%%%%
$$\framebox{$\rotor\magvector_\ttt
=\parder\tau\ttt\eledisp_\tau
+\elecurrttt
+\rotor(\eledispttt\times\Bv_{\moto,\ttt})
+\chargescalarttt\,\Bv_{\moto,\ttt}\,,$}$$
%%%%%%%%%%%%%%%%%%%%%%%%%%%%%%%%%%%
which should be compared with the customary one,
in which the velocity vanishes,
e.g. \cite[eq.$\,9.23\,$]{Sadiku2010}:
%%%%%%%%%%%%%%%%%%%%%%%%%%%%%%%%%%%
$$\,\rotor\magvector_\ttt=\parder\tau\ttt\eledisp_\tau+\elecurrttt\,.$$
%%%%%%%%%%%%%%%%%%%%%%%%%%%%%%%%%%%

\subsection{Galilei invariance of Ampre law}
\label{sec: GalileiAmpere}

Two \Galilei\ observers will measure velocity fields differing by
a time-independent translational velocity field $\,\Bv_\gamma\in\cont^1\di{\EU\sp\TEU}\,$. 
The \Galilei\ invariance of \Ampere\ law follows from the
 \Galilei\ invariance of the involved fields and from Lemma
 \ref{lm: invarianceconv} ensuring the  invariance of
 the convective time-derivative of invariant tensors.

\section{Electromagnetic constitutive relations}
\label{sec: constitutive}

When expressed in terms of differential forms,
the laws of electromagnetic induction do not involve
neither the chosen orientation nor
the metric properties of the physical space.
The constitutive laws expressing the
\emph{electric permittivity} and the \emph{magnetic permeability}
of a medium in terms of differential forms,
are independent of the metric properties of the space
but depend on the choice of a volume form.

In the classical volume manifold $\,\coppia\EU\volform\,$,
the \emph{electric permittivity} 
$\,\Perm_\ele\in\cont^1\di{\TEU\sp\CTEU}\,$
is a pointwise relation between
the electric circulation one-form $\,\elefieldttt\,$ and the electric flux two-form $\,\elefluxttt\,$.
The \emph{odd} electric flux $\,\elefluxttt\,$ 
is in one-to-one linear correspondence with
the \emph{even} electric displacement vector field $\,\eledispttt\,$
according to the relation $\,\elefluxttt=\volform\punto\eledispttt\,$.
The separating duality induced by the pairing 
$\,\scalar{\elefieldttt}{\eledispttt}\,$,
between dual \emph{even} geometrical fields,
leads to the following electric constitutive equation:
%%%%%%%%%%%%%%%%%%%%%%%%%%%%%%%%%%%
$$\, \elefieldttt=\Perm_\ele\di{\eledispttt}\,.$$
%%%%%%%%%%%%%%%%%%%%%%%%%%%%%%%%%%%.
Analogously, the \emph{magnetic permeability} $\,\Perm_\mag\in\cont^1\di{\CTEU\sp\TEU}\,$
is a pointwise relation between the magnetic winding \emph{odd} one-form $\,\magwindttt\,$
and the magnetic \emph{odd} vector field $\, \magindttt\,$ which is in one-to-one 
linear correspondence with the \emph{even} magnetic vortex two-form $\,\magcurlttt\,$
according to the relation $\, \magcurlttt=\volform\punto\magindttt\,$.
The duality pairing
$\,\scalar{\magwindttt}{\magindttt}\,$,
between dual \emph{odd} geometrical fields,
leads to the following magnetic constitutive equation:
%%%%%%%%%%%%%%%%%%%%%%%%%%%%%%%%%%%
$$\, \magwindttt=\Perm_\mag\di{\magindttt}\,.$$
%%%%%%%%%%%%%%%%%%%%%%%%%%%%%%
The \emph{empty space} is assumed to be massless and to have 
have linear, uniform and isotropic electromagnetic constitutive properties. 
The \emph{electric permittivity}  and the \emph{magnetic permeability}
are then fields of linear maps between dual spaces,
which can be represented by scalar fields.
Indeed, in the standard \Euclid\ space $\,\coppia{\EU}{\metric}\,$
the non-singular metric tensor leads to the one-to-one correspondences:
%%%%%%%%%%%%%%%%%%%%%%%%%%%%%%%%%%%
$$\,\elefieldttt=\metric\punto\elevector_\ttt\,,
\quad\magcurlttt=\volform \punto\magindttt\,,
\quad\magwindttt=\metric\punto\magvector_\ttt\,,$$
%%%%%%%%%%%%%%%%%%%%%%%%%%%%%%%%%%%
and in \emph{empty space} we may set:
%%%%%%%%%%%%%%%%%%%%%%%%%%%%%%%%%%%
$$\,\elevector_\ttt=\perm_\ele\,\eledispttt\,,\quad
\magvector_\ttt=\perm_\mag\,\magindttt\,,$$
%%%%%%%%%%%%%%%%%%%%%%%%%%%%%%%%%%%
with $\,\perm_\ele,\perm_\mag:\EU\mapsto\Re\,$ constant scalar fields.
Due to the uniformity and isotropy of its electro\-magnetic constitutive properties,
no motion of the massless \emph{empty space} can be detected
and the laws of induction in the \emph{empty space} reduce to the standard ones
with the partial derivatives in place of the \Lie\ time-derivatives.

\subsection{Poynting vector}
\label{sec: Poynting}

The total electric and magnetic power expended, per unit volume in
a control window in \emph{empty space}, is the \emph{even} scalar field given by the formula:
%%%%%%%%%%%%%%%%%%%%%%%%%%%%%%%%%%%
$$\,\scalar{\elefieldttt}{\elecurrttt+\dot\eledisp_\ttt}+\scalar{\magwindttt}{\dot\magind_\ttt}\,,$$
%%%%%%%%%%%%%%%%%%%%%%%%%%%%%%%%%%%
where $\,\dot\eledisp_\ttt\equaldef\parder\tau\ttt\eledisp_\tau\,$
and $\,\dot\magind_\ttt\equaldef\parder\tau\ttt\magind_\tau\,$.

On the other hand, we have the identity:
%%%%%%%%%%%%%%%%%%%%%%%%%%%%%%%%%%%
$$\,\scalar{\elefieldttt}{\rotor\magvector_\ttt}
-\scalar{\magwindttt}{\rotor\elevector_\ttt}
=-\diverg(\elevector_\ttt\times\magvector_\ttt)\,.$$
%%%%%%%%%%%%%%%%%%%%%%%%%%%%%%%%%%%
\Faraday\ and \Ampere\ laws of induction:
%%%%%%%%%%%%%%%%%%%%%%%%%%%%%%%%%%%
$$\vcenter{\halign{
\hfil$#$&$#$\hfil&$#$\hfil&$#$\hfil\cr
\rotor\magvector_\ttt&\,=\elecurrttt+\dot\eledisp_\ttt\,,
\vspace{8pt}\cr
-\rotor\elevector_\ttt&\,=\dot \magind_\ttt
\,,\cr}}$$
%%%%%%%%%%%%%%%%%%%%%%%%%%%%%%%%%%%
substituted in the identity above, yield \Poynting\ relation:
%%%%%%%%%%%%%%%%%%%%%%%%%%%%%%%%%%%
$$\,\scalar{\elefieldttt}{\elecurrttt+\dot\eledisp_\ttt}+\scalar{\magwindttt}{\dot\magind_\ttt}
=-\diverg(\elevector_\ttt\times\magvector_\ttt)\,,$$
%%%%%%%%%%%%%%%%%%%%%%%%%%%%%%%%%%%
whose integral version pertaining to a $\,3$D control window $\,\controlt\,$ writes:
%%%%%%%%%%%%%%%%%%%%%%%%%%%%%%%%%%%
$$\,\integrale{\controltout}{}\scalar{\elefieldttt}{\elecurrttt+\dot\eledisp_\ttt}+\scalar{\magwindttt}{\dot\magind_\ttt}\volform
+\ointegrale{\partial \controltout}{}\volform\punto(\elevector_\ttt\times\magvector_\ttt)=0\,.$$
%%%%%%%%%%%%%%%%%%%%%%%%%%%%%%%%%%%
The introduction of the \emph{even} vector field $\,\elevector_\ttt\times\magvector_\ttt\,$
is due to \persone{John Henry} \Poynting\
in \citep{Poynting1884} and to \persone{Oliver} \Heaviside\ in the same year,
see \cite[ch.II, p.132]{Stratton1941}.
The relation may be read as follows:
The total electric and magnetic power expended,
per unit volume of a control window in \emph{empty space},
is equal to the incoming flux of the \Poynting\ vector field through
its boundary.

\section{Formulation in affine space-time manifold}
\label{sec: Fourdimensional}

Let $\,\MMM^4\,$ be a four-dimensional affine manifold
(a flat space-time)
with model linear space $\,\VV^4\,$, see e.g. \citep{Cartan1924}.
In classical space-time $\,\MMM^4\,$, each observer defines a field
of world-lines directed along a constant time-like $\,4$-vector field 
$\,\Bu\in\cont^1\di{\EU\sp\TM^4}\,$.
These world-lines induce a diffeomorphism 
$\,\obs\in\cont^1\di{\EU\times\II\sp\MMM^4}\,$ given by:
%%%%%%%%%%%%%%%%%%%%%%%%%%%%%%%%%%%
$$\,\obs\coppia\Bx\ttt\equaldef\ttt\Bu\di\Bx+\obs\coppia\Bx0\in\MMM^4\,,
\qquad\Bx\in\EU\,,\ttt\in\II\,,$$
%%%%%%%%%%%%%%%%%%%%%%%%%%%%%%%%%%%
with the time-like field assumed to be the same for all \Euclid\ observers.

The inverse map 
$\,\inv\obs\in\cont^1\di{\MMM^4\sp\EU\times\II}\,$
assigns, to any event in $\,\MMM^4\,$, the location and the time-instant
as detected by the observer.

Defining the injective immersion $\,\obs_\ttt\in\cont^1\di{\EU\sp\MMM^4}\,$
by $\,\obs_\ttt\di\Bx\equaldef\obs\coppia\Bx\ttt\,$,
the range $\,\NNN_\ttt\equaldef\obs_\ttt\di\EU\,$ collects the events 
that are judged as simultaneous at time $\,\ttt\in\II\,$ by the observers.

Time instant and spatial position
are extracted from the pair $\,\coppia\Bx\ttt\in\EU\times\II\,$ by cartesian projectors
$\,\projE\in\cont^1\di{\EU\times\II\sp\EU}\,$ and
$\,\projI\in\cont^1\di{\EU\times\II\sp\II}\,$ which are affine maps
defined by:
%%%%%%%%%%%%%%%%%%%%%%%%%%%%%%%%%%%
$$\,\projE\coppia{\Bx}{\ttt}=\Bx\,,
\quad
\projI\coppia{\Bx}{\ttt}=\ttt\,.$$
%%%%%%%%%%%%%%%%%%%%%%%%%%%%%%%%%%%
The differentials 
$\,\TT\projE\in\cont^0\di{\TEU\times\TI\sp\TEU}\,$ and
$\,\TT\projI\in\cont^0\di{\TEU\times\TI\sp\TI}\,$
are constant maps which are linear in the increments
$\,\coppia{\delta\Bx}{\delta\ttt}\in\TANG_\Bx\EU\times\TANG_\ttt\II\,$:
%%%%%%%%%%%%%%%%%%%%%%%%%%%%%%%%%%%
$$\vcenter{\halign{
\hfil$#$&$#$\hfil&$#$\hfil&$#$\hfil\cr
&\TT\projE\di{\Bx,\ttt}\punto\coppia{\delta\Bx}{\delta\ttt}\equaldef\delta\Bx\,,
\vspace{8pt}\cr
&\TT\projI\di{\Bx,\ttt}\punto\coppia{\delta\Bx}{\delta\ttt}\equaldef\delta\ttt
\,.\cr}}$$
%%%%%%%%%%%%%%%%%%%%%%%%%%%%%%%%%%%
By the identifications $\,\TANG_\ttt\II\equiv\TANG_0\II\equiv\II\,$, we will consider
$\,\TT\projI\in \EXForms^1\di{\EU\times\II\sp\Re}\,$ as a one-form.

The next Lemma shows that 
a $\,k$-form on the space-time manifold $\,\MMM^4\,$ is seen
by an observer as equivalent to a pair of forms,
respectively of degree $\,k\,$ and $\,k-1\,$, 
in \Euclid\ space.
This basic result enables to compare formulations of electrodynamics 
in the four dimensional space-time with the standard one in the \Euclid\ space.
%%%%%%%%%%%%%%%%%%%%%%%%%%%%%%%%%%%
\begin{lemma}[Split of exterior forms]\label{lm: fourtothree}
For any $\,k$-form $\,\Exter\in\EXForms^k\di{\MMM^4\sp\Re}\,$, $\,k\le{4}\,$,
being $\,\obs_\ttt\pull\Boo^k\in\EXForms^k\di{\EU\sp\Re}\,$ and 
$\,\obs_\ttt\pull(\Exter\punto\Bu)\in\EXForms^{(k-1)}\di{\EU\sp\Re}\,$,
we have the split formula:
%%%%%%%%%%%%%%%%%%%%%%%%%%%%%%%%%%%
$$\,\obs\pull\Exter=
\projE\pull(\obs_\ttt\pull\Exter)
+\TT\projI\wedge\projE\pull\obs_\ttt\pull(\Exter\punto\Bu)\,.$$
%%%%%%%%%%%%%%%%%%%%%%%%%%%%%%%%%%%
\end{lemma}\proof
%%%%%%%%%%%%%%%%%%%%%%%%%%%%%%%%%%%
Assuming for simplicity $\,k=2\,$, we have that:
%%%%%%%%%%%%%%%%%%%%%%%%%%%%%%%%%%%
$$\,\obs\push\coppia{\delta\Bx_i}{\delta\ttt_i}=\delta\ttt_i\,\Bu+\obs_\ttt\push{\delta\Bx_i}\in\TM^4\,,$$
%%%%%%%%%%%%%%%%%%%%%%%%%%%%%%%%%%%
with $\,\coppia{\delta\Bx_i}{\delta\ttt_i}\in\TANG_\Bx\EU\times\TANG_\ttt\II\,$ for $\,i=1,2\,$,
and
%%%%%%%%%%%%%%%%%%%%%%%%%%%%%%%%%%%
$\,\TT\projI\di{\Bx,\ttt}\punto\coppia{\delta\Bx_i}{\delta\ttt_i}=\delta\ttt_i\,$.
%%%%%%%%%%%%%%%%%%%%%%%%%%%%%%%%%%%
The definition of exterior product gives:
%%%%%%%%%%%%%%%%%%%%%%%%%%%%%%%%%%%
$$\vcenter{\halign{
\hfil$#$&$#$\hfil&$#$\hfil&$#$\hfil\cr
&\,(\TT\projI\wedge\projE\pull\obs_\ttt\pull(\Exter\punto\Bu)
\punto\coppia{\coppia{\delta\Bx_1}{\delta\ttt_1}}{\coppia{\delta\Bx_2}{\delta\ttt_2}}
\vspace{8pt}\cr
&\,=\obs_\ttt\pull(\Exter\punto\Bu)\punto{\delta\Bx_2}\,\delta\ttt_1
-\obs_\ttt\pull(\Exter\punto\Bu)\punto{\delta\Bx_1}\,\delta\ttt_2
\,.\cr}}$$
%%%%%%%%%%%%%%%%%%%%%%%%%%%%%%%%%%%
Hence:
%%%%%%%%%%%%%%%%%%%%%%%%%%%%%%%%%%%
$$\vcenter{\halign{
\hfil$#$&$#$\hfil&$#$\hfil&$#$\hfil\cr
&\,(\obs\pull\Exter)\punto\coppia{\coppia{\delta\Bx_1}{\delta\ttt_1}}{\coppia{\delta\Bx_2}{\delta\ttt_2}}
\vspace{8pt}\cr
&\,=\obs\pull\bigl(
\Exter\punto\coppia{\obs\push\coppia{\delta\Bx_1}{\delta\ttt_1}}{\obs\push\coppia{\delta\Bx_2}{\delta\ttt_2}}
\bigr)
\vspace{8pt}\cr
&\,=\obs\pull\bigl(
\Exter\punto\coppia{\delta\ttt_1\,\Bu+\obs_\ttt \push{\delta\Bx_1}}{\delta\ttt_2\,\Bu+\obs_\ttt \push{\delta\Bx_2}}
\bigr)
\vspace{8pt}\cr
&\,=\obs\pull\bigl(
\Exter\punto\coppia{\obs_\ttt \push{\delta\Bx_1}}{\obs_\ttt \push{\delta\Bx_2}}
+\Exter\punto\coppia{\Bu}{\obs_\ttt \push{\delta\Bx_2}}\,\delta\ttt_1
-\Exter\punto\coppia{\Bu}{\obs_\ttt \push{\delta\Bx_1}}\,\delta\ttt_2
\bigr)
\,.\cr}}$$
%%%%%%%%%%%%%%%%%%%%%%%%%%%%%%%%%%%
Observing that:
%%%%%%%%%%%%%%%%%%%%%%%%%%%%%%%%%%%
$\,\obs_\ttt\push\delta\Bx=\obs\push\coppia{\delta\Bx}{0_\ttt}\,$
and hence
$\,\obs\pull(\obs_\ttt\push\delta\Bx)=\coppia{\delta\Bx}{0_\ttt}\,$,
%%%%%%%%%%%%%%%%%%%%%%%%%%%%%%%%%%%
we may write:
%%%%%%%%%%%%%%%%%%%%%%%%%%%%%%%%%%%
$$\vcenter{\halign{
\hfil$#$&$#$\hfil&$#$\hfil&$#$\hfil\cr
&\,\obs\pull\Exter\punto\coppia{\coppia{\delta\Bx_1}{\delta\ttt_1}}{\coppia{\delta\Bx_2}{\delta\ttt_2}}
=\obs\pull\Exter\punto\coppia{\coppia{\delta\Bx_1}{0_\ttt}}{\coppia{\delta\Bx_2}{0_\ttt}}
\vspace{8pt}\cr
&\,+\,\,\obs\pull(\Exter\punto\Bu)\punto\coppia{\delta\Bx_2}{0_\ttt}\,\delta\ttt_1
+\obs\pull(\Exter\punto\Bu)\punto\coppia{\delta\Bx_1}{0_\ttt}\,\delta\ttt_2
\vspace{8pt}\cr
&\,=(\obs_\ttt\pull\Exter)\punto\coppia{\delta\Bx_1}{\delta\Bx_2}
+\obs_\ttt\pull(\Exter\punto\Bu)\punto{\delta\Bx_2}\,\delta\ttt_1
-\obs_\ttt\pull(\Exter\punto\Bu)\punto{\delta\Bx_1}\,\delta\ttt_2
\vspace{8pt}\cr
&\,=(\projE\pull(\obs_\ttt\pull\Exter)
+\TT\projI\wedge\projE\pull\obs_\ttt\pull(\Exter\punto\Bu))
\punto\coppia{\coppia{\delta\Bx_1}{\delta\ttt_1}}{\coppia{\delta\Bx_2}{\delta\ttt_2}}
\,,\cr}}$$
%%%%%%%%%%%%%%%%%%%%%%%%%%%%%%%%%%%
which is the result.
\endprova
%%%%%%%%%%%%%%%%%%%%%%%%%%%%%%%%%

\subsection{Space-time formulations}
\label{sec: FourForms}

The expressions of electric and magnetic induction rules,
according to \Faraday\ and \Ampere\ laws, 
take their most concise form in the space-time manifold $\,\MMM^4\,$
when expressed in terms of the
\Faraday \ and \Ampere\
electromagnetic two-forms 
$\,\Faraddue,\Amper\in\EXForms^2\di{\TANG\MMM^4\sp\Re}\,$.

These forms are referred to as 
electromagnetic \emph{field strength} and 
electromagnetic \emph{excitation}, respectively,
see \citep{HehlObukhov2003},
or electromagnetic \emph{field} and
electromagnetic \emph{induction},
see \citep{Marmo2005}.

The split relations of \Faraday\ and \Ampere\ electromagnetic two-forms
are expressed by:
%%%%%%%%%%%%%%%%%%%%%%%%%%%%%%%%%%%
$$\vcenter{\halign{
\hfil$#$&$#$\hfil&$#$\hfil&$#$\hfil\cr
\obs\pull\Faraddue&\,=\projE\pull\magcurlttt-\TT\projI\wedge\projE\pull\elefieldttt\,,
\vspace{8pt}\cr
\obs\pull\Amper&\,=\projE\pull\elefluxttt+\TT\projI\wedge\projE\pull\magwindttt
\,.\cr}}$$
%%%%%%%%%%%%%%%%%%%%%%%%%%%%%%%%%%%

The formulation of \Faraday\ induction law is expressed by the closedness of
\Faraday\ \emph{even} two-form $\, \Faraddue\,$, equivalent to
vanishing of its integral on the boundary of any three-dimensional submanifold 
$\,\surf^3_\MMM\subset\MMM^4\,$:
%%%%%%%%%%%%%%%%%%%%%%%%%%%%%%%%%%%
$$\,\ointegrale{\partial\surf^3_\MMM}{}\Faraddue
=\integrale{\surf^3_\MMM}{}\der\Faraddue
\equi\der\Faraddue=\Magnet
\,.$$
%%%%%%%%%%%%%%%%%%%%%%%%%%%%%%%%%%%
In the same way, \Ampere\ induction law is expressed,
in terms of the \emph{odd} two-form $\,\Amper\,$ by the condition:
%%%%%%%%%%%%%%%%%%%%%%%%%%%%%%%%%%%
$$\,\ointegrale{\partial\surf^3_\MMM}{}\Amper
=\integrale{\surf^3_\MMM}{}\Current
\equi\der\Amper=\Current\,,$$
%%%%%%%%%%%%%%%%%%%%%%%%%%%%%%%%%%%
with $\,\Current\,$ called the $\,4$-current.
The manifold $\,\MMM^4\,$ being star-shaped, according to \Poincare\ Lemma
these conditions are equivalent to the closedness properties:
%%%%%%%%%%%%%%%%%%%%%%%%%%%%%%%%%%%
$$\,\der\Current=0\,,\quad\der\Magnet=0\,,$$
%%%%%%%%%%%%%%%%%%%%%%%%%%%%%%%%%%%
which are expressions of the conservation of electric and magnetic charges, respectively.
To esplicate the relation between these conditions and the standard
ones in the three-dimensional \Euclid\ space,
we resort to the split induced by an \Euclid\ observer.

\subsection{Space-time formulation of Faraday law}
\label{sec: FourFaraday}

Let $\,\Bv_{\moto,\ttt}\equaldef\parder\tau\ttt\moto_{\tau,\ttt}\in\TEU\,$
be the spatial velocity of the body, as measured at time $\,\ttt\in\II\,$ by the observer,
we consider the four-velocity field
$\,\Bu_\moto\equaldef\obs\push\coppia{\Bv_{\moto}}{1}\in\TM^4\,$
corresponding to the spatial velocity of the body and to a unit time-velocity.
The electric field and magnetic vortex in the body in motion
are then defined by the following pull-backs, to the space manifold, of the
electromagnetic two-form $\,\Faraddue\,$ in the space-time manifold:
%%%%%%%%%%%%%%%%%%%%%%%%%%%%%%%%%%%
$$\vcenter{\halign{
\hfil$#$&$#$\hfil&$\qquad#$\hfil&$#$\hfil\cr
\elefield
&\,\equaldef-\obs\pull(\Faraddue\punto\Bu_\moto)\in\EXForms^1\di{\TEU\times\TI\sp\Re}\,,
\vspace{8pt}\cr
\magcurl
&\,\equaldef\obs\pull\Faraddue\in\EXForms^2\di{\TEU\times\TI\sp\Re}
\vspace{8pt}\cr
\elefieldttt
&\,\equaldef-\obs_\ttt\pull(\Faraddue\punto\Bu_\moto)\in\EXForms^1\di{\TEU\sp\Re}\,,
\vspace{8pt}\cr
\magcurlttt
&\,\equaldef\obs_\ttt\pull\Faraddue\in\EXForms^2\di{\TEU\sp\Re}
\,.\cr}}$$
%%%%%%%%%%%%%%%%%%%%%%%%%%%%%%%%%%%
Moreover, being $\,\Bu=\obs\push\coppia\Bo{1}\,$
and $\,\obs_\ttt\push\Ba=\obs\push\coppia\Ba{0_\ttt}\,$,
with $\,\Ba\in\TANG_\Bx\EU\,$, and:
%%%%%%%%%%%%%%%%%%%%%%%%%%%%%%%%%%%
$$\,(\obs_\ttt\pull\ff)\di\Bx=(\ff\circ\obs_\ttt)\di\Bx=(\ff\circ\obs)\coppia\Bx\ttt
=(\obs\pull\ff)\coppia\Bx\ttt=(\obs\pull\ff)_\ttt\di\Bx\,,$$
%%%%%%%%%%%%%%%%%%%%%%%%%%%%%%%%%%%
for any $\,\ff:\MMM\mapsto\Re\,$, we have that:
%%%%%%%%%%%%%%%%%%%%%%%%%%%%%%%%%%%
$$\vcenter{\halign{
\hfil$#$&$#$\hfil&$#$\hfil&$#$\hfil\cr
\obs_\ttt\pull\Lieder_{\Bu_\moto}\,\Faraddue\punto\coppia{\Ba}{\Bb}
&\,=\obs_\ttt\pull(\Lieder_{\Bu_\moto}\,\Faraddue\punto\coppia{\obs_\ttt\push\Ba}{\obs_\ttt\push\Bb})
\vspace{8pt}\cr
&\,=(\obs\pull((\Lieder_{\Bu_\moto}\,\Faraddue)\punto\coppia{\obs\push\coppia\Ba{0_\ttt}}{\obs\push\coppia\Bb{0_\ttt}}))_\ttt
\vspace{8pt}\cr
&\,=(\obs\pull(\Lieder_{\Bu_\moto}\,\Faraddue)\punto\coppia{\coppia\Ba{0_\ttt}}{\coppia\Bb{0_\ttt}})_\ttt
\vspace{8pt}\cr
&\,=((\Lieder_{\obs\pull{\Bu_\moto}}\,\obs\pull\Faraddue)\punto\coppia{\coppia\Ba{0_\ttt}}{\coppia\Bb{0_\ttt}})_\ttt
\vspace{8pt}\cr
&\,=((\Lieder_{\coppia{\Bv_\moto}{1}}\,\obs\pull\Faraddue)\punto\coppia{\coppia\Ba{0_\ttt}}{\coppia\Bb{0_\ttt}})_\ttt
\vspace{8pt}\cr
&\,=\Lieder_{\moto,\ttt}\,\magcurltau\punto\coppia\Ba\Bb
\,.\cr}}$$
%%%%%%%%%%%%%%%%%%%%%%%%%%%%%%%%%%%
Let us now recall the basic commutativity property:
$\,\der\circ\obs_\ttt\pull=\obs_\ttt\pull\circ\der\,$
stated in Lemma \ref{lm: extderpushes}.
To simplify the notations, we will denote with the same symbol $\,\der\,$
the exterior derivatives in different manifolds. 
Then the previous formula for the \Lie\ derivative, 
taking account of the homotopy formula of Section \ref{sec: manifolds}:
%%%%%%%%%%%%%%%%%%%%%%%%%%%%%%%%%%%
$$\,\der\Faraddue\punto\Bu_\moto
=\Lieder_{\Bu_\moto}\,\Faraddue-\der(\Faraddue\punto\Bu_\moto)\,,$$
%%%%%%%%%%%%%%%%%%%%%%%%%%%%%%%%%%%
implies that the closedness condition
$\,\der\Faraddue=0\,$ is equivalent to the pair of conditions:
%%%%%%%%%%%%%%%%%%%%%%%%%%%%%%%%%%%
$$\left\{\vcenter{\halign{
\hfil$#$&$#$\hfil&$#$\hfil&$#$\hfil\cr
\obs_\ttt\pull(\der\Faraddue)
&\,=\der(\obs_\ttt\pull\Faraddue)=\der\magcurlttt=0\,,
\vspace{8pt}\cr
\obs_\ttt\pull(\der\Faraddue\punto\Bu_\moto)
&\,=\obs_\ttt\pull(\Lieder_\Bu\,\Faraddue-\der(\Faraddue\punto\Bu_\moto))
%\vspace{8pt}\cr
&\,=\parder\tau\ttt\magcurltau+\der\elefieldttt=0
\,.\cr}}\right.$$
%%%%%%%%%%%%%%%%%%%%%%%%%%%%%%%%%%%
We have thus recovered \Gauss\ law for the magnetic vortex
and \Faraday\ law:
%%%%%%%%%%%%%%%%%%%%%%%%%%%%%%%%%%%
$$\vcenter{\halign{
\hfil$#$&$#$\hfil&$#$\hfil&$#$\hfil\cr
\obs_\ttt\pull(\der\Faraddue\punto\Bu_\moto)
&\,=\obs_\ttt\pull(\Lieder_{\Bu_\moto}\,\Faraddue-\der(\Faraddue\punto\Bu_\moto))
\vspace{8pt}\cr
&\,=\Lieder_{\moto,\ttt}\,\magcurl+\der\elefieldttt=0
\,.\cr}}$$
%%%%%%%%%%%%%%%%%%%%%%%%%%%%%%%%%%%
The previous treatment extends 
classical results, as exposed in \citep[p. 17-19]{Cartan1924},
which dealt with to absence of motion.
In this respect, we underline that partial time derivatives, 
such as the one appearing in the equation of continuity for electric charges,
may be not defined, due to abrupt changes, with respect to time, of the electric charge
at a spatial point crossed by an electrically charged body.

\subsection{Space-time formulation of Ampere law}
\label{sec: FourAmpere}

Turning to the \Ampere\ induction law,
the magnetic winding $\,\magwindttt\,$, 
the electric flux $\,\elefluxttt\,$,
the electric current flux $\,\elecurrformttt\,$, 
the electric charge $\,\elechargeformttt\,$,
and the corresponding time-dependent fields in a body in motion, 
are defined by the pull-backs:
%%%%%%%%%%%%%%%%%%%%%%%%%%%%%%%%%%%
$$\vcenter{\halign{
\hfil$#$&$#$\hfil&\qquad\hfil$#$&$#$\hfil\cr
\magwindttt
&\,=\obs_\ttt\pull(\Amper\punto\Bu_\moto)\,,
&\magwind&\,=\obs\pull(\Amper\punto\Bu_\moto)\,,
\vspace{8pt}\cr
\elefluxttt
&\,=\obs_\ttt\pull\Amper\,,
&\eleflux&\,=\obs\pull\Amper\,,
\vspace{8pt}\cr
-\elecurrformttt
&\,=\obs_\ttt\pull(\Current\punto\Bu_\moto)\,,
&-\elecurrform&\,=\obs\pull(\Current\punto\Bu_\moto)\,,
\vspace{8pt}\cr
\elechargeformttt
&\,=\obs_\ttt\pull\Current\,,
&\elechargeform&\,=\obs\pull\Current
\,.\cr}}$$
%%%%%%%%%%%%%%%%%%%%%%%%%%%%%%%%%%%
By the homotopy formula we have that:
%%%%%%%%%%%%%%%%%%%%%%%%%%%%%%%%%%%
$$\,\der\Current\punto\Bu_\moto
=\Lieder_{\Bu_\moto}\,\Current-\der(\Current\punto\Bu_\moto)\,.$$
%%%%%%%%%%%%%%%%%%%%%%%%%%%%%%%%%%%
Then, observing that $\,\der\circ\obs\pull=\obs\pull\circ\der\,$, 
the pull-back of the exterior derivatives at the r.h.s. may be written as:
%%%%%%%%%%%%%%%%%%%%%%%%%%%%%%%%%%%
$$\left\{\vcenter{\halign{
\hfil$#$&$#$\hfil&$#$\hfil&$#$\hfil\cr
\obs_\ttt\pull(\Lieder_{\Bu_\moto}\,\Current)
&\,=(\obs\pull\Lieder_{\Bu_\moto}\,\Current)_\ttt
=(\Lieder_{\obs\pull\Bu_\moto}\,\obs\pull\Current)_\ttt\,,
\vspace{8pt}\cr
\obs_\ttt\pull\der(\Current\punto\Bu_\moto)
&\,=(\obs\pull\der(\Current\punto\Bu_\moto))_\ttt
=(\der\,\obs\pull(\Current\punto\Bu_\moto))_\ttt
\,.\cr}}\right.$$
%%%%%%%%%%%%%%%%%%%%%%%%%%%%%%%%%%%
Hence, according to Lemma \ref{lm: fourtothree}, the condition $\,\der\Current=0\,$
is equivalent to the pair of conditions:
%%%%%%%%%%%%%%%%%%%%%%%%%%%%%%%%%%%
$$\left\{\vcenter{\halign{
\hfil$#$&$#$\hfil&$#$\hfil&$#$\hfil\cr
\obs_\ttt\pull(\der\Current)
&\,=\der\,(\obs_\ttt\pull\Current)=\der\elechargeformttt=0\,,
\vspace{8pt}\cr
\obs_\ttt\pull(\der\Current\punto\Bu_\moto)
&\,=(\Lieder_{\obs\pull\Bu_\moto}\,\obs\pull\Current)_\ttt
-(\der\,\obs\pull(\Current\punto\Bu_\moto))_\ttt
\vspace{8pt}\cr
&\,=(\Lieder_{\coppia{\Bv_{\moto}}{1}}\,\obs\pull\Current)_\ttt
-(\der\, \elecurrform)_\ttt
\vspace{8pt}\cr
&\,=\Lieder_{\moto,\ttt}\, \elechargeform +\der \elecurrformttt =0
\,.\cr}}\right.$$
%%%%%%%%%%%%%%%%%%%%%%%%%%%%%%%%%%%
The former is a trivial condition,
because the charge form $\,\elechargeformttt\,$ is of maximal order in $\,\EU\,$, 
while the latter is the proper differential expression of the charge conservation law
at each particle of a moving body.

By Lemma \ref{lm: fourtothree}, the \Ampere\ law
$\,\der\Amper=\Current\,$ is equivalent to the pair of conditions:
%%%%%%%%%%%%%%%%%%%%%%%%%%%%%%%%%%%
$$\left\{\vcenter{\halign{
\hfil$#$&$#$\hfil&$#$\hfil&$#$\hfil\cr
\obs_\ttt\pull(\der\Amper)
&\,=\der(\obs_\ttt\pull\Amper)=\der\elefluxttt=\elechargeformttt=\obs_\ttt\pull\Current\,,
\vspace{8pt}\cr
\obs_\ttt\pull(\der\Amper\punto\Bu_\moto)
&\,=(\obs\pull\Lieder_{\Bu_\moto}\,\Amper-\der(\Amper\punto\Bu_\moto))_\ttt
\vspace{8pt}\cr
&\,=(\Lieder_{\coppia{\Bv_{\moto}}{1}}\,\eleflux
-\der\obs\pull(\Amper\punto\Bu_\moto))_\ttt
\vspace{8pt}\cr
&\,=\Lieder_{\moto,\ttt}\,\eleflux-\der\magwindttt
=\obs_\ttt\pull(\Current\punto\Bu_\moto)=-\elecurrformttt
\,.\cr}}\right.$$
%%%%%%%%%%%%%%%%%%%%%%%%%%%%%%%%%%%
The former is \Gauss\ law for the electric displacement flux,
and the latter is \Ampere\ law of induction in \Euclid\ $\,3$-space.

\subsection{Electromagnetic potentials in space-time formulation}
\label{sec: Fourpotential}

In conclusion, we see that the laws of electrodynamic induction
are written and discussed in the simplest way, from the geometric point of view,
when formulated in a $\,4$-dimensional
space-time manifold $\,\MMM^4\,$.
The physical interpretation is however more cryptic than in the standard
$\,3$-dimensional treatment, 
since the familiar picture, provided by the everyday space-time splitting, is lost.

The mathematical expressions
of magnetic and electric charge balance laws in the space-time manifold
are respectively given by:
%%%%%%%%%%%%%%%%%%%%%%%%%%%%%%%%%%%
$$\left\{\vcenter{\halign{
\hfil$#$&$\displaystyle#$\hfil&$#$\hfil&$#$\hfil\cr
\der \Magnet&\,=0
\equi\ointegrale{\partial\BOmega^4_\MMM}{} \Magnet=0\,,
\vspace{8pt}\cr
\der \Current&\,=0
\equi\ointegrale{\partial\BOmega^4_\MMM}{} \Current =0\,,
\cr}}\right.$$
%%%%%%%%%%%%%%%%%%%%%%%%%%%%%%%%%%%
to hold for all $\,4$-dimensional submanifold $\,\BOmega^4_\MMM\subset\MMM^4\,$.

These closedness properties are equivalent to assume that 
absence of bulk sources of magnetic or electric charges is found
by any observer testing the charge balance laws.

By \Poincare\ Lemma, the closedness conditions above are equivalent to the potentiality
requirements:
%%%%%%%%%%%%%%%%%%%%%%%%%%%%%%%%%%%
$$\left\{\vcenter{\halign{
\hfil$#$&$#$\hfil&$#$\hfil&$#$\hfil\cr
\Magnet&\,=\der\Faraddue\,,
\vspace{8pt}\cr
\Current&\,=\der\Amper\,,
\cr}}\right.$$
%%%%%%%%%%%%%%%%%%%%%%%%%%%%%%%%%%%
which have been previously shown to be equivalent to the observer-dependent
formulation of the differential expression of \Faraday\ and \Ampere\ induction laws.
The integral expression are given by:
%%%%%%%%%%%%%%%%%%%%%%%%%%%%%%%%%%%
$$\left\{\vcenter{\halign{
\hfil$\displaystyle#$&$\displaystyle#$\hfil&$\displaystyle#$\hfil&$#$\hfil\cr
\integrale{\BOmega^3_\MMM}{}\Magnet
=\ointegrale{\partial\BOmega^3_\MMM}{}\Faraddue\,,
\vspace{8pt}\cr
\integrale{\BOmega^3_\MMM}{}\Current
=\ointegrale{\partial\BOmega^3_\MMM}{}\Amper\,,
\cr}}\right.$$
%%%%%%%%%%%%%%%%%%%%%%%%%%%%%%%%%%%
to hold for all $\,3$-dimensional submanifold $\,\BOmega^3_\MMM\subset\MMM^4\,$.

It is usually assumed that $\,\Magnet=0\,$, due to the fact that magnetic monopoles
and magnetic currents are still undiscovered.

\Faraday\ law of electromagnetic induction may accordingly be written as:
%%%%%%%%%%%%%%%%%%%%%%%%%%%%%%%%%%%
$$\vcenter{\halign{
\hfil$\displaystyle#$&$#$\hfil&$#$\hfil&$#$\hfil\cr
\ointegrale{\partial\BOmega^3_\MMM}{} \Faraddue =0
\equi 0&\,=\der\Faraddue\equi\Faraddue=\der\Faraduno\,,
\cr}}$$
%%%%%%%%%%%%%%%%%%%%%%%%%%%%%%%%%%%
with the potential one-form $\,\Faraduno\in\EXForms^1\di{\TANG\MMM^4\sp\Re}\,$,
called \emph{electromagnetic potential},
related to the spatial \Faraday\ potential one-form $\,\faradpotttt\in\EXForms^1\di{\TEU\sp\Re}\,$
and to the scalar potential $\,\VV_{\elevector,\ttt}\,$
by the pull-backs:
%%%%%%%%%%%%%%%%%%%%%%%%%%%%%%%%%%%
$$\vcenter{\halign{
\hfil$#$&$#$\hfil&$#$\hfil&$#$\hfil\cr
\faradpotttt&\,=\obs_\ttt\pull\Faraduno\,,
\vspace{8pt}\cr
\VV_{\elevector,\ttt}&\,=\obs_\ttt\pull(\Faraduno\punto\Bu_\moto)
\,.\cr}}$$
%%%%%%%%%%%%%%%%%%%%%%%%%%%%%%%%%%%

\subsection{Galilei invariance}
\label{sec: invGalilei}

In the \Euclid\ $\,3$D space,
invariance of the electric and magnetic forms:
$\,\elefieldttt,\magcurlttt\,$ and $\,\elefluxttt,\magwindttt\,$,
under a change of \Galilei\ observer, 
follows from the space-time representation by observing that 
such a change leaves the map $\,\obs_\ttt\pull\,$ invariant.
Indeed, denoting by $\,\Bw\in\cont^1\di{\EU\sp\TEU}\,$ the uniform relative spatial velocity field
between two \Galilei\ observers,
being the time-like $\,4$-vector field $\,\Bu:\EU\mapsto\TM^4\,$ common to both observers,
the time origin may be assumed to be the same.
Hence the affine sets 
$\,\NNN_\ttt=\obs_{1\ttt}\di\EU=\obs_{2\ttt}\di\EU\,$
of simultaneous events at time $\,\ttt\in\II\,$ are also the same.
Then, setting $\,\rho\coppia\Bx\ttt\equaldef\coppia{\Bx+\ttt\Bw\di\Bx}{\ttt}\,$,
the diffeomorphisms induced by the observers are related by:
%%%%%%%%%%%%%%%%%%%%%%%%%%%%%%%%%%%
$$\,\obs_2=\obs_1\circ\rho\,.$$
%%%%%%%%%%%%%%%%%%%%%%%%%%%%%%%%%%%
The tangent map
$\,\TT_\Bx\rho_\ttt\in\Linmap{\TANG_\Bx\EU\sp\TANG_\Bx\EU}\,$ 
is the identity, being the field $\,\Bw\in\cont^1\di{\EU\sp\TEU}\,$ independent of $\,\Bx\in\EU\,$.
Hence:
%%%%%%%%%%%%%%%%%%%%%%%%%%%%%%%%%%%
$$\,\TT_\Bx\obs_{2\ttt}
=\TT_{\rho_\ttt\di{\Bx}}\obs_{1\ttt}\circ\TT_\Bx\rho_\ttt
=\TT_{\rho_\ttt\di{\Bx}}\obs_{1\ttt}\,,$$
%%%%%%%%%%%%%%%%%%%%%%%%%%%%%%%%%%%
that is $\,\obs_{2\ttt}\pull=\obs_{1\ttt}\pull\,$.
The same argument yields also the \Galilei\ invariance of electric current
and charge forms, $\,\elecurrformttt,\elechargeformttt\,$.

\section{Examples of applications of Faraday law}
\label{sec:EAFL}

\subsection{Material body translating in a uniform magnetic field}
\label{sec: CMSMF}

Let a material body in a translational motion 
$\,\moto\in\cont^1\di{\CORPO\times\II\sp\EU}\,$
with respect to an observer be
crossing a region with a constant value of the spatial magnetic vortex,
according to the standard \Euclid\ connection, so that:
%%%%%%%%%%%%%%%%%%%%%%%%%%%%%%%%%%%
$$\,\nabla\magcurlttt=0\,.$$
%%%%%%%%%%%%%%%%%%%%%%%%%%%%%%%%%%%
Let us first exeven the idea in discursive terms.
The vector potential $\,\potmag_\ttt\,$ associated with
the odd-vector of magnetic vortex $\,\magindttt\,$ may be assumed to have 
transversal circular envelope lines
around the point of a longitudinal axis with the direction of the magnetic field.
Then, at any istant of time, 
the vector potential intensity is linearly varying along any straight line.
Let the body velocity be orthogonal to the magnetic vortex odd vector.
Then, the parallel derivative of the vector potential, along the motion velocity,
will have the direction of the vector potential
and intensity given by the product of half the intensity of the rotor times the intensity
of the velocity.
Taking into account the usual orientations, 
and evaluating the parallel derivative of the magnetic flux potential,
the electric field due to magnetic induction is given by 
one-half the standard expression of the \Lorentz\ force (per unit electric charge):
%%%%%%%%%%%%%%%%%%%%%%%%%%%%%%%%%%%
$$\,\unmezzo\Bv_{\moto,\ttt}\times\magindttt\,.$$
%%%%%%%%%%%%%%%%%%%%%%%%%%%%%%%%%%%
To see this result expressed in formulae, we rely on the expression 
of the \Lie\ derivative of a spatial tensor field in terms of parallel derivatives,
which for a covariant tensor field writes \citep{RomanoDiff2007}:
%%%%%%%%%%%%%%%%%%%%%%%%%%%%%%%%%%%
$$\vcenter{\halign{
\hfil$#$&$#$\hfil&$#$\hfil&$#$\hfil\cr
\Lieder_{\Bv}\,\Baa^\COV
&\,=\nabla_{\Bv}\,\Baa^\COV
+\Baa^\COV\circ\nabla\Bv
+(\nabla\Bv)^*\circ\Baa^\COV
\,,\cr}}$$
%%%%%%%%%%%%%%%%%%%%%%%%%%%%%%%%%%%
and on the following results.
%%%%%%%%%%%%%%%%%%%%%%%%%%%%%%%%%%%
\begin{lemma}[Linear Faraday potential]\label{lem: potlin}
A magnetic vortex field which is spatially constant,
according to the standard connection of the \Euclid\ space,
admits a linear \Faraday\ potential one-form
$\,\faradpotttt\in\EXForms^1\di{\TEU\sp\Re}\,$, that is 
$\,\der\faradpotttt=\magcurlttt\,$ with:
%%%%%%%%%%%%%%%%%%%%%%%%%%%%%%%%%%%
$$\,\faradpotttt
\equaldef\unmezzo\volform\punto\magindttt\punto\identvec
=\unmezzo\magcurlttt\punto\identvec\,,$$
%%%%%%%%%%%%%%%%%%%%%%%%%%%%%%%%%%%
where $\,\volform\,$ is the standard volume form and $\,\identvec\di\Bx\equaldef\Bx\,$.
\end{lemma}
\proof
Being 
$\,\nabla\magcurlttt=0\,$,
$\,\nabla\identvec=\BI\,$, $\,\nabla^*\identvec=\dual\BI\,$,
and recalling that $\,\der\magcurlttt=0\,$,
the homotopy formula (see Section \ref{sec: manifolds})
and the above quoted expression of the \Lie\ derivative in terms of parallel derivative,
give:
%%%%%%%%%%%%%%%%%%%%%%%%%%%%%%%%%%%
$$\,\der\,(\magcurlttt\punto\identvec)
=\Lieder_{\identvec}\,\magcurlttt
=\nabla_{\identvec}\,\magcurlttt
+\magcurlttt\circ\nabla\identvec
+\nabla^*\identvec\circ\magcurlttt
=2\,\magcurlttt\,,$$
%%%%%%%%%%%%%%%%%%%%%%%%%%%%%%%%%%%
which is the formula to be proved.
\endprova
%%%%%%%%%%%%%%%%%%%%%%%%%%%%%%%%%%%
%%%%%%%%%%%%%%%%%%%%%%%%%%%%%%%%%%%
\begin{proposition}[Electric field in a translating body]\label{prop: unmezzo}
A body with a translational motion, across a region of spatially uniform magnetic vortex,
experiences an electric field whose spatial description is given by:
%%%%%%%%%%%%%%%%%%%%%%%%%%%%%%%%%%%
$$\framebox{$-\elefieldttt
=\parder\tau\ttt\faradpottau
+\unmezzo\magcurlttt\punto\Bv_{\moto,\ttt}
+\der\VV_{\elevector,\ttt}\,.$}$$
%%%%%%%%%%%%%%%%%%%%%%%%%%%%%%%%%%%
Let the electric zero-form $\,\VV_{\elevector,\ttt}\,$
have a null gradient. Then, a \Galilei\ observer which measures a time-independent
\Faraday\ one-form $\, \faradpot\,$ in space, will detect, in the translating body,
an electric field which admits a potential and is given by the formula:
%%%%%%%%%%%%%%%%%%%%%%%%%%%%%%%%%%%
$$\vcenter{\halign{
\hfil$#$&$#$\hfil&$#$\hfil&$#$\hfil\cr
\elefieldttt&\,=-\unmezzo\magcurlttt\punto\Bv_{\moto,\ttt}
=\der(\faradpotttt\punto\Bv_{\moto,\ttt})\equi
%\vspace{8pt}\cr
\elevector_\ttt&\,=\unmezzo\,(\Bv_{\moto,\ttt}\times\magindttt)
\,.\cr}}$$
%%%%%%%%%%%%%%%%%%%%%%%%%%%%%%%%%%%
\end{proposition}
\proof
%%%%%%%%%%%%%%%%%%%%%%%%%%%%%%%%%%%
Let us consider a \Galilei\ observer which sees the translational motion 
$\,\moto\in\cont^1\di{\CORPO\times\II\sp\EU}\,$ and measures its velocity 
$\,\Bv_{\moto,\ttt}\equaldef\parder\tau\ttt\moto_{\tau,\ttt}\,$,
which is a uniform field: $\,\nabla\Bv_{\moto,\ttt}=0\,$.
From the formula for the \Lie\ derivative
in terms of parallel derivatives, we get:
%%%%%%%%%%%%%%%%%%%%%%%%%%%%%%%%%%%
$$\vcenter{\halign{
\hfil$#$&$#$\hfil&$#$\hfil&$#$\hfil\cr
\Lieder_{\Bv_{\moto,\ttt}}\,\faradpotttt
&\,=\nabla_{\Bv_{\moto,\ttt}}\,\faradpotttt
+\faradpotttt\circ\nabla\Bv_{\moto,\ttt}
+(\nabla\Bv_{\moto,t})^*\circ\faradpotttt
=\nabla_{\Bv_{\moto,\ttt}}\,\faradpotttt
\,.\cr}}$$
%%%%%%%%%%%%%%%%%%%%%%%%%%%%%%%%%%%
Being $\,\nabla\magcurlttt=0\,$, from Lemma \ref{lem: potlin} we infer that
$\,\faradpotttt=\unmezzo\magcurlttt\punto\identvec\,$
and hence:
%%%%%%%%%%%%%%%%%%%%%%%%%%%%%%%%%%%
$$\vcenter{\halign{
\hfil$#$&$#$\hfil&$#$\hfil&$#$\hfil\cr
\Lieder_{\Bv_{\moto,\ttt}}\,\faradpotttt
&\,=\nabla_{\Bv_{\moto,\ttt}}\,\faradpotttt
=\unmezzo\magcurlttt\punto\Bv_{\moto,\ttt}
\,.\cr}}$$
%%%%%%%%%%%%%%%%%%%%%%%%%%%%%%%%%%%
Then the electric field is given by:
%%%%%%%%%%%%%%%%%%%%%%%%%%%%%%%%%%%
$$\vcenter{\halign{
\hfil$#$&$#$\hfil&$#$\hfil&$#$\hfil\cr
-\elefieldttt
&\,=\Lieder_{\moto,\ttt}\,\faradpotttt+\der\VV_{\elevector,\ttt}
\vspace{8pt}\cr
&\,=\parder\tau\ttt\faradpottau
+\Lieder_{\Bv_{\moto,\ttt}}\,\faradpotttt
+\der\VV_{\elevector,\ttt}
\vspace{8pt}\cr
&\,=\parder\tau\ttt\faradpottau
+\unmezzo\magcurlttt\punto\Bv_{\moto,\ttt}
+\der\VV_{\elevector,\ttt}
\,.\cr}}$$
%%%%%%%%%%%%%%%%%%%%%%%%%%%%%%%%%%%
The term
%%%%%%%%%%%%%%%%%%%%%%%%%%%%%%%%%%%
$\,-\unmezzo\magcurlttt\punto\Bv_{\moto,\ttt}
=-\unmezzo\volform\punto\magindttt \punto\Bv_{\moto,\ttt}
= \unmezzo\metric\punto(\Bv_{\moto,\ttt}\times\magindttt)\,$
%%%%%%%%%%%%%%%%%%%%%%%%%%%%%%%%%%%
is the one-form providing the velocity-dependent part of the electric field
(that is, \emph{force} per unit electric charge)
as detected by an observer sitting on the electromagnets.
The \Faraday\ potential is then seen to vary in time 
according to the time-schedule of the electric current in the electromagnets.
To see that the electric field admits a potential,
we make a comparison between the homotopy formula:
%%%%%%%%%%%%%%%%%%%%%%%%%%%%%%%%%%%
$$\vcenter{\halign{
\hfil$#$&$#$\hfil&$#$\hfil&$#$\hfil\cr
\Lieder_{\Bv_{\moto,\ttt}}\,\faradpotttt
&\,=(\der\faradpotttt)\punto\Bv_{\moto,\ttt}
+\der(\faradpotttt\punto\Bv_{\moto,\ttt})
\vspace{8pt}\cr
&\,=\magcurlttt\punto\Bv_{\moto,\ttt}
+\der(\faradpotttt\punto\Bv_{\moto,\ttt})
\,,\cr}}$$
%%%%%%%%%%%%%%%%%%%%%%%%%%%%%%%%%%%
and the formula
$\,\Lieder_{\Bv_{\moto,\ttt}}\,\faradpotttt=\unmezzo\magcurlttt\punto\Bv_{\moto,\ttt}\,$
which together yield the potentiality property:
%%%%%%%%%%%%%%%%%%%%%%%%%%%%%%%%%%%
$\,-\unmezzo\magcurlttt\punto\Bv_{\moto,\ttt}=\der(\faradpotttt\punto\Bv_{\moto,\ttt})\,$.
%%%%%%%%%%%%%%%%%%%%%%%%%%%%%%%%%%%
\endprova
%%%%%%%%%%%%%%%%%%%%%%%%%%%%%%%%%%%
%%%%%%%%%%%%%%%%%%%%%%%%%%%%%%%%%%%
\begin{remark}\label{rem: galrem}
Let us now consider a starred \Galilei\ observer $\,()^*\,$,
drifted by the translational motion of the material body.
The position vector $\,\Br^*_\ttt\,$ is no more fixed in time but moving with velocity
$\,\Bv_{\relat,\ttt}=-\Bv_{\moto,\ttt}\,$ and the 
\Galilei-invariant magnetic flux potential
is given by:
%%%%%%%%%%%%%%%%%%%%%%%%%%%%%%%%%%%
$$\,\faradpotttt=(\faradpotttt)^*
\equaldef\unmezzo\volform\punto\magindttt\punto\identvec^*_\ttt
=\unmezzo\magcurlttt\punto\identvec^*_\ttt\,.$$
%%%%%%%%%%%%%%%%%%%%%%%%%%%%%%%%%%%
Then 
$\,\nabla_{\Bv_{\relat,\ttt}}\,(\faradpotttt)^*$\,\,$
=-\unmezzo\magcurlttt\punto\Bv_{\moto,\ttt}\,$
and, by Lemma \ref{lm: timerelmot}, the partial time derivative evaluates to:
%%%%%%%%%%%%%%%%%%%%%%%%%%%%%%%%%%%
$$\vcenter{\halign{
\hfil$#$&$#$\hfil&$#$\hfil&$#$\hfil\cr
(\parder\tau\ttt\faradpottau)^*
&\,=\parder\tau\ttt\faradpottau
-\Lieder_{\Bv_{\relat,\ttt}}\,(\faradpotttt)^*
\vspace{8pt}\cr
&\,=\parder\tau\ttt\faradpottau
-\nabla_{\Bv_{\relat,\ttt}}\,(\faradpotttt)^*
\vspace{8pt}\cr
&\,=\parder\tau\ttt\faradpottau
+\unmezzo\magcurlttt\punto\Bv_{\moto,\ttt}
\,,\cr}}$$
%%%%%%%%%%%%%%%%%%%%%%%%%%%%%%%%%%%
Hence, being $\,(\Bv_{\moto,\ttt})^*=0\,$, the electric field takes the expression:
%%%%%%%%%%%%%%%%%%%%%%%%%%%%%%%%%%%
$$\vcenter{\halign{
\hfil$#$&$#$\hfil&$#$\hfil&$#$\hfil\cr
(\elefieldttt)^*
&\,=-\parder\tau\ttt(\faradpottau)^*
+\der\VV_{\elevector,\ttt}
\vspace{8pt}\cr
&\,=-\parder\tau\ttt\faradpottau
-\unmezzo\magcurlttt\punto\Bv_{\moto,\ttt}
+\der\VV_{\elevector,\ttt}
=\elefieldttt
\,.\cr}}$$
%%%%%%%%%%%%%%%%%%%%%%%%%%%%%%%%%%%
This explicit calculation is in accord with the general result about
\Galilei\ invariance of \Faraday\ law of magnetic induction.
\end{remark}
%%%%%%%%%%%%%%%%%%%%%%%%%%%%%%%%%%%
\begin{remark}\label{rem: thomson}
It is manifest that the so-called \Lorentz\ force law is 
contradicted by the previous calculation which instead agrees with the $\,1881\,$ findings
by \persone{J.J.} \Thomson.
His result was subsequently modified by \persone{Oliver} \Heaviside\ in $\,1885-1889\,$
and by \persone{Hendrik Antoon} \Lorentz\ in $\,1892\,$, who eliminated
the factor one-half.
These historical notes, taken from \citep{Darrigol2000}, came to the attention of the author
just after the present theory had been independently developed.
Reading the original papers should help in discovering what reasonings 
were originarily made to get the formula with the one-half factor and for its subsequent elimination.
\end{remark}
%%%%%%%%%%%%%%%%%%%%%%%%%%%%%%%%%%%
%%%%%%%%%%%%%%%%%%%%%%%%%%%%%%%%%%%

\subsection{Bar sliding on rails under a uniform magnetic vortex}
\label{sec: sliding}

Let us consider the problem
concerning the electromotive force \emph{(emf)}
generated in a conductive bar sliding on two fixed
parallel rails under the action of a magnetic vortex which is spatially uniform,
time-independent and complanar.
An observer sitting on the rails measures a time independent \Faraday\ potential field 
and may thus evaluate the \emf\ due to the electric field
distributed along the bar is found by integration
along the line from $\,\Bx_1\,$ to $\,\Bx_2\,$:
%%%%%%%%%%%%%%%%%%%%%%%%%%%%%%%%%%%
$$\,\elefieldttt\punto\Bl=-\unmezzo\magcurlttt\punto\Bv_{\moto,\ttt}\punto\Bl\,.$$
%%%%%%%%%%%%%%%%%%%%%%%%%%%%%%%%%%%
On the other hand, by the integral formula of \Faraday, the total
\emf\ in a circuit, obtained by closing the loop
by another transversal bar fixed to the rails, is evaluated to be:
%%%%%%%%%%%%%%%%%%%%%%%%%%%%%%%%%%%
$$\,\ointegrale{}{}\elefieldttt
=-\ointegrale{}{}\magcurlttt\punto\Bv_{\moto,\ttt}
=-\magcurlttt\punto\Bv_{\moto,\ttt}\punto\Bl\,.$$
%%%%%%%%%%%%%%%%%%%%%%%%%%%%%%%%%%%
So one-half of the total \emf\ is lost as a result of
our previous evaluation of the contribution provided by 
the electric field distributed along the bar.
To resolve this puzzling result we have to consider that,
in this thought experiment, the velocity field is no more uniform in space.
Moreover, being uniform in the bar and vanishing in the rails, it presents two
points of jump discontinuities at the sliding contacts.
Then, the observer sitting on the rails measures the distributed electric field
in the bar, as evaluated before,
plus two impulses of \emf\ concentrated at the sliding contacts, 
whose sum is given by:
%%%%%%%%%%%%%%%%%%%%%%%%%%%%%%%%%%%
$$\,(\faradpotttt\di{\Bx_1}-\faradpotttt\di{\Bx_2})\punto\Bv_{\moto,\ttt}
=-\unmezzo\magcurlttt\punto\Bv_{\moto,\ttt}\punto\Bl\,$$
%%%%%%%%%%%%%%%%%%%%%%%%%%%%%%%%%%%
where $\,\Bx_1,\Bx_2\,$ are the positions of the sliding contacts
and $\,\Bl=\Bx_2-\Bx_1\,$.
Indeed the velocity jumps, in going from $\,1\,$ to $\,2\,$, are 
$\,\Bv_{\moto,\ttt}\,$ and $\,-\Bv_{\moto,\ttt}\,$, respectively.
Thus, the two impulses of \emf\ concentrated at the sliding contacts provide
just the lost one-half of the total \emf\ in the circuit, 
which therefore amounts to
$\,-\magcurlttt\punto\Bv_{\moto,\ttt}\punto\Bl\,$
and is equal to the one previously computed in one stroke by the integral rule of \Faraday.
The instructive problem illustrated above is discussed in 
\cite[C. Moving Loop in Time-Varying Field, Example 9.1, p.$\,375\,$]{Sadiku2010},
by tacitly assuming a \Galilei\ observer sitting on the rails and adopting the \Lorentz\ force
expression. 
The same problem with one bar fixed and the other one translating
is discussed in \cite[II.17.1, fig.17.1]{Feynman1964}
both in terms of the flux rule and in terms of the \Lorentz\ force
(also with a tacit choice of the suitable \Galilei\ observer).
Both analyses, and similar ones in literature,
make no distinction between distributed and concentrated contributions to the \emf\,
and are based on the non-invariant \Lorentz\ force expression.
The right value of the total \emf\ in the circuit is however found,
because the doubled value of the distributed electric field
is equivalent to the addition of the impulses of \emf\ at the sliding contacts.
%%%%%%%%%%%%%%%%%%%%%%%%%%%%%%%%%%%

\subsection{Faraday's paradox}
\label{sec: FaradPar}

\Faraday\ disk:
the classical device is constructed from a brass or copper
disk that can rotate in front of a circular magnet.
The induction EM force between the center of the disk and a point on its rim is measured
by closing the circuit with the aid of brush contacts.
%%%%%%%%%%%%%%%%%%%%%%%%%%%%%%%%%%%
\begin{itemize}\item[-]
$1^{st}$ experiment:
The magnet is held to prevent it from rotating, while the disc is spun on its axis.
The result is that the galvanometer registers a direct current.
\item[-]
$2^{nd}$ experiment:
The disc is held stationary while the magnet is spun on its axis.
The result is that the galvanometer registers no current.
\item[-]
$3^{rd}$ experiment:
The disc and magnet are spun together.
The galvano\-meter registers a current, as it did in step 1.
\end{itemize}
%%%%%%%%%%%%%%%%%%%%%%%%%%%%%%%%%%%
These experiments are commonly referred to as a \emph{paradox} 
as it violates the standard spatial version 
of \Faraday's law of electromagnetic induction.

In fact, according to \cite[II.17.2]{Feynman1964}:
\emph{as the disc rotates, the "circuit",
in the sense of the place in space where the currents are,
is always the same. But the part of the "circuit" in the disc is in material which is moving.
Although the flux through the "circuit" is constant, there is still an EMF, 
as can be observed by the deflection of the galvanometer.
Clearly, here is a case where the $\,\Bv\times\magind\,$ 
force in the moving disc gives rise to an EMF 
which cannot be equated to a change of flux.}
The conviction that there are evidences of failure of
\Faraday's flux rule has been taken for granted in literature
as witnessed by the recent comments in
\cite[6.1.4. p.349]{Lehner2010}.
A perfectly similar situation is provided by the experiment of the \emph{homopolar generator}
where a cylindrical magnet itself is spinning around its axis and two brush contacts,
at the axel and on the rim, are placed to close the conducting circuit.
These and others, real or thought, experiments have repeatedly been proposed in literature
to confirm the possible failure of \Faraday's flux rule.
What really emerges from these examples
is the inadequacy of the standard formulation of the induction law in
which the motion of the material circuit is not taken into account.
\persone{Hering}'s experiment, discussed in \cite[6.1.4. p.349]{Lehner2010},
can be interpreted according to \Faraday's flux rule by observing that there is a circuit
including the galvanometer through which the magnetic flux is vanishing at all times
during the opening phase of the experiment.
An \emf\ is induced between the sliding contacts but this
gives rise to eddy currents in the magnet and not in the controlled circuit.
All these experiments are thus not adducing evidences against \Faraday's flux rule
but rather they warn for a correct interpretation of it.

Let us discuss the \emph{paradox} by applying the formula
for the spatial description of the induced electric field, 
illustrated in Section \ref{sec: elepot}:
%%%%%%%%%%%%%%%%%%%%%%%%%%%%%%%%%%%
$$\vcenter{\halign{
\hfil$#$&$#$\hfil&$#$\hfil&$#$\hfil\cr
\elefieldttt
&\,=-\parder\tau\ttt\faradpottau
-\der(\faradpotttt\punto\Bv_{\moto,\ttt})
-\magcurlttt\punto\Bv_{\moto,\ttt}
+\der \VV_{\elevector,\ttt}
\,.\cr}}$$
%%%%%%%%%%%%%%%%%%%%%%%%%%%%%%%%%%%
In \Faraday\ experiments
the spatial description of the magnetic vortex is time-independent,
when measured by the \Galilei\ observer sitting on 
the support of the disk axis. The same observer will measure also
a time-independent \Faraday\ potential, so that:
$\,\parder\tau\ttt\faradpottau=0\,$ and
a velocity field of the spinning disk which, in terms of the angular velocity
antisymmetric tensor $\,\BW_\ttt=\omega_\ttt\,\BR\,$, is given by:
%%%%%%%%%%%%%%%%%%%%%%%%%%%%%%%%%%%
$$\,\Bv_{\moto,\ttt}\di\Bx=\BW_\ttt\punto\identvec\di\Bx
=\omega_\ttt\,\BR\punto\identvec\di\Bx\,$$
%%%%%%%%%%%%%%%%%%%%%%%%%%%%%%%%%%%
with $\,\BR\,$ rotation of $\,\pi/2\,$ in the disk plane,
$\,\Bx\,$ a radius vector with origin at the disk axis
and $\,\identvec\di\Bx\equaldef\Bx\,$.
Then 
$\,\nabla\Bv_{\moto,\ttt}=\BW_\ttt\,$.
Assuming that the magnetic flux $\,\magcurlttt\,$ 
is spatially constant in the disk,
i.e. $\,\nabla\,\magcurlttt=0\,$,
from Lemma \ref{lem: potlin}
we know that the \Faraday\ potential is given by:
$\,\faradpotttt=\unmezzo\magcurlttt\punto\identvec\,$, 
so that:
%%%%%%%%%%%%%%%%%%%%%%%%%%%%%%%%%%%
$$\vcenter{\halign{
\hfil$#$&$#$\hfil&$#$\hfil&$#$\hfil\cr
\Lieder_{\Bv_{\moto,\ttt}}\,\faradpotttt
&\,=\nabla_{\Bv_{\moto,\ttt}}\faradpotttt
+\faradpotttt\circ\nabla\Bv_{\moto,\ttt}
\vspace{8pt}\cr
&\,=\nabla_{\Bv_{\moto,\ttt}}\,\faradpotttt
+\unmezzo(\magcurlttt\punto\identvec)\circ\BW_\ttt
\,.\cr}}$$
%%%%%%%%%%%%%%%%%%%%%%%%%%%%%%%%%%%
The parallel derivative of the magnetic potential,
being $\,\nabla\,\magcurlttt=0\,$ by assumption, evaluates to:
%%%%%%%%%%%%%%%%%%%%%%%%%%%%%%%%%%%
$$\,2\,\nabla_{\Bv_{\moto,\ttt}}\,\faradpotttt
=\magcurlttt\punto\Bv_{\moto,\ttt}
+\nabla_{\Bv_{\moto,\ttt}}\magcurlttt\punto\identvec
=\magcurlttt\punto\Bv_{\moto,\ttt}\,.$$
%%%%%%%%%%%%%%%%%%%%%%%%%%%%%%%%%%%
For an arbitrary spatial vector field $\,\vecf\,$ in the disk plane, we have that:
%%%%%%%%%%%%%%%%%%%%%%%%%%%%%%%%%%%
$$\vcenter{\halign{
\hfil$#$&$#$\hfil&$#$\hfil&$#$\hfil\cr
2\,\scalar{\Lieder_{\Bv_{\moto,\ttt}}\,\faradpotttt}{\vecf}
&\,=2\,\scalar{\nabla_{\Bv_{\moto,\ttt}}\,\faradpotttt}{\vecf}
+\magcurlttt\di{\identvec,\BW_\ttt\punto\vecf}
\vspace{8pt}\cr
&\,=\magcurlttt\di{\BW_\ttt\punto\identvec,\vecf}
+\magcurlttt\di{\identvec,\BW_\ttt\punto\vecf}
=0
\,,\cr}}$$
%%%%%%%%%%%%%%%%%%%%%%%%%%%%%%%%%%%
being $\,\BR^T=\inv\BR\,$ and hence:
%%%%%%%%%%%%%%%%%%%%%%%%%%%%%%%%%%%
$$\,\magcurlttt\di{\BW_\ttt\punto\identvec,\vecf}
=\omega_\ttt\,\magcurlttt\di{\BR\punto(\BR\punto\identvec),\BR\punto\vecf}
=-\magcurlttt\di{\identvec,\BW_\ttt\punto\vecf}\,.$$
%%%%%%%%%%%%%%%%%%%%%%%%%%%%%%%%%%%
%%%%%%%%%%%%%%%%%%%%%%%%%%%%%%%%%%%
The analysis reveals that the magnetically induced electric vector field in the disk
vanishes identically, when the magnetic vortex in the disk is spatially uniform.
However, to compute the electromotive force in the circuit we must take into account the
jump discontinuity of the velocity at the axis and at the rib brush contacts.
These provide concentrated contributions to the \emf\ whose sum is equal to:
%%%%%%%%%%%%%%%%%%%%%%%%%%%%%%%%%%%
$$\vcenter{\halign{
\hfil$#$&$#$\hfil&$#$\hfil&$#$\hfil\cr
&\,-\faradpotttt\di{\Bx_1}\punto(\BW_\ttt\punto\Bx_1)
+\faradpotttt\di{\Bx_2}\punto(\BW_\ttt\punto\Bx_2)
\vspace{8pt}\cr
=&\,-\unmezzo\magcurlttt\punto\Bx_1 \punto(\BW_\ttt\punto\Bx_1)
+\unmezzo\magcurlttt\punto\Bx_2\punto(\BW_\ttt\punto\Bx_2)
\,.\cr}}$$
%%%%%%%%%%%%%%%%%%%%%%%%%%%%%%%%%%%
The global \emf\ so evaluated is coincident with the one provided by the integral
formula of \Faraday\ for moving bodies, see Section \ref{sec: MatFAR},
when the spinning velocity of the disk radius closing the circuit is taken into account.
Indeed the expression above is exactly equal to the rate at which the area
is spanned by the rotating radius times the magnetic induction.
This formula is evaluated also in \cite[6.1.4. p. 350]{Lehner2010}
where however a doubtful conclusion is drawn about 
whether a fixed or a spinning radius should be considered,
thus sharing the previously quoted opinion of \Feynman.

\section{Discussion}
\label{sec: Disc}

According to our treatment, in both induction laws,
the motion of material particles could be measured by any \Galilei\ observer,
without changing the evaluation of the electric field and of the magnetic winding.
In this respect, confusions are still made in the recent literature,
when dealing with the general laws of electromagnetic induction,
as can be verified by inspecting several exposition of the fundamentals of electromagnetism.

The treatment of \emph{Galileian Electromagnetism}  
by \cite{LeBellacLeblond1972} considers 
two nonrelativistic limits (electric and magnetic)
with arguments based on a non covariant formulation of the 
laws of electromagnetism.

In the introduction and survey of \citep[p.3]{Jackson1999} it is said:
\emph{Also essential for consideration of charged particle motion is the Lorentz force equation, 
$\,\mathbf{F}=q(\elevector+\Bv\times\magind)\,$, 
which gives the force acting on a point charge q in the presence of electromagnetic fields.}
In \Faraday's law of induction \citep[p.209]{Jackson1999}
the electric field is denoted by $\,\elevector'\,$ which is so described, ibid. p.210:
\emph{It is important to note, however, that the electric field $\,\elevector'\,$
is the electric field at $\, \der\textbf{l}\,$ \emph{(an infinitesimal piece of circuit)}
in the coordinate system or medium in which $\,\der\textbf{l}\,$ 
is at rest, since is that field that causes current to flow if a circuit is actually present.}
Then, ibid. p.21,1 in writing:
$\,\elevector'=\elevector+\Bv\times\magind\,$ it is said that
$\,\elevector\,$ is \emph{the electric field in the laboratory} and 
$\,\elevector'\,$ is \emph{the electric field at $\,\der\textbf{l}\,$ in its rest frame of coordinates}.
So an infinite number of observers would be needed to measure $\,\elevector'\,$
in a material circuit in arbitrary motion.
Moreover, how to define univocally the rest frame of reference for an infinitesimal piece of  of circuit?
The same formula is reported in 
\citep[p.71-72]{Post1962}, \citep[p.73]{MisnerThorneWheeler1973},
 \citep[p.88]{Barut1980} and \citep[p.43]{Wegner2003}.
In all these treatments, no convincing strategy is envisaged to choose the 
observer measuring the velocity which appears in the expression of the \Lorentz\ force.

The formula providing the spatial description of \Faraday\ law for mobile circuits is 
reported, without motivations, in \cite[eq.$\,9.16\,$]{Sadiku2010} but
a similar extension to mobile circuits is not considered for \Ampere\ law.
Moreover, ibid. ch. 9.5, the general form of 
\Maxwell\ equations is written according to the classical formulation, 
corresponding to a vanishing material velocity,
and it is literally said:
\emph{it is worthwhile to mention other equations that go hand in hand with Maxwell's equations.
The \Lorentz\ force equation 
$\,\mathbf{F}=q(\elevector+\Bv\times\magind)\,$
is associated with Maxwell's equations. Also the equation of continuity
is implicit in Maxwell's equations.}

In \cite[p.475]{Griffiths1999}, introductory remark to
\emph{Electrodynamics and Relativity}, it is affirmed that:
\emph{Does it \emph{(\Galilei\ principle of relativity)} also apply
to the laws of electrodynamics?
At first glance the answer would seem to be no.}
A discussion, on the effect of relative motion between a conducting loop 
moving with a train and a magnet fixed on the rails, follows, but the whole analysis
contains unmotivated affirmations.

In \cite[p.12-14]{Thide2010},
the electromotive force induced by a magnetic field
on a moving (translating) circuit, is evaluated by means of
the material time-derivative,
(i.e. the sum of the partial time-derivative 
plus the parallel derivative at frozen time
of the spatial description), according to the formula
(in our notations):
%%%%%%%%%%%%%%%%%%%%%%%%%%%%%%%%%%%
$$\vcenter{\halign{
$\hfil\displaystyle#$&$\displaystyle#$\hfil&$\quad\textrm{#}$\hfil\cr
-\der\elefieldttt
=\parder\tau\ttt\magcurltau
+\nabla_{\Bv_{\moto,\ttt}}\, \magcurlttt
\,.\cr}}$$
%%%%%%%%%%%%%%%%%%%%%%%%%%%%%%%%%%% 
In his treatise on Space-Time-Matter (\emph{Raum-Zeit-Materie}) \citep[p.191-192]{Weyl1922},
\persone{Hermann} \Weyl\ attributes this formula to
\persone{Heinrich} \Hertz,
who is credited to have formulated it in \citep{Hertz1892}, see also \citep{Darrigol2000}.
There \Hertz\ formulation is described however as an \emph{ad hoc} modification of 
\Maxwell\ equations motivated by the aim of recovering \Galilei\ invariance.
In fact \Hertz\ modification consisted in substituting the partial time-derivative
with the parallel time-derivative along the motion.
This trick works right in the special instance of \Galilei\ invariance,
since \Lie\ time-derivatives and parallel time-derivatives are coincident
for translational motions, but the formula cannot be assumed as a general
expression of the induction law.
A similar procedure has been reported in \citep{Phipps1993},
who claims to give a proof of the rule,
and in \citep[p.9]{Schwinger1998}.
In this last  the continuity equation for the electric charge is based on the equality
$\,\Bv_{\moto,\ttt}\,\nabla\rho_{\moto,\ttt}=\nabla(\rho_{\moto,\ttt}\Bv_{\moto,\ttt})\,$ 
which is imputed to follow from the property that $\,\Bv_{\moto,\ttt}\,$ is constant in space, 
an unmotivated assertion.

\persone{Richard Phillips} \Feynman\ in
\emph{The Feynman Lectures on Physics} 
\cite[II.17-1]{Feynman1964}, while illustrating \Faraday\ law of induction, says:
\emph{We know of no other place in physics where such a simple and accurate general principle 
requires for its real understanding an analysis in terms of two different phenomena.
Usually such a beautiful generalization  is found to stem from a single deep underlying principle.
Nevertheless, in this case there does not appear to be any such profound implication.
We have to understand the rule as the combined effect of two quite separate phenomena.}
Moreover, ibid. ch. II.17-2, as a comment to the paradoxes of \Faraday\ disk
and of the circuit with rocking contacts, envisaged for discussing the applicability of 
\Faraday\ law of magnetic induction (referred to as the \emph{flux rule}), it is said that:
\emph{The "flux rule" does not work in this case. 
It must be applied to circuits in which the material of the circuit remains the same. 
When the material of the circuit is changing, we must return to the basic laws.
The correct physics is always given by the two basic laws
$\,\mathbf{F}=q(\elevector+\Bv\times\magind)\,$
and $\,\rotor\elevector_\ttt=-\parder\tau\ttt\magind_\tau\,$.}
%%%%%%%%%%%%%%%%%%%%%%%%%%%%%%%%
On the contrary, according to the point of view exposed in the present paper paper, 
neither one of the previous laws can be considered as a basic law of magnetic induction.
The expressions of the \Lorentz\ force law
(with a correction factor one-half) and
of the induction law in terms of partial time-derivative
is simply evaluations of the electric field according to \Faraday\ law,
made by a special observer in special circumstances.
The basic position in the theory is reserved to \Faraday\ law and to the consequent
expression of the electric field in terms of the magnetic potential.
When dealing with \emph{the relativity of magnetic and electric fields}
in \cite[II.13-6]{Feynman1964} it is written:
\emph{When we said that the magnetic force on a charge was proportional to its velocity, 
you may have wondered:
"What velocity? With respect to which reference frame?"
It is, in fact, clear from the definition of $\,\magind\,$ given at the beginning of this chapter 
that what this vector is will depend on what we choose as a reference frame 
for our specification of the velocity of charges.
But we have said nothing about which is the proper frame for specifying the magnetic field.}
\Feynman's answer to the question is based on a subtle relativity argument,
which has however imputed of contradicting conservation of electric charge
\citep{Field2006}.
A relativity argument is also resorted to in the treatment developed in 
\citep[ch.5]{Purcell1985}.
The same approach is taken in a recent book by \cite{Crowell2010}.
Anyway, it is hardly acceptable that experiments in 
classical electrodynamics should require relativistic
arguments for their interpretation. 
Our treatment shows that
the \Galilei\ invariant formulation, the one naturally set up in the present paper,
does the job, without any recourse to special relativity.
\Feynman\ definition of $\,\magind\,$ is based on the 
\Lorentz\ \emph{force} law exerted on an electrically charged body in motion,
a magnetic force which, as he says, has a \emph{strange directional character} 
\cite[II.13-1]{Feynman1964}.
The same approach is taken in \citep[ch.6]{Purcell1985}.
In this respect the treatments, of moving conductors or dielectrics
in magnetic fields, performed in \citep{{LandauLifshits1984}} should also be consulted.
These views concerning the \Lorentz\ \emph{force} law originate from the treatment
given by \persone{Hermann} \Weyl\ in his treatise 
on \emph{Raum-Zeit-Materie} \citep[p.191-192]{Weyl1922}.

The recent treatment of classical electrodynamics in \citep{HehlObukhov2003}
is performed in terms of differential forms and adopts the elegant
and synthetic geometric approach in the $\,4$-dimensional space-time manifold.
However body motions are still ignored and,
in the expression of the induction laws,
partial time derivatives at fixed points in the \Euclid\ space are considered 
instead of \Lie\ time-derivatives along the body motion, 
with the consequence that the laws of induction are not covariant
and hence \Galilei\ invariance does not follows.

In \citep[sec.$\,8\,$]{Kovetz2000}, when illustrating \Faraday\ law, 
the magnetic induction flux is considered through a \emph{fixed}, \emph{open} surface.
An \emph{open} surface probably there stands for a surface with boundary,
but the meaning of \emph{fixed} is not (and could hardly be) clarified.
In \citep[ch.$\,8.2\,$]{Sadiku2010} the force acting on an electrically 
charged particle is said to be the sum of two terms. 
The former is the electric field and the second is the \Lorentz\ force due
to the magnetic induction and to the charged body velocity.
But the electric field is just defined as the field providing the force acting on 
the unit point charge, so that a contradiction is apparent. 
The only way of picking the electric field out of the total force would indeed be
to consider a fixed charged body, but again fixed with respect to what \Galilei\ observer?
A critical discussion on \Lorentz\ force is reported by \cite{ThomasSmid},
although in somewhat na\"ive terms.
The intrinsic strangeness of \Lorentz\ law and the unanswered question about what 
\Galilei\ observer is measuring the body velocity,
both quoted by \Feynman,
may be overcome, as illustrated in this paper,
by considering the correct form of the magnetic induction law
for moving material circuits.
The electric field in a body in motion in a magnetic field 
is found to be independent of the \Galilei\ observer. 
This formulation results in a confirmation of the classical treatments provided by
\cite{Maxwell1861} and \cite{Thomson1893},
but neglected in the subsequent literature.
The usual introduction, on an experimental basis, 
of the \Lorentz\ \emph{force} law appears to be untenable, being non \Galilei-invariant.
The evaluation of the electric field acting on a charged particle in motion
through a region of uniform magnetic vortex, 
performed in Section \ref{sec: CMSMF} on the basis of \Faraday\ law of induction, 
leads to conclusion that the standard expression must be corrected by a multiplicative factor one-half 
and completed by the addition of the negative time-rate of the magnetic vortex potential.
Thus a \Galilei-invariant expression of the electric field is got, 
to within the differential of a \Galilei-invariant electric scalar potential.
This analysis shows that the evaluation of the electric field generated by magnetic induction
cannot be expressed by a simple pointwise formula 
(like the \Lorentz\ \emph{force} law)
but requires instead the determination of the \Faraday\ potential field 
and of the scalar electric potential field,
a much harder task, in general.
The analysis,
performed in Sections \ref{sec: sliding} and \ref{sec: FaradPar},
of two well-known examples of a magnetically induced \emf\, puts moreover into evidence
that due attention to jump discontinuities of the velocity field
must be paid, to evaluate concentrated impulses of the induction 
\emf\ there located.

The $\,4$D formulation of the electromagnetic induction laws,
in terms of conservation laws of two basic tensor fields,
was proposed by \cite{Bateman1910} on the basis of earlier work by \cite{Hargreaves1908}.
The theory is illustrated in detail in \citep[Ch. F]{TruesdellToupin1960}.
A formulation in terms of differerential forms was provided in \citep{MisnerThorneWheeler1973}
and has recently been revisited in the textbook 
\citep{HehlObukhov2003} on the foundations of classical electrodynamics.
A geometry treatment of electromagnetism in space-time, 
with a careful distinction between even and odd forms,
has been contributed in
\citep{Marmo2005,MarmoTulczyjew2006}.
\textcolor{red}{A new feature of the approach developed in Sect. \ref{sec: Fourdimensional} 
of the present paper,
is that a generalized \Lorentz\ \emph{force} relation is 
introduced as a further assumption,
see the second of the six axioms, in \citep[B2 p.121]{HehlObukhov2003}
}while no additional law is introduced in the present theory,
wherein only the induction laws and the constitutive relations are considered as basic.
Constitutive relations in the four-dimensional formalism have been recently considered in 
\citep{Marmo2005} and in \citep{Lindell2006}.

\section{Conclusions}
\label{sec: Co}

The fundamentals of electromagnetism
have been revisited by a proper formulation of the electromagnetic induction laws 
for material bodies in motion.
We emphasize that considering the motion of a body is an unavoidable
task since the absence of motion would imply a restriction to consideration of a 
translating body as seen by an observer sitting on it.
Then, bodies in relative translational motion and, more in general, deforming bodies,
which are dealt with in everyday engineering applications of electromagnetism,
would be ruled out.
The differential geometric approach, performed in terms of integrals of exterior forms,
leads to a formulation involving the \Lie\ time-derivative, along the spatial motion, 
of the magnetic vortex (\Faraday) 
and of the electric displacement flux (\Ampere).
The well-posedness conditions 
(independence of the considered surfaces and of their spatial motion)
have been investigated and explicated in terms of balance laws. 
\Galilei\ invariance of the new form of the induction laws
is discussed and assessed in terms of \Lie\ time-derivatives along the motions.
The \Lorentz\ force law, concerning the non-\Galilei\ invariant 
\emph{force} acted by a magnetic vortex 
upon a moving electrically charged particle, usually introduced
as an independent axiom motivated by experience,
has been critically addressed.
The Galilei\ invariant formula, which differs by a one-half 
multiplicative factor in the velocity dependent term and by an additional term
expressing the time-rate of the magnetic vortex potential, has been deduced as
a direct consequence of \Faraday\ law, when applied to the
detection of the electric field induced in a body translating 
in a region of spatially uniform magnetic vortex.
Constitutive relations have been
briefly discussed in the \Euclid\ framework.
The formulation of electromagnetism in the four-dimensional space-time affine manifold
has been extended to moving bodies, thus providing the most clear picture of
the following fundamental result.
The balance laws for the electric and the magnetic charges,
expressed by the closedness conditions on two electromagnetic $\,3$-forms,
are equivalent to the laws of electromagnetic induction which state
the existence of corresponding potential $\,2$-forms.
Motions of the involved bodies have been taken into account by 
considering the description provided by an observer. 
\Galilei\ invariance of the induction laws is a natural consequence of the 
observer-independent space-time formulation.

%\newpage

\end{document}